\documentclass[final,5p,twocolumn]{elsarticle}

\usepackage{amssymb}
\usepackage{amsmath}
\usepackage[caption=true]{subfig}
\usepackage{multirow}
\usepackage{comment}
\usepackage{booktabs}
\usepackage{fancyvrb}
\usepackage{enumerate}
\usepackage{algpseudocode}
\usepackage{float}
\usepackage[colorlinks=true,linkcolor=black, citecolor=blue, urlcolor=blue]{hyperref}
\usepackage{tikz,pgfplots}
\usepackage{breqn}
\usepackage{bbm}
\usepackage{cleveref}
\usepackage{mathtools}
\usepackage{rotating}
\usepackage[linesnumbered,ruled,vlined]{algorithm2e}
\usepackage{balance}
\pgfplotsset{%
    compat=newest, 
    tick label style={font=\footnotesize},
    label style={font=\small},
    legend style={font=\small},
    axis x line = center,
    axis y line = center,
    every axis/.style={pin distance=1ex},
    trim axis left
    } 

\numberwithin{equation}{section}
\newtheorem{remark}{Remark}

\journal{ }

\usepackage{pifont}

\numberwithin{theorem}{section}

\begin{document}

\begin{frontmatter}

\title{Comparing the market microstructure between two South African exchanges}

\author[uct-sta]{Ivan Jericevich}
\ead{jrciva001@myuct.ac.za}
\author[uct-sta]{Patrick Chang}
\ead{chnpat005@myuct.ac.za}
\author[uct-sta]{Tim Gebbie}
\ead{tim.gebbie@uct.ac.za}
\address[uct-sta]{Department of Statistical Sciences, University of Cape Town, Rondebosch 7700, South Africa}

\begin{abstract}
We consider shared listings on two South African equity exchanges: the Johannesburg Stock Exchange (JSE) and the A2X Exchange. A2X is an alternative exchange that provides for both shared listings and new listings within the financial market ecosystem of South Africa. From a science perspective it provides the opportunity to compare markets trading similar shares, in a similar regulatory and economic environment, but with vastly different liquidity, costs and business models. A2X currently has competitive settlement and transaction pricing when compared to the JSE, but the JSE has deeper liquidity. In pursuit of an empirical understanding of how these differences relate to their respective price response dynamics, we compare the distributions and auto-correlations of returns on different time scales; we compare price impact and master curves; and we compare the cost of trading on each exchange. This allows us to empirically compare the two markets. We find that various stylised facts become similar as the measurement or sampling time scale increase. However, the same securities can have vastly different price responses irrespective of time scales. This is not surprising given the different liquidity and order-book resilience. Here we demonstrate that direct costs dominate the cost of trading, and the importance of competitively positioning cost ceilings. Universality is crucial for being able to meaningfully compare cross-exchange price responses, but in the case of A2X, it has yet to emerge in a meaningful way due to the infancy of the exchange --- making meaningful comparisons difficult.  
\end{abstract}


\end{frontmatter}

\tableofcontents

\section{Introduction}\label{sec:intro}

The price response of a trade has long been central to the study of market microstructure \cite{BGPW2004,Gatheral2010,HKG2020,Kyle1985}. Price response, or specifically market impact, continues to have practical interest because of the centrality it holds within the Almgren and Chriss framework \cite{AC2000} for managing the cost of trading. However, it is also important for understanding universal properties that may be shared across markets, and across assets within markets \cite{LFM2003}. 

The key phenomenological tool used to frame potential universality in the context of market impact is via the ``master curve" --- a single curve that aims to describe, and summarise, the dominant relationships between trade related price changes, and the volume of those trades, across sectors, securities and markets. Master curves have been observed in many different markets where the aim has been to try calibrate or identify signatures of universality by comparing the power law exponents for these liquidity adjusted aggregated price impact curves (see \cite{HHGD2016,LC2005,WCBH2015,Zhou2012}). Irrespective of whether this is a theoretically meaningful exercise or not; there remain surprising similarities across very different markets, and very different assets within markets; these continue to suggest some degree of universality. The question of what constitutes sufficient liquidity or trading volume at which such comparisons become possible (if at all) remains unclear; or at what level of market maturity does apparent universality emerge? 

Including market impact directly into portfolio optimisation in order to combine execution and portfolio risk within a single real-time dynamic portfolio analysis framework has increasingly become necessary given the increased dependence on algorithmic execution and trading. However, in practice much of the focus has been on the indirect costs of trading often approximated using the convenient square-root formula (see \citet{Gatheral2010}). 

Another common approach has been to assume a fixed basis point (bps.) per day lost due to combined costs, often without accounting for ceiling costs \cite{NMTG2019}, market feed-backs and other dynamics. An accurate account for the full cost is of particular importance when testing the efficacy of trading strategies when trying to adjust for the size of orders because of the feed-backs experienced when trading in real markets. Moreover, it becomes more complex when one can buy and sell the same security on different exchanges with different direct costs --- then accounting for the full cost of trading is critical {\it e.g.} when pre-training (or batch training) reinforcement learning algorithms for order execution \cite{HW2014ARL}. First, one would like meaningful ``rules-of-thumb'' to allow one to error check and make informed decisions that do not necessitate complete market simulation prior to live-trading. Second, one would like to be able to aggregate price response data to better account for measurement errors and noise. Using master curve techniques is one such approach as it may allow one to extend price impact modelling from liquid to less liquid assets. Third, one would like to better understand the applicability of arguments about universal features when comparing markets and assets when traded by informed and purposeful strategic agents.

Here we focus on price impact, which is the immediate price response, which we will then try to use to compare master curves between two exchanges in the South Africa market --- markets that trade similar shares within a similar regulatory and economic environment. The master curve approach taken here is similar to that of \citet{HHGD2016}. Master curves have also briefly been contrasted in prior work comparing BRICs markets \cite{Nonyane2019} as an extension to an initial analysis of the Johannesburg Stock Exchange (JSE) \cite{DuPreez2015}. The aim here is to better understand, or gain insight, into the minimum liquidity requirements that may be pre-requisites before one may reasonably identify universality in and across markets. The work here demonstrates the lack of maturity, in the sense of liquidity, within the A2X market at the current time, but also demonstrates how quickly the market has matured from being a juvenile market to one that is evolving to share key properties with a mature markets. Although this work focuses primarily on self response, we keep in mind the interest that has grown for cross-correlated price response functions. We argue that at this stage of A2X's development, the cross-correlations between the stocks in this market will not be strong enough to have any detectable effect due to the lack of liquidity. Hence, our focus here relates to visualising price response in two very different exchanges. Nonetheless, we are mindful of the extent of the work carried out in mature markets \cite{BMEB2017,HKG2020,WSG2016b,WSG2016a}.

Many studies of price response have disregarded the interactions between different order flows and focused on the dynamics of single stocks to address the self-responses (due to the high self-correlation of order-flow). These approximations potentially lead to the under-estimation of trading costs and are generally not realistic since investors typically have diversified portfolios of many assets that are traded simultaneously. Recent studies investigate the price response of one stock to the trades of other stocks in a correlated market to address the cross-responses \cite{WSG2016a}. \citet{BMEB2017} also argues in favour of this by introducing a multivariate linear propagator model to describe such a structure and account for the significant fraction of the covariance of stock returns. 

Here transactions are found to mediate a significant part of the correlation between different instruments. \citet{WSG2016b} studies this further by performing different averages to identify active and passive cross-responses. Average cross–responses for a given stock are evaluated either with respect to the whole market or to different sectors. Related to all these, \citet{HKG2020} provides a comparison of different measures of price response and how they modify the results. More specifically, they compare responses calculated on the trade time scale and the physical time scale by making the assumption that mid-prices in the trade time scale are the same as that in the physical time scale. Results are found to be qualitatively similar for the two definitions of time scales. Additionally, the dominating contribution of immediate price response directly after a trade is made clear as they find that delayed response are suppressed. Without trade liquidity and market maturity these types of comparisons are not empirically feasible without additional theoretical assumptions. We do not consider the fundamental issue of cross-impact and the related issues of time scales further. 

Here we consider South African market microstructure of two related exchanges, the \href{https://www.jse.co.za/}{JSE} and \href{https://www.a2x.co.za/}{A2X}; central to the structure of any market are the regulatory framework, IT infrastructure, and fee structure.

The JSE was established in 1887 and has grown to be the largest stock exchange in Africa with 342 listings and a combined market cap of over R14trn as of August 2020. On the JSE, market participants began to industrialise algorithmic trading and then more widely adopt high-frequency trading methods when the JSE matching engine moved from London to Johannesburg in 2012. A2X, on the other hand, is a relatively new exchange founded in 2014.\footnote{The founders were Sean Melnick, Ashley Mendelowitz and Kevin Brady in October 2014.} A2X went live three years later in 2017. The stated aim of A2X was to: try improve execution, lower transaction and clearing fees, and try to push local market structure towards a federated multi-venue framework that may narrow spreads, allow crossing or facilitation trades to occur at lower costs, while aspiring to increase market activity. As of June 2020, A2X has 37 listed securities with a combined market cap of over R2.2trn \cite{a2xcosts,a2xrules,jserules,jsecosts}.

The regulatory framework that determines the market microstructure in South Africa is summarised in the Financial Markets Act (2012) \cite{FMA2012}, the JSE rules and directives \cite{JSEequitydirectives,JSEequityrules}, and the Financial Intelligence Centre (FIC) Act (2001) \cite{FICA2001}. Regulatory oversight is maintained by the Financial Sector Conduct Authority (FSCA)\footnote{The FSCA was established in 2018 \cite{FSCAurl}.} which is responsible for enhancing market conduct by regulating and supervising institutions {\it e.g.} to identify market abuse such as insider trading and market manipulation in the financial sector of South Africa. The South African Reserve Bank (SARB)\cite{SARBurl} is responsible for maintaining price and financial stability. Within this ecosystem, \href{https://www.a2x.co.za/faq/}{A2X} acts as a secondary listing venue and therefore relies on these bodies as well as partly on the \href{https://www.jse.co.za/services/market-regulation}{JSE} to provide the regulation on the issuers and traders of securities. 

On the technology side: the JSE uses the MillenniumIT trading systems \cite{MilleniumITurl} following the approach of the London Stock Exchange with the BDA broker-dealer clearing system \cite{BDAurl}, while A2X has licensed the systems for its core exchange platform from Aquis Exchange \cite{Aquisurl}. The core systems include the matching engine, the surveillance system, and an integrated clearing platform. Aquis Exchange PLC is authorised and regulated by the UK Financial Conduct Authority (FCA) \cite{FCAurl}. Matching engines allow for a high number of fast simultaneous trading connections, high transaction capacity, and are designed to process a high volume of messages per day. The latency of both A2X and JSE is lower than many existing globally important exchanges. A2X only provides the ability to trade in equities but boasts lower fees and greater transparency. 

In terms of fee structuring: A2X has transaction fees as well as clearing and settlement fees; these are lower than those of the JSE. A2X has 0.29 bps settlement fees subject to a ceiling of R154 per trade (excluding VAT\footnote{Value Added Tax.}). Passive and aggressor transaction fees of 0.2bps and 0.4bps per trade, respectively. Both being subject to a ceiling limit of R355 per trade (excluding VAT) \cite{a2xcosts}. In contrast, the JSE offers transaction fees with 0.48 bps and a ceiling limit of R420.4 per trade (excluding VAT). Clearing and settlement fees have 0.36bps with a ceiling of R180 per trade (excluding VAT) \cite{jsecosts}.

Simulating the entire financial market ecosystem for trade strategy testing and risk management is appealing because of the mechanistic complexity of the market structure and costs and the nonlinear feed-backs and interactions that multiple interacting agents bring to the market ecosystem. This type of simulation can be both a costly, as well as computational expensive exercise; it is tractable, and widely used for system verification {\it e.g.} in vendor provided test market venues. However, a key component remains the realism of the underlying trading agents and their interactions within test markets. 

The realism and applicability of the counter-party order-flows, the simulated order-book, and trade-event scenarios that algorithms face in many simulated test environments remains highly questionable. Hence the importance of reasonable, but sufficiently representative, low costs approaches to understanding strategy behaviour and impact. Researchers developing trading strategies do not want to expose their strategies to competitors in shared testing environments, while sufficiently realistic multi-agents simulation environments remain illusive, particularly for low liquidity and collective behaviour based risk-event scenarios. It is for these reasons that we think better understanding signatures of universality and the boundary of their applicability remain an important question --- both from a risk management and strategy development perspective.

Considering, for example, the interplay between the JSE, A2X and the various regulations, costs, business models and technology dependencies, it is worthwhile to have a mechanistic understanding of each market, their similarities and their differences. This makes the repeated cataloguing of empirical findings and market stylised facts as they evolve and change through time both interesting and useful, from both a theoretical and practical standpoint. They provide a minimalist catalogue describing the relationship between liquidity, market design and cost structures, all in the presence of actual strategic agents governed by various regulations and technical mechanisms that simulated environments would need to be able to faithfully reproduce to make strategy simulation statistically realistic. This type of data-science activity is also crucial to help the regulators concerned with the organisation of liquidity in electronic markets and various issues raised by high-frequency trading \cite{abergel2012market}. Additionally, understanding cross-market universality and dependencies is necessary for risk management when trading on multiple trading venues, as well as managing optimal trade execution services. To this end, we compare stylised facts for the same securities listed on both exchanges and compare the two exchanges as a whole. The exchanges are compared through their respective price response dynamics and the cost associated with trading.

The paper is structured as follows: \Cref{sec:SF} compares the stylised facts of microprice returns for the same security between the two exchanges at different time scales. \Cref{sec:comp} compares the two exchanges as a whole by comparing the immediate price impact and master curves, and by comparing the various cost components and total cost of trading on each exchange. \Cref{sec:conc} ends with some concluding remarks. Additionally, the appendices are structured as follows: \ref{app:A2X} and \ref{app:JSE} explains how the datasets are constructed using raw message data and commercial data vendor datasets respectively. \ref{app:tickers} lists the various securities used when comparing the markets. \ref{sec:class} investigates the efficacy of various trade classification rules. \ref{app:vis} investigates the lack of liquidity in the A2X order book and finally, \ref{app:comp} compares the order-flow auto-correlation and intraday seasonality between the exchanges.

\section{Comparing stylised facts} \label{sec:SF}

\begin{figure*}[htb]
    \centering
    \subfloat[A2X]{\label{fig:MPRetACF:a}\includegraphics[width=0.48\textwidth]{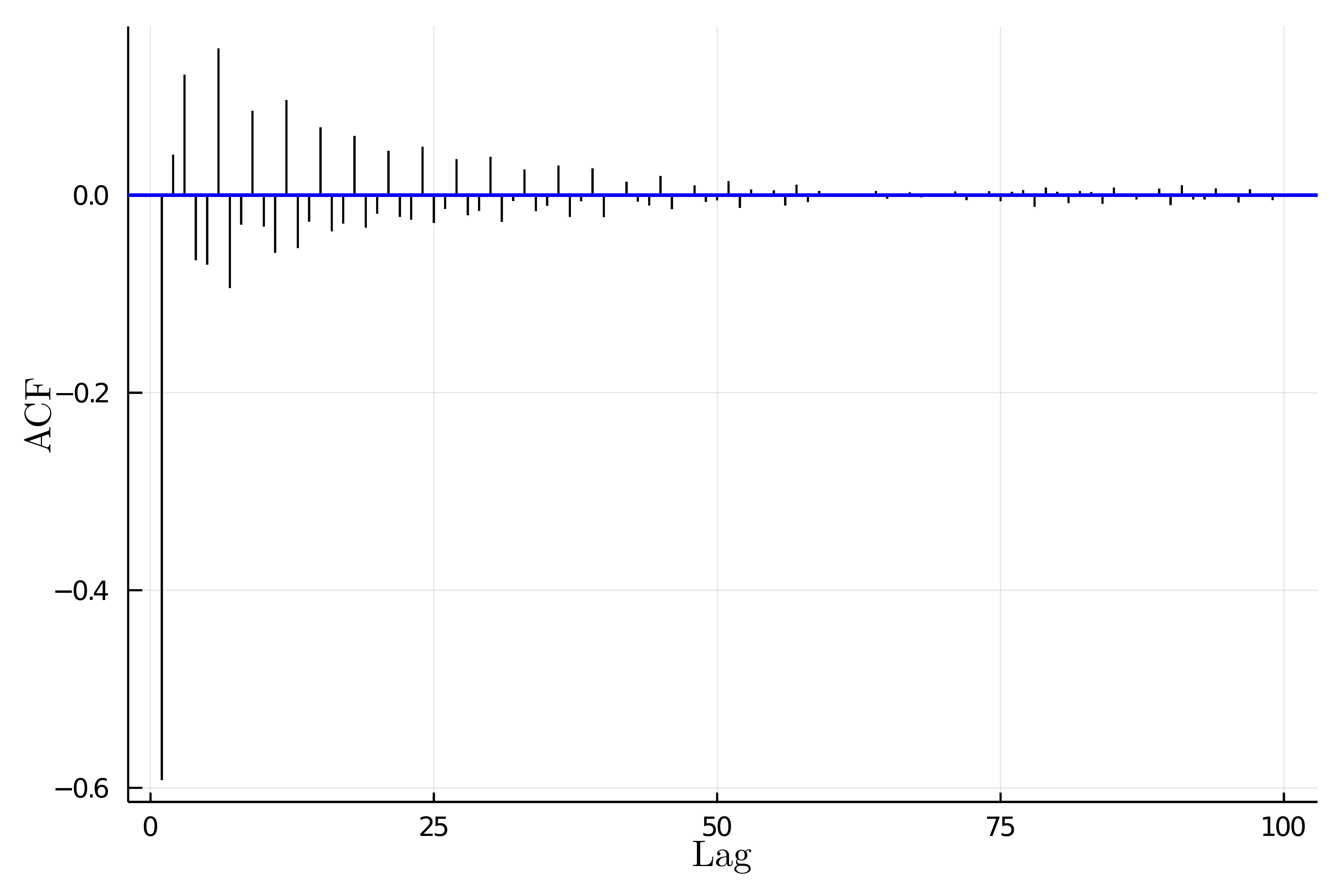}}
    \subfloat[JSE]{\label{fig:MPRetACF:b}\includegraphics[width=0.48\textwidth]{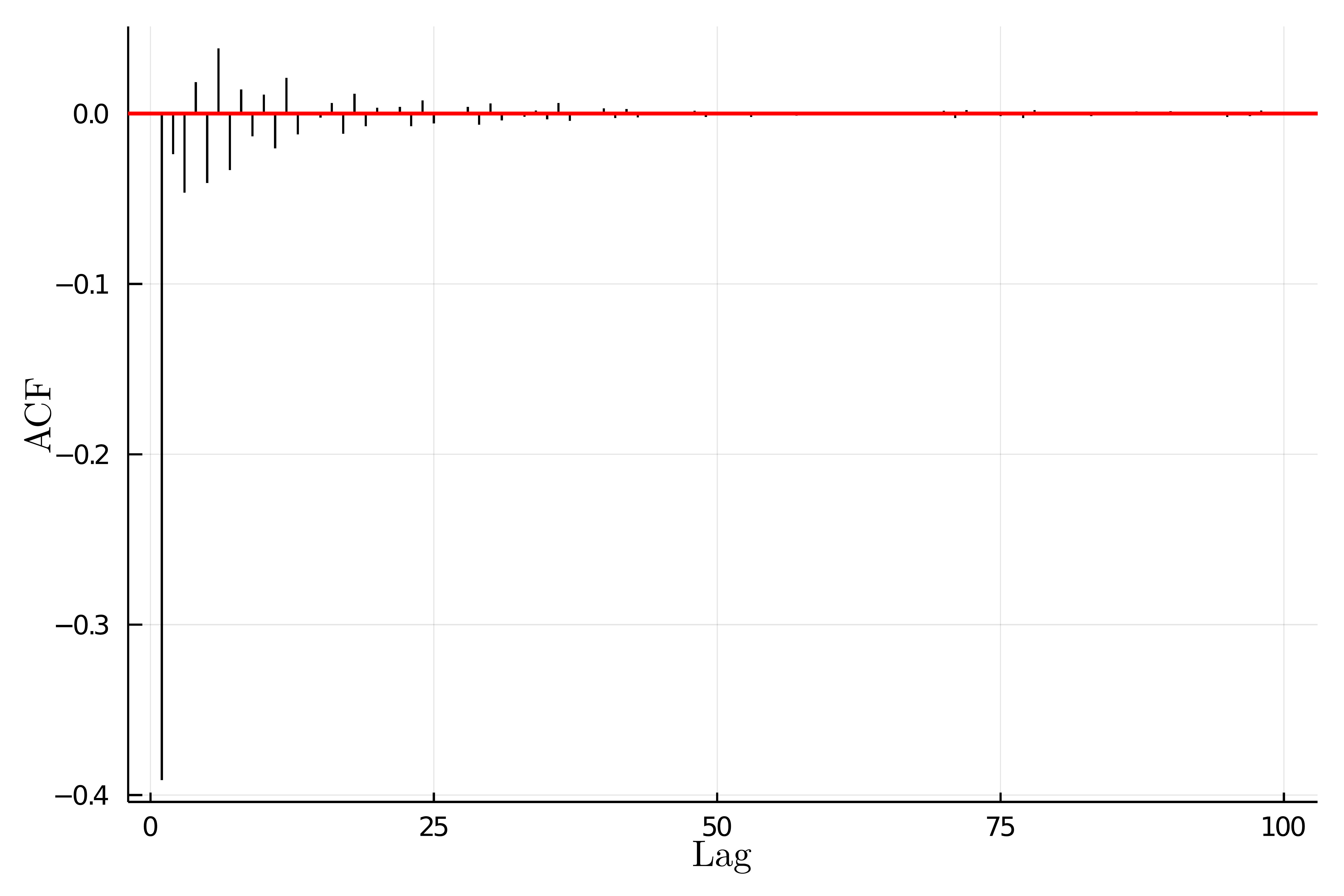}} 
    \caption{Tick-by-tick microprice return auto-correlation for Naspers on the (a) A2X exchange and the (b) Johannesburg Stock Exchange. We compute up to 100 lags.}
\label{fig:MPRetACF}
\end{figure*}

\begin{figure*}[htb]
    \centering
    \subfloat[Left tail]{\label{TickDist:a}\includegraphics[width=0.33\textwidth]{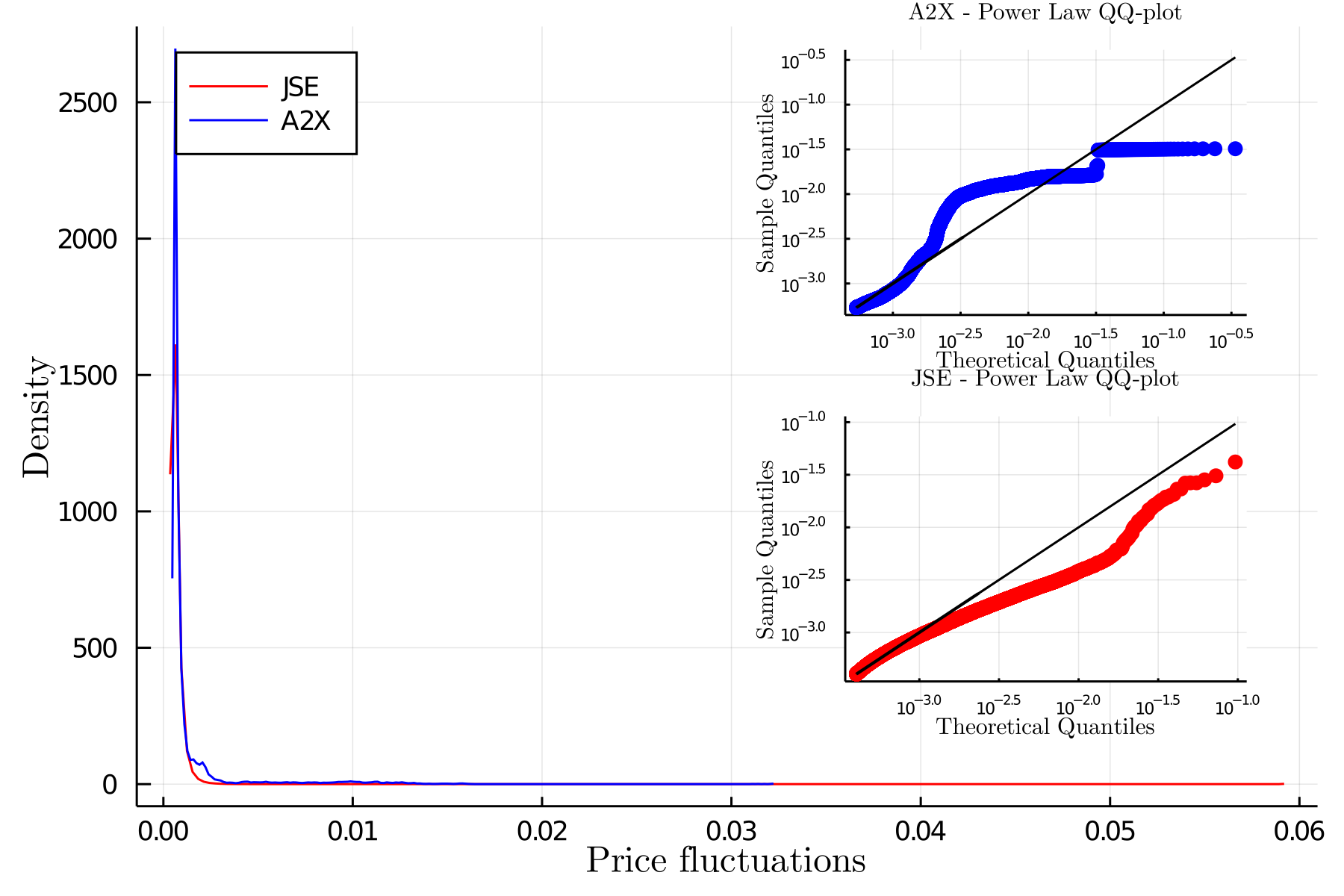}}
    \subfloat[Full distribution]{\label{TickDist:b}\includegraphics[width=0.33\textwidth]{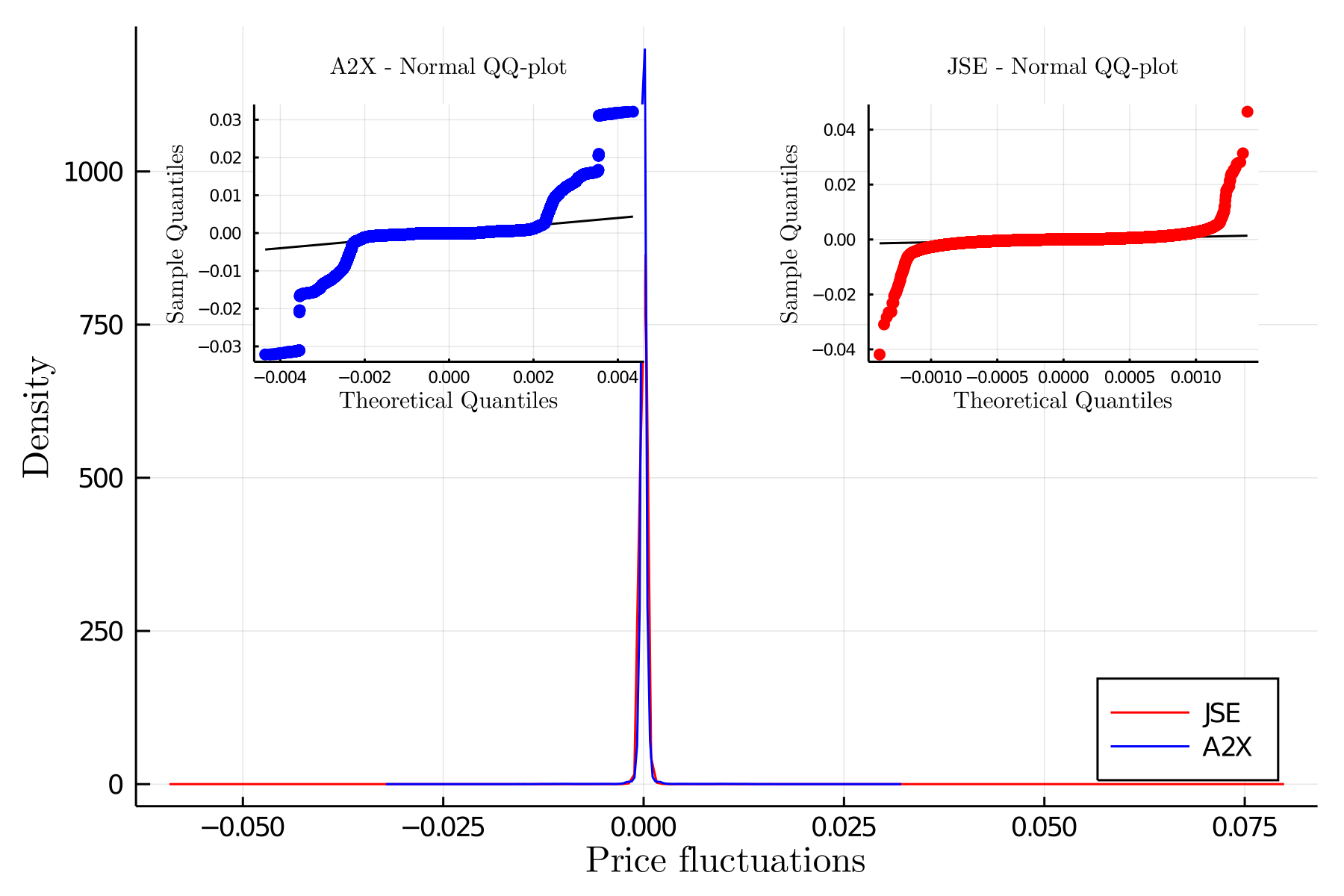}}
    \subfloat[Right tail]{\label{TickDist:c}\includegraphics[width=0.33\textwidth]{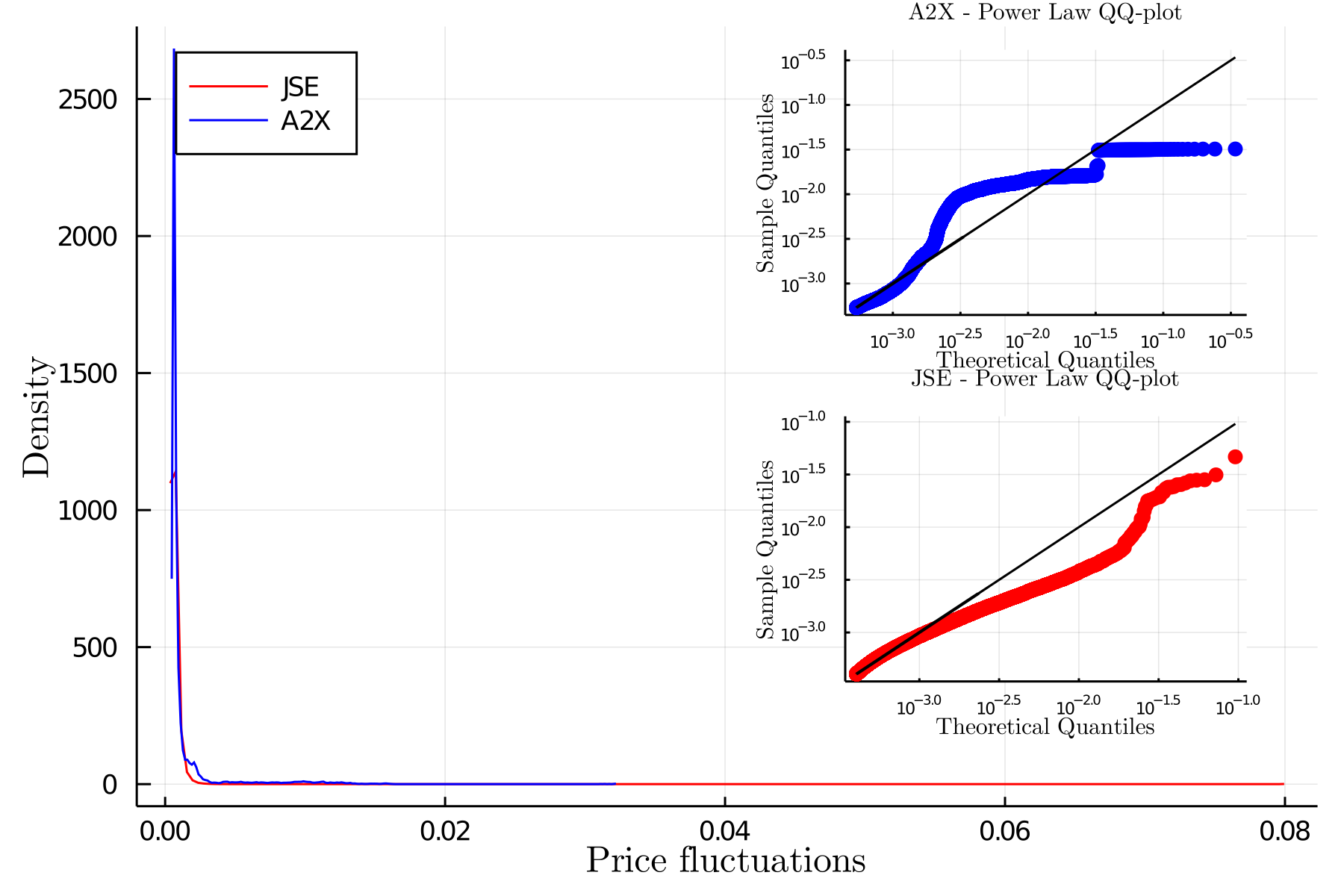}}
    \caption{Tick-by-tick microprice return distributions for Naspers. The left and right tails have QQ-plots fitted to a power-law distribution provided as insets on a log-log scale whereas the full distribution has QQ-plots fitted to a Normal distribution provided as insets. A2X is given in blue and JSE in red.}
\label{fig:TickDist}
\end{figure*}

Empirical studies on financial time series often suggest seemingly different time series share statistical similar characteristics that are consistent through time, across markets and securities \cite{SC2019}. Our interest here are less ambitious, here we directly compare stylised facts by considering the same security that is listed on different exchanges. We will compare the microprice returns and stylised facts relating to this, such as strong mean reversion and heavy/fat tails for the distributions as compared to a normal distribution.

To avoid repetition, the results presented in this section are for Naspers Ltd (NPN) over the period starting on 2019-01-02 until 2019-07-15; chosen so that our two data sets overlap. Returns are computed on a tick-by-tick scale as well as on bar data at 1, 10 and 20 minute granularities. Tail distribution plots are obtained by plotting the histogram of returns above the 95th percentile (upper tail) and below the 5th percentile (lower tail).

The stylised facts included are by no means exhaustive, but rather are indicative. There are several restrictions preventing a more comprehensive comparison, {\it e.g.} raw A2X message data is recorded at the nanosecond time scale, while the JSE data is aggregated data captured from Bloomberg PRO and is down-sampled in time to seconds. This means that we are unable to compare the distributions of inter-arrivals between the exchanges. A like-for-like analyses below one second would require access to the JSE UDP raw market feed message data.\footnote{It is here that we must thank A2X for their openness and transparency in providing us with access to raw message data snap-shots for non-commercial research purposes --- this allows us to access the correct time-stamps associated with transactions as they actually occurred, even though we cannot easily map these transactions consistently to actual trader ID's.}

\subsection{Tick-by-tick data}\label{ssec:TAQ}

Here we consider the returns of tick-by-tick microprices. The microprice is updated whenever an event changes the best bid or ask (see \ref{app:A2X} and \ref{app:JSE}). The returns are computed as price fluctuations:
\begin{equation}\label{eq:MPrets}
    r_{t_k} = \log\left( S_{t_{k+1}} \right) - \log\left( S_{t_{k}} \right),
\end{equation}
where $S_{t_{k}}$ is the microprice as time ${t_{k}}$.

We expect to see strong negative first order auto-correlations --- these are an indication of a strong mean reverting component \cite{PCJ2015}. \Cref{fig:MPRetACF} plots the auto-correlation of the microprice returns for both exchanges. 

We investigate up to 100 lags. We find that the first lag presents the largest magnitude in auto-correlation, confirming what is seen in prior literature. Of interest here is that a large portion of the lags are significant. There appears to be no strong theoretical reasons to observe such patterns, and hence they may be sample specific. We also note a difference in magnitude of the first order auto-correlation between the two exchanges, with the magnitude in A2X being larger than that in the JSE. We again speculate that this is due to the significant difference in relative liquidity and order book resilience. 

Next, we investigate the distribution of the tick-by-tick returns. \Cref{fig:TickDist} plots the full distribution of the returns with quantile to quantile (QQ) plots fitted to a normal distribution provided as insets. The figure also includes the distributions of the left tail (below the 5th percentile) and the right tail (above the 95th percentile) with QQ-plots fitted to a power-law distribution provided as insets (presented on a log-log scale). The distributions are fitted using maximum likelihood estimation (MLE). This means the parameters for the normal distribution are:
\begin{equation}
    \begin{aligned}
    & \hat{\mu} = \frac{\sum_{i=1}^n x_i}{n} = \bar{x},  
    & \hat{\sigma}^2 = \frac{1}{n} \sum_{i=1}^n \left( x_i - \bar{x} \right)^2,
    \end{aligned}
\end{equation}
where $x_i$ are the price fluctuations.

The power-law distribution uses the specification provided by \citet{CSR2009} where the probability density function is as:
\begin{equation}
    p(x) = \frac{\alpha - 1}{x_{\text{min}}} \left( \frac{x}{x_{\text{min}}} \right)^{-\alpha},
\end{equation}
where $x \geq x_{\text{min}} > 0$. The complementary cumulative distribution is:
\begin{equation}
    1 - F(x) = \left( \frac{x}{x_{\text{min}}} \right)^{-\alpha + 1}.
\end{equation}
Here we only estimate $\alpha$ because $x_{\text{min}}$ is set as the cutoff percentile. The MLE for $\alpha$ is given as:
\begin{equation}
    \hat{\alpha} = 1+n\left[\sum_{i=1}^{n} \ln \frac{x_{i}}{x_{\min }}\right]^{-1}.
\end{equation}
We note that there are better approaches to pick $x_{\text{min}}$. One approach is to pick $x_{\text{min}}$ such that the distance between the probability distribution of the measured data and the fitted power law is minimised, whereby the distance is quantified using the Kolmogorov-Smirnov statistic. For more details we refer the interested reader to \citet{CSR2009}.

Considering \Cref{fig:TickDist}, we note that the full distribution is highly leptokurtic with very heavy tails. This is consistent with what is found with \citet{PCJ2015}. However, the tail distributions between the exchanges are very different. We see that the JSE returns seem to closely follow a power-law distribution, as expected in the literature \cite{PCJ2015}. However, the returns from A2X seem to reach a maximum deviation and flattens out at the extreme tails --- again, we speculate that this is due to the lack of order-book resilience and limited number of trading agents interacting through the illiquid order-book. Tick-by-tick microprices do not seem to have the same properties, even for the same equity, on the different exchanges.

\subsection{Bar-data}\label{sec:bar}

\begin{figure*}[p]
    \centering
    \subfloat[A2X 1 min]{\label{bar:a}\includegraphics[width=0.33\textwidth]{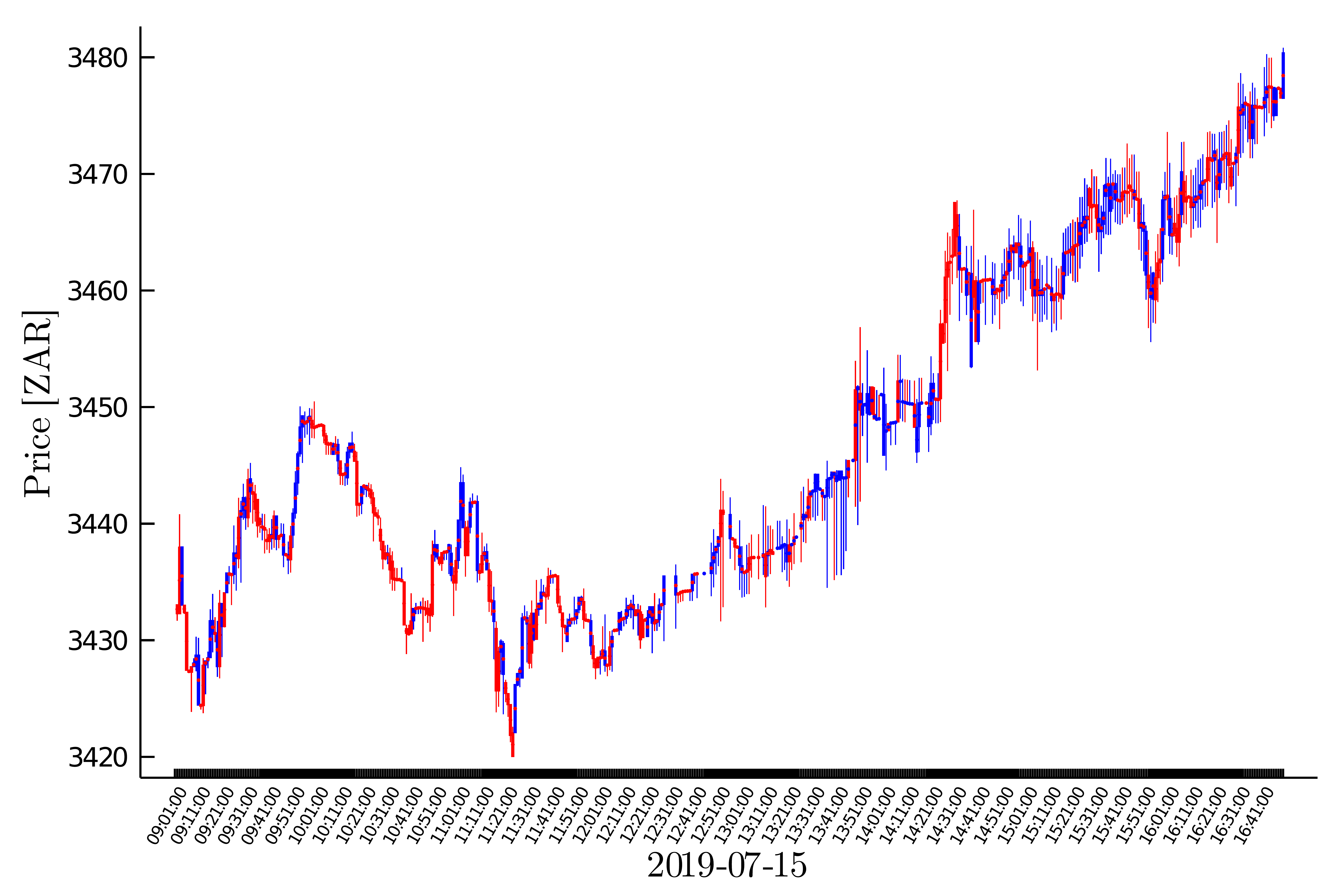}}
    \subfloat[A2X 10 min]{\label{bar:b}\includegraphics[width=0.33\textwidth]{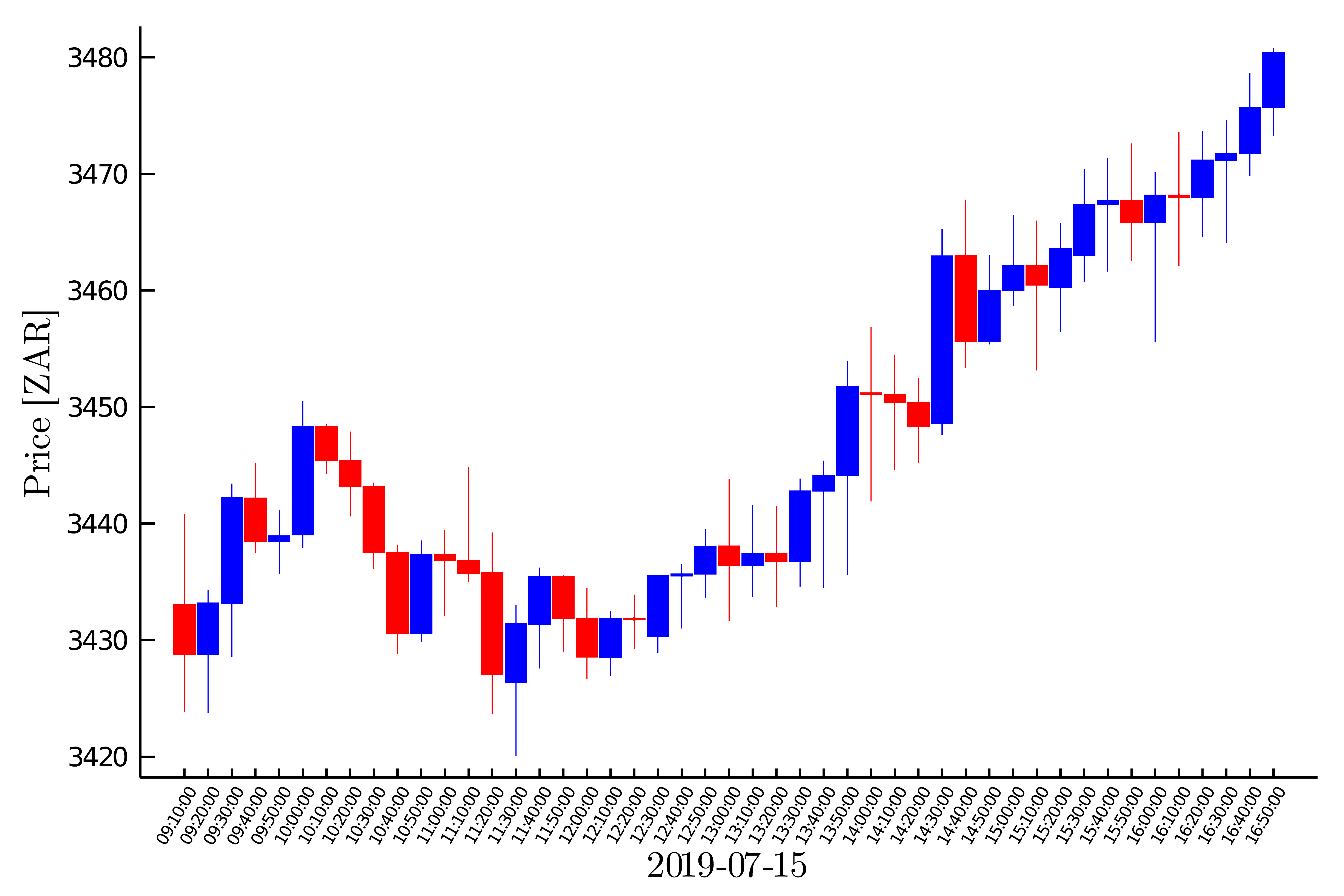}}
    \subfloat[A2X 20 min]{\label{bar:c}\includegraphics[width=0.33\textwidth]{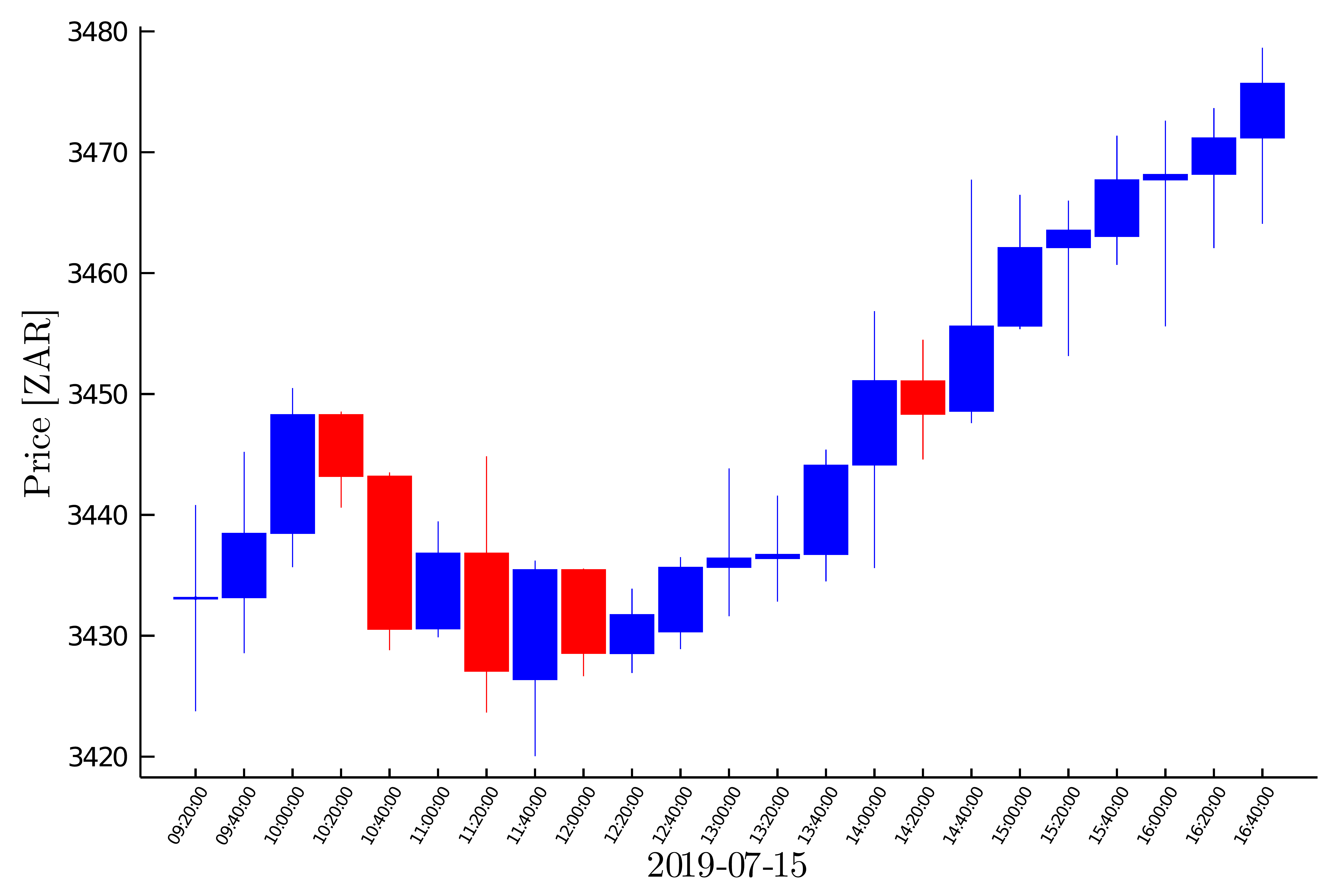}}    \\
    \subfloat[JSE 1 min]{\label{bar:d}\includegraphics[width=0.33\textwidth]{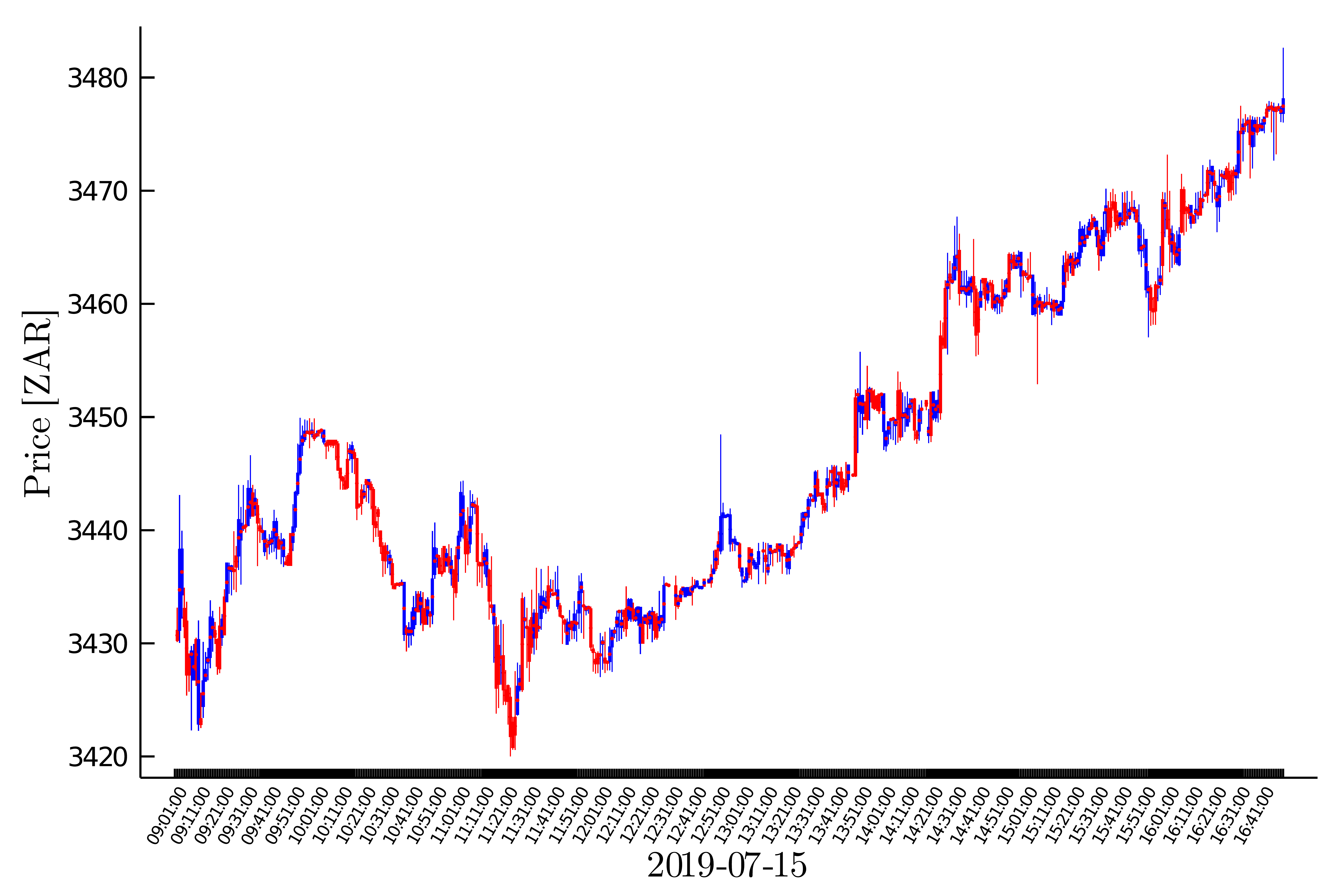}}
    \subfloat[JSE 10 min]{\label{bar:e}\includegraphics[width=0.33\textwidth]{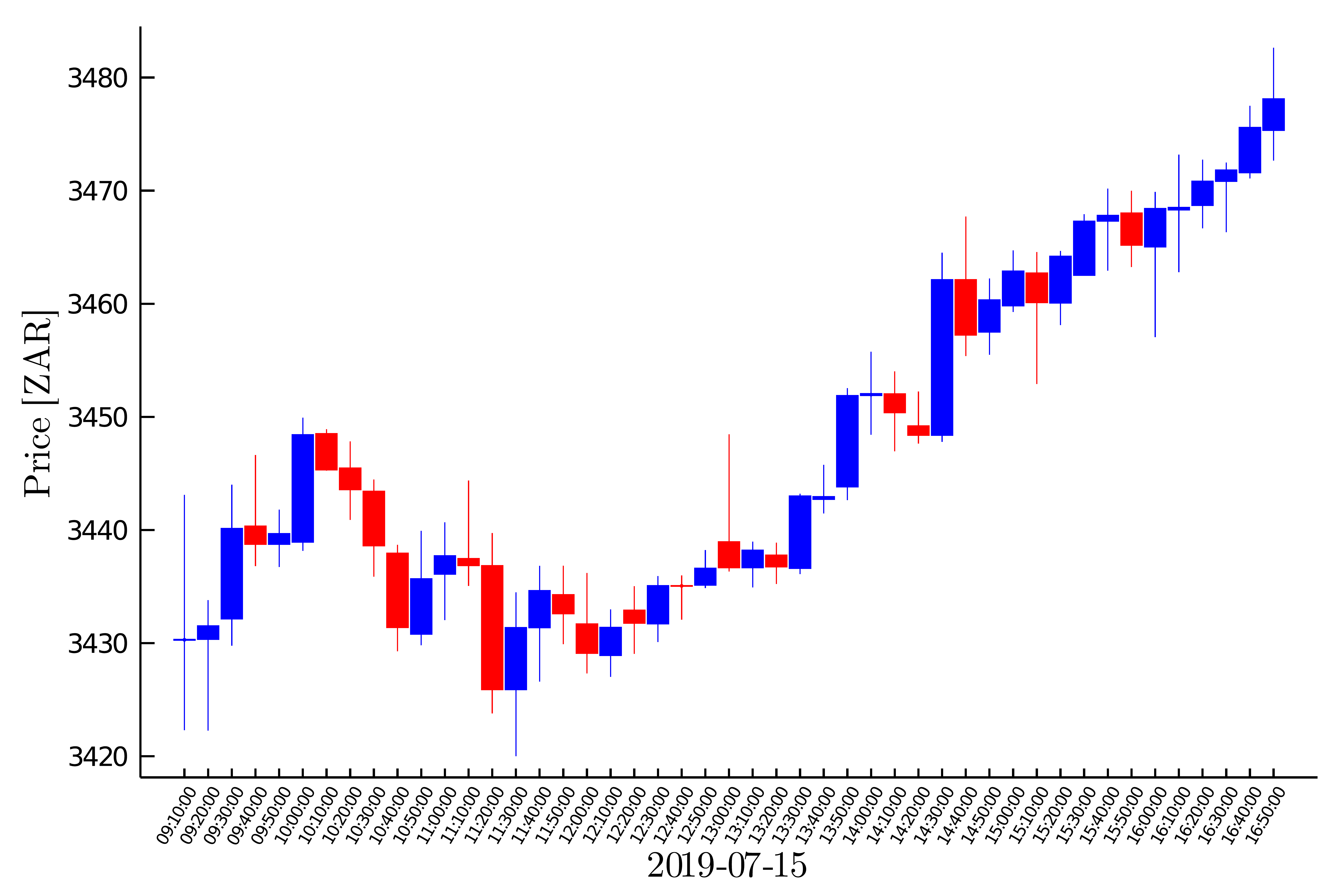}}
    \subfloat[JSE 20 min]{\label{bar:f}\includegraphics[width=0.33\textwidth]{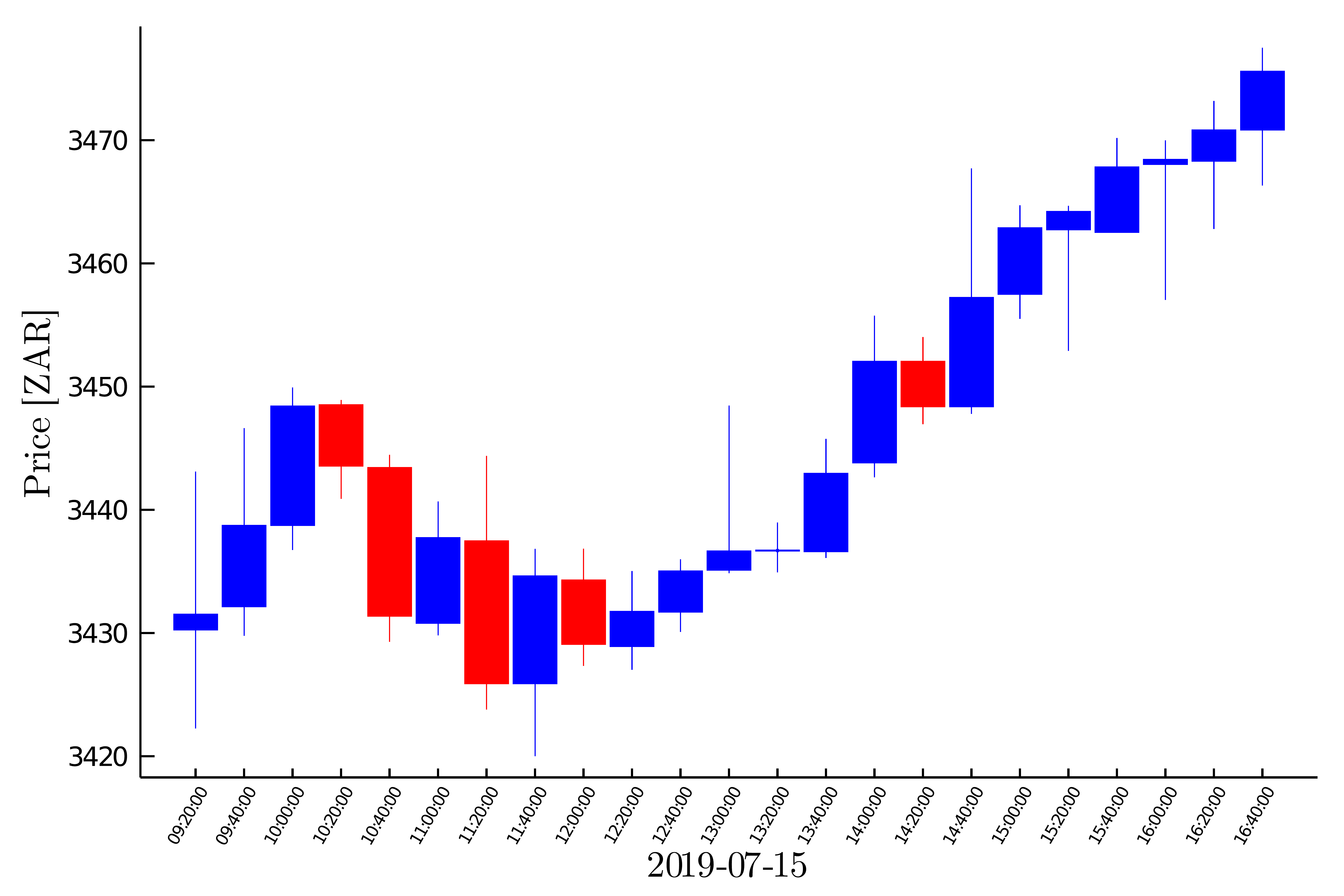}}
    \caption{Candlestick charts for Naspers microprice bar data over different bar sizes and different exchanges on 2019-07-15. The first row is the A2X exchange while the second row is the JSE. From the first to third column we have 1-minute, 10-minute and 20-minute bars respectively.}
\label{fig:bar}
\end{figure*}

\begin{figure*}[p]
    \centering
    \subfloat[A2X 1 min]{\label{baracf:a}\includegraphics[width=0.33\textwidth]{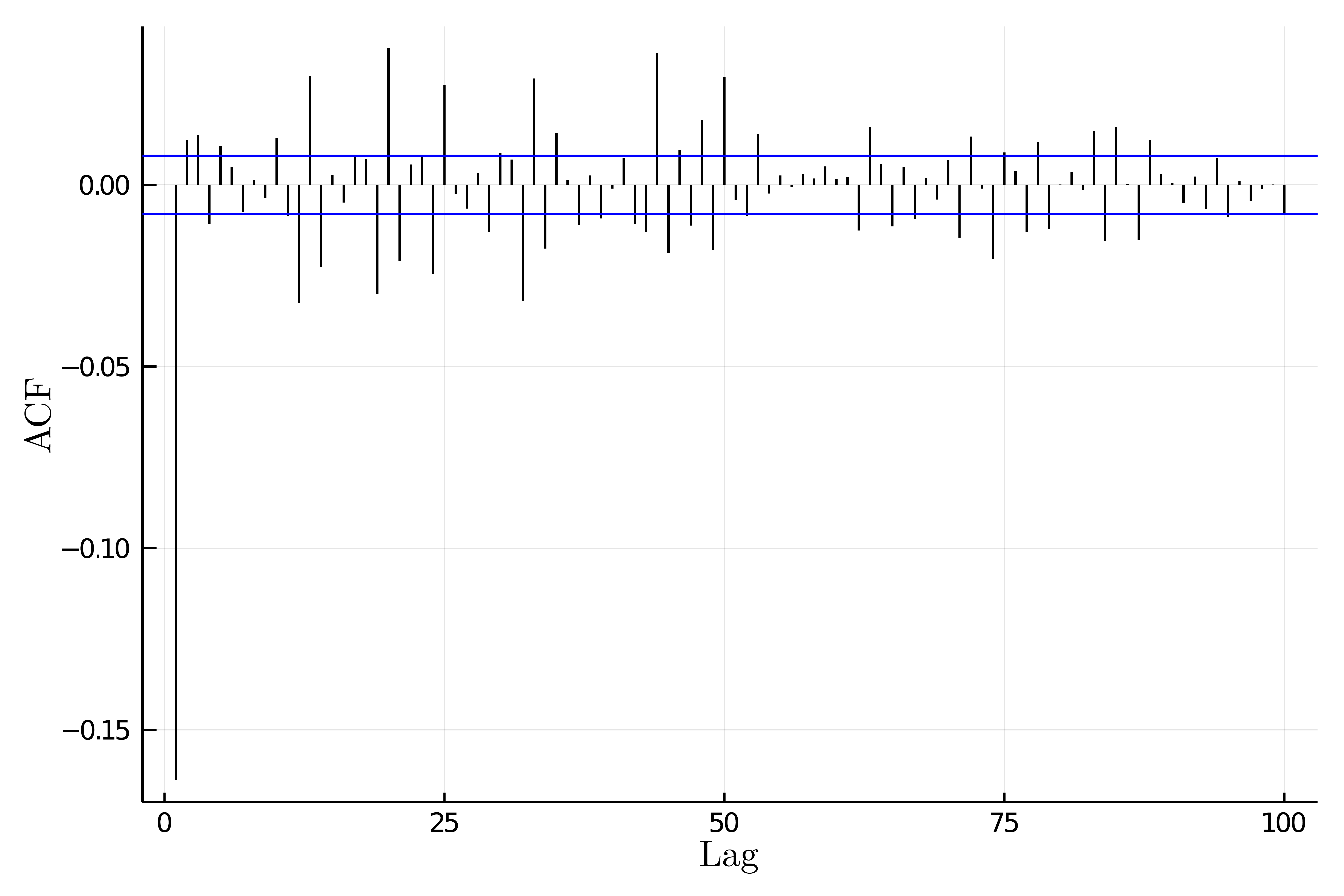}}
    \subfloat[A2X 10 min]{\label{baracf:b}\includegraphics[width=0.33\textwidth]{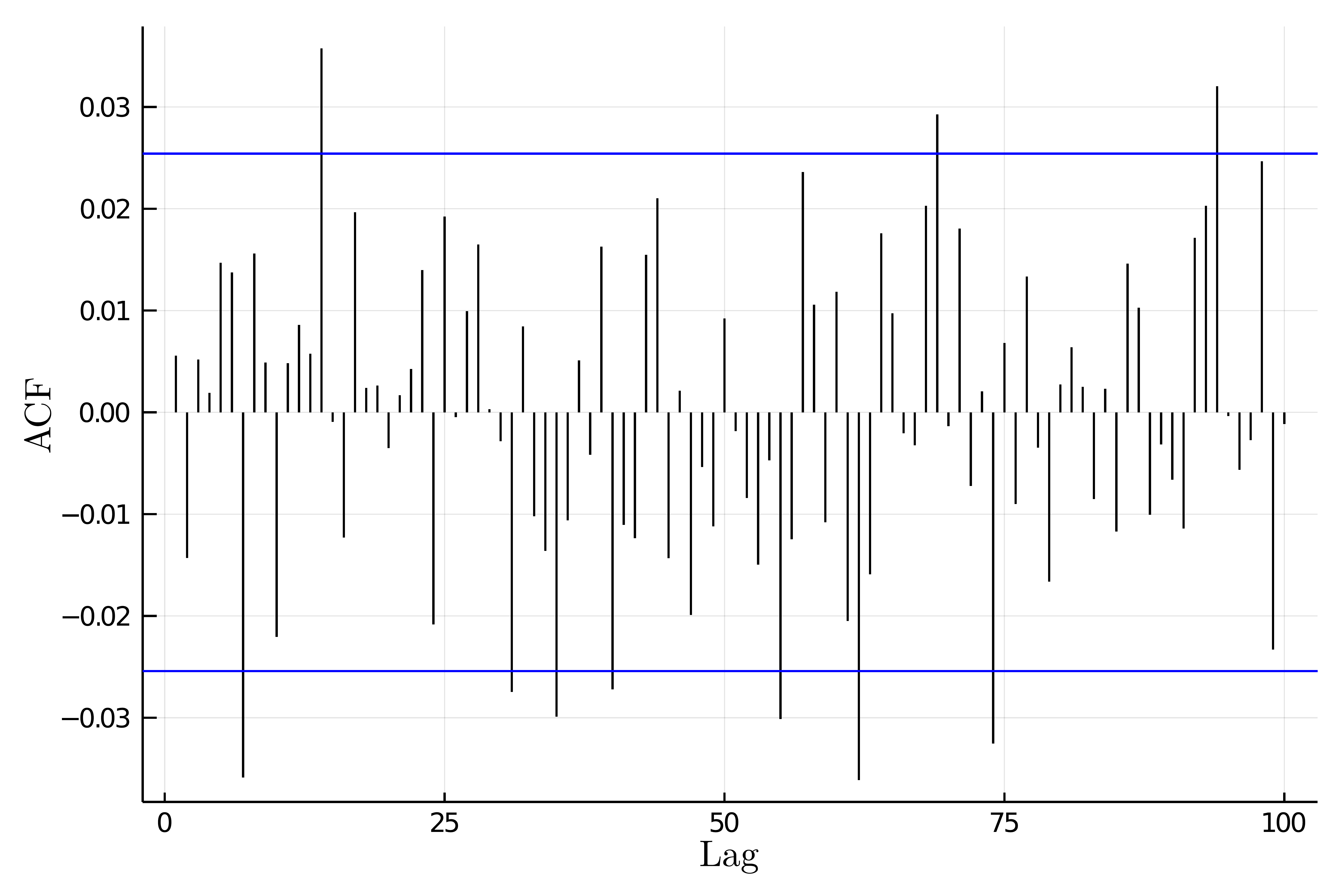}}
    \subfloat[A2X 20 min]{\label{baracf:c}\includegraphics[width=0.33\textwidth]{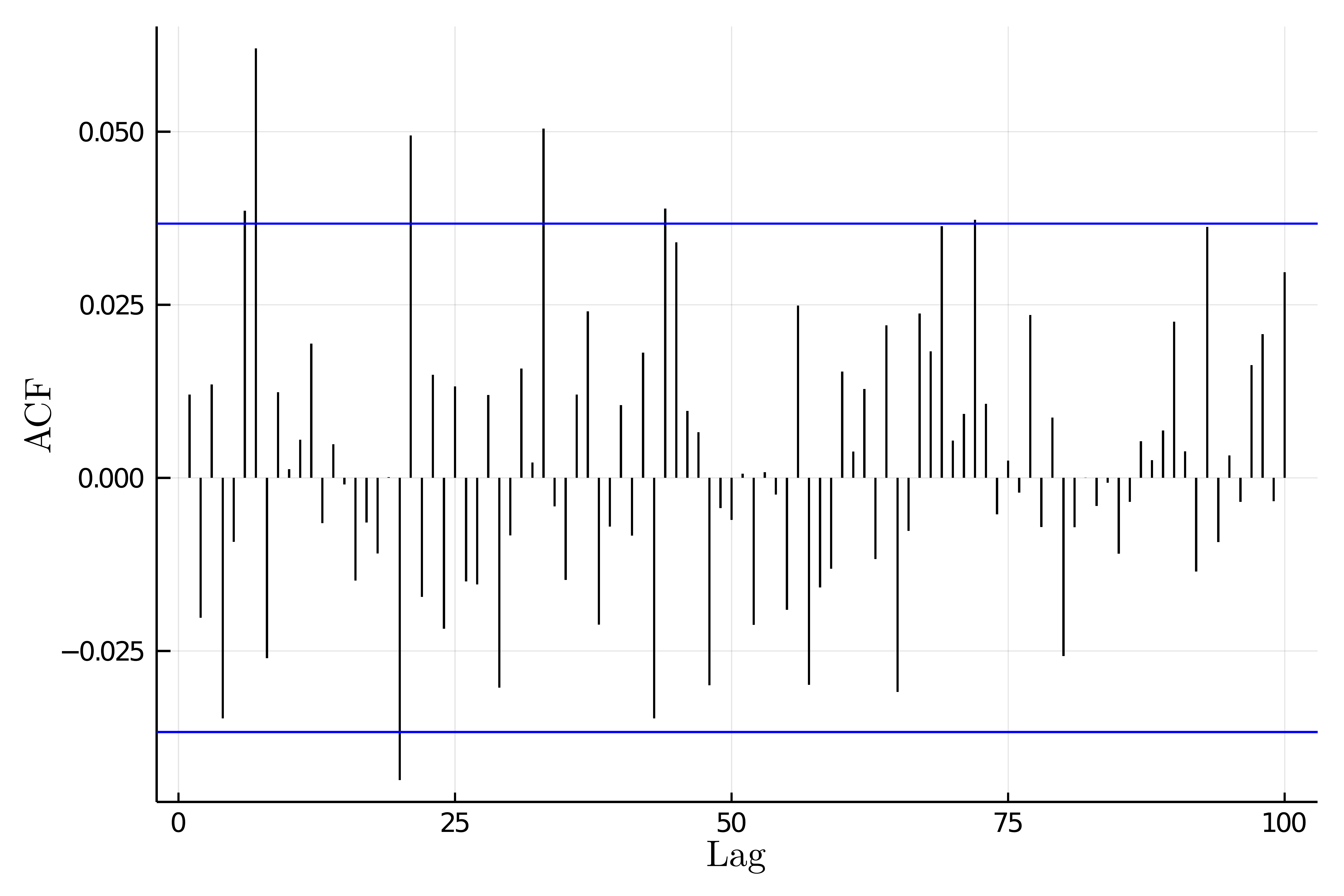}}    \\
    \subfloat[JSE 1 min]{\label{baracf:d}\includegraphics[width=0.33\textwidth]{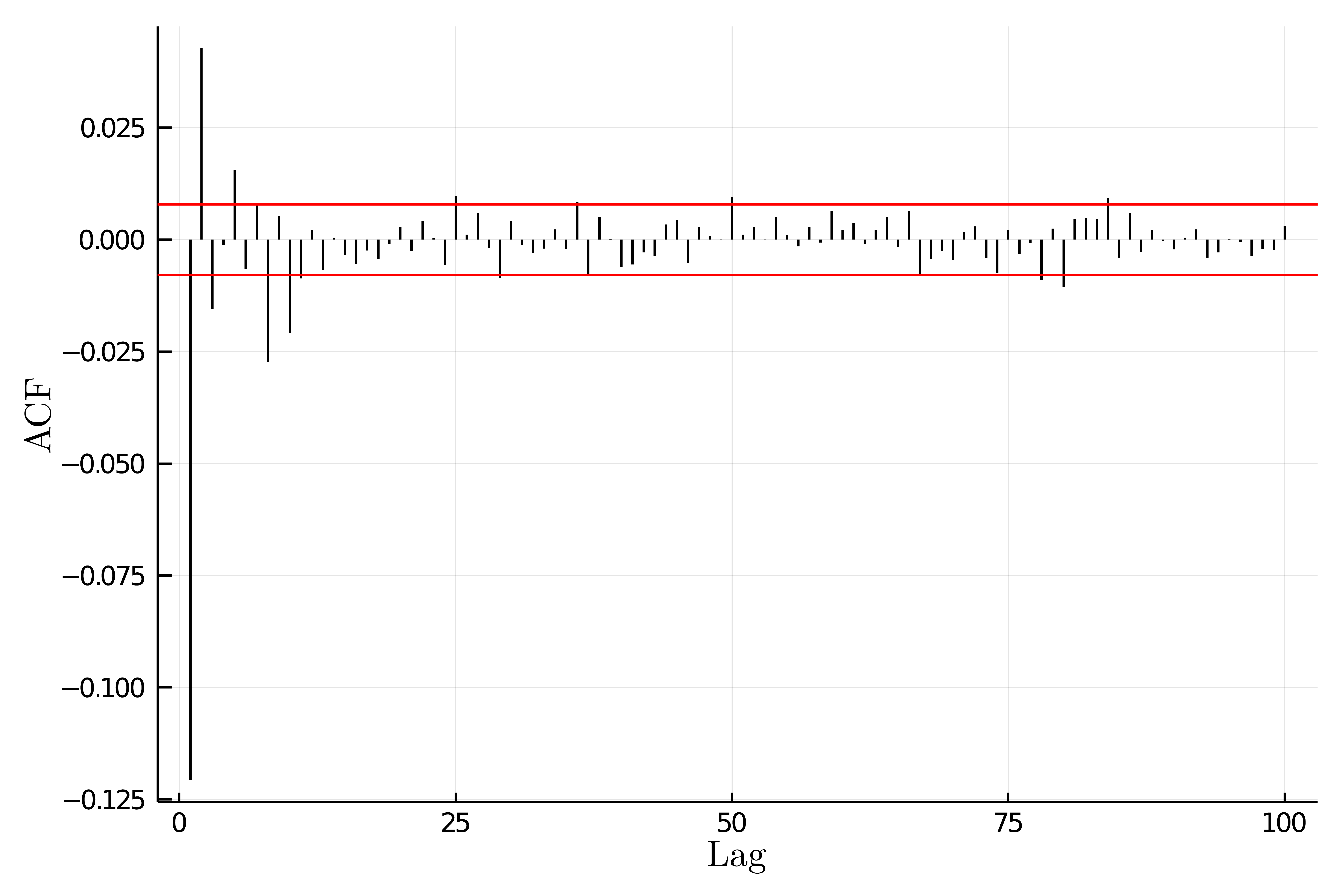}}
    \subfloat[JSE 10 min]{\label{baracf:e}\includegraphics[width=0.33\textwidth]{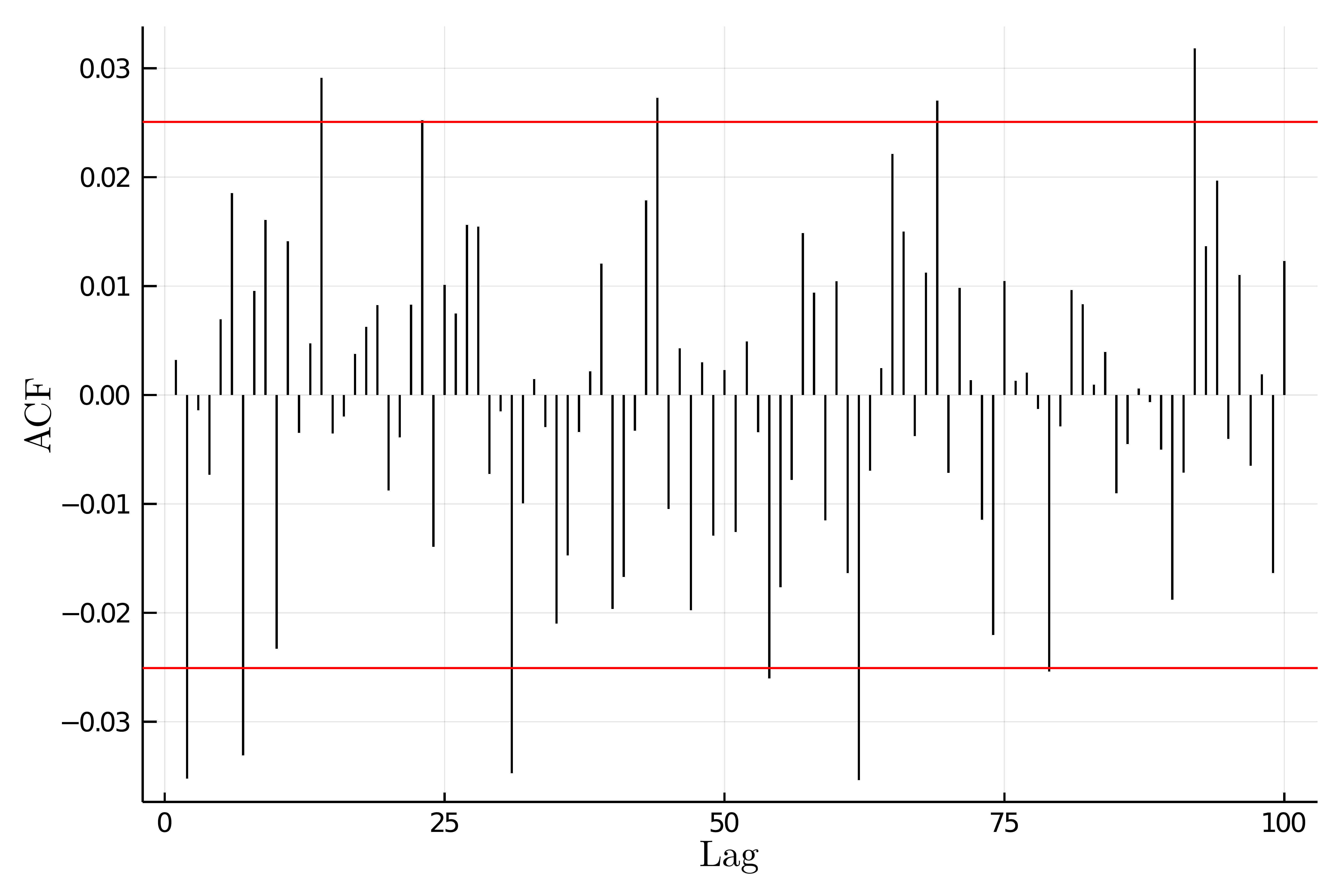}}
    \subfloat[JSE 20 min]{\label{baracf:f}\includegraphics[width=0.33\textwidth]{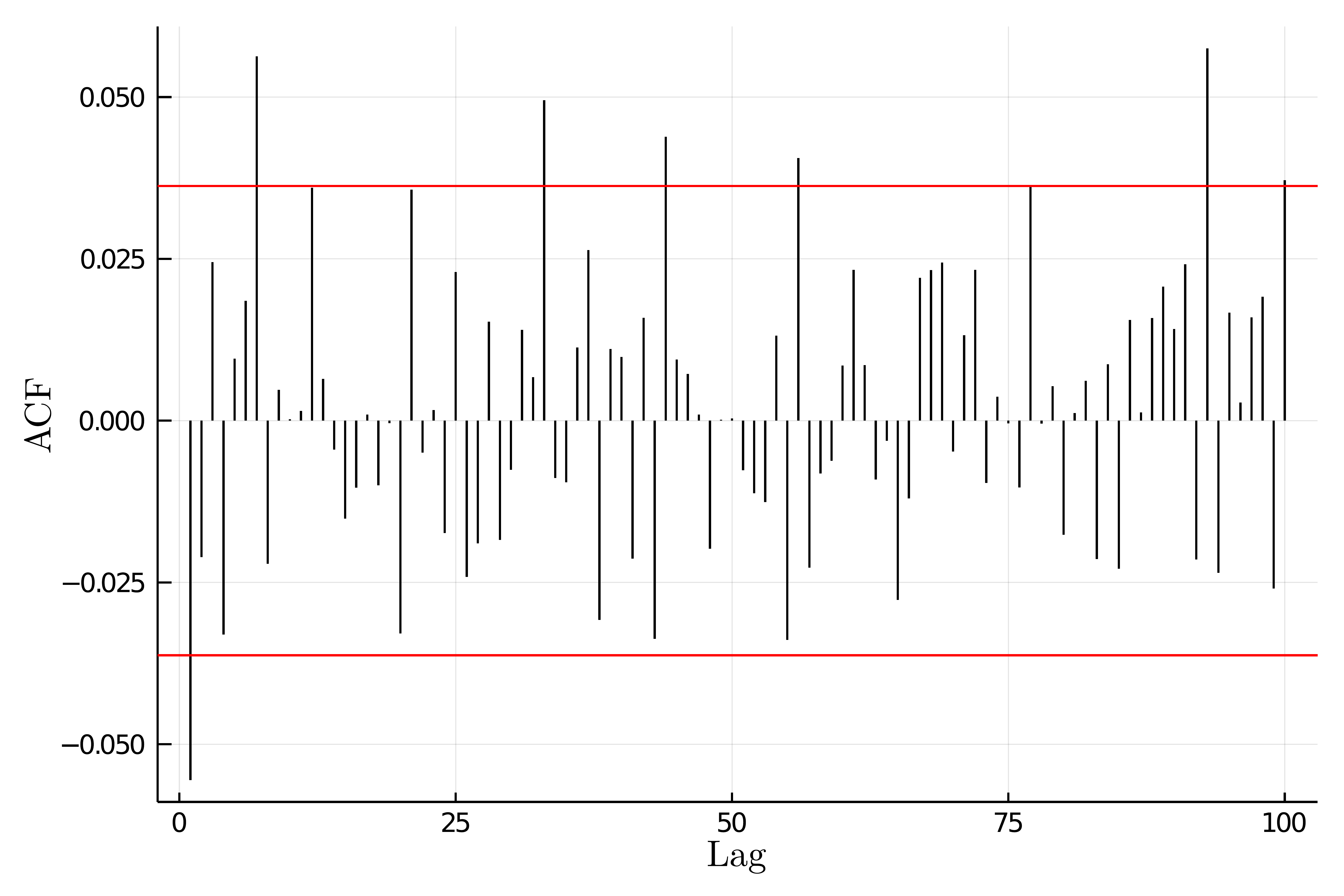}}
    \caption{Auto-correlations of microprice bar returns for Naspers over different bar sizes and different exchanges. The first row is the A2X exchange while the second row is the JSE. From the first to third column we have 1-minute, 10-minute and 20-minute bars respectively.}
\label{fig:baracf}
\end{figure*}

\begin{figure*}[htb]
    \centering
    \subfloat[Left tail 1 min]{\label{Dist:a}\includegraphics[width=0.33\textwidth]{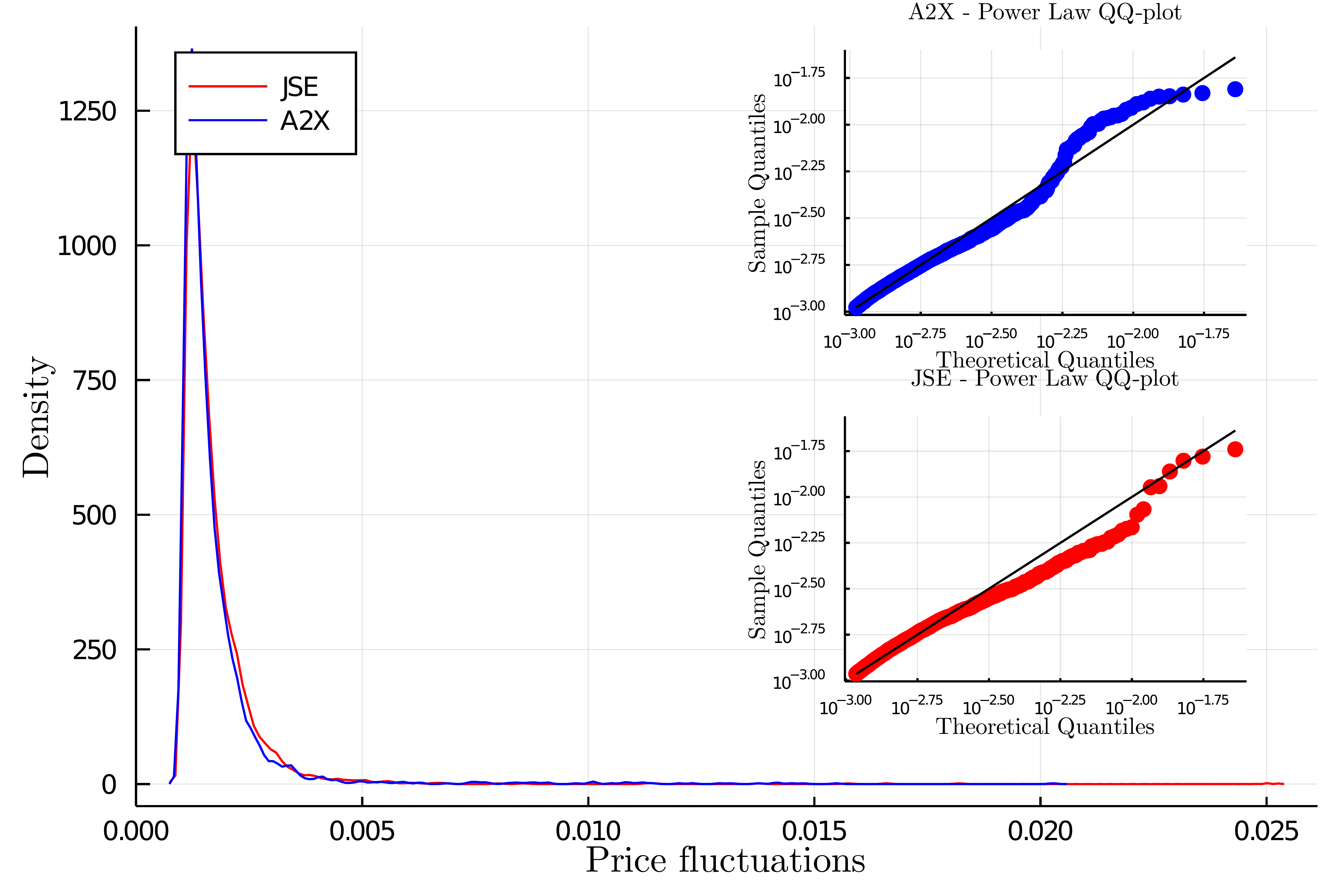}}
    \subfloat[1 min]{\label{Dist:b}\includegraphics[width=0.33\textwidth]{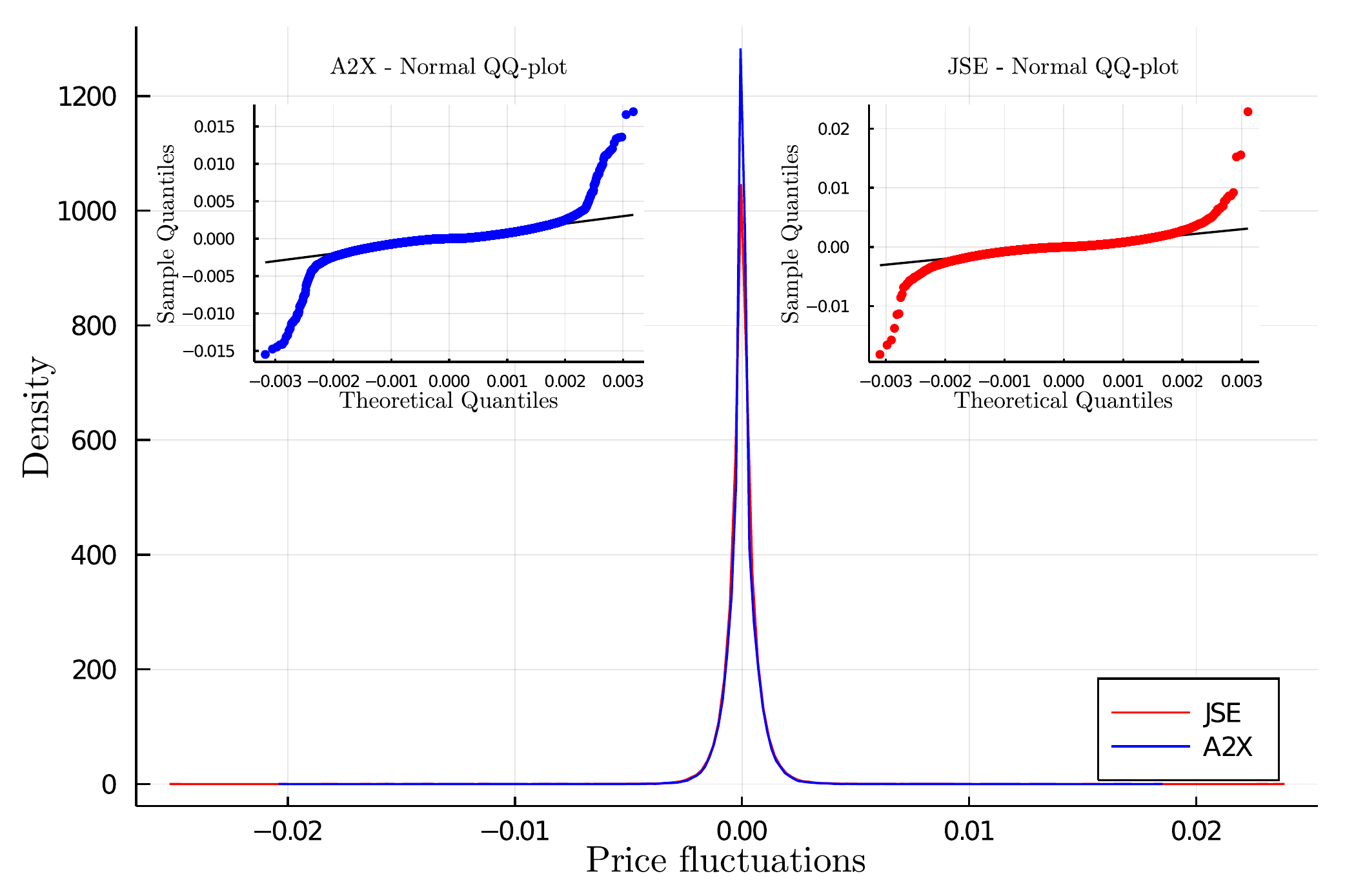}}
    \subfloat[Right tail 1 min]{\label{Dist:c}\includegraphics[width=0.33\textwidth]{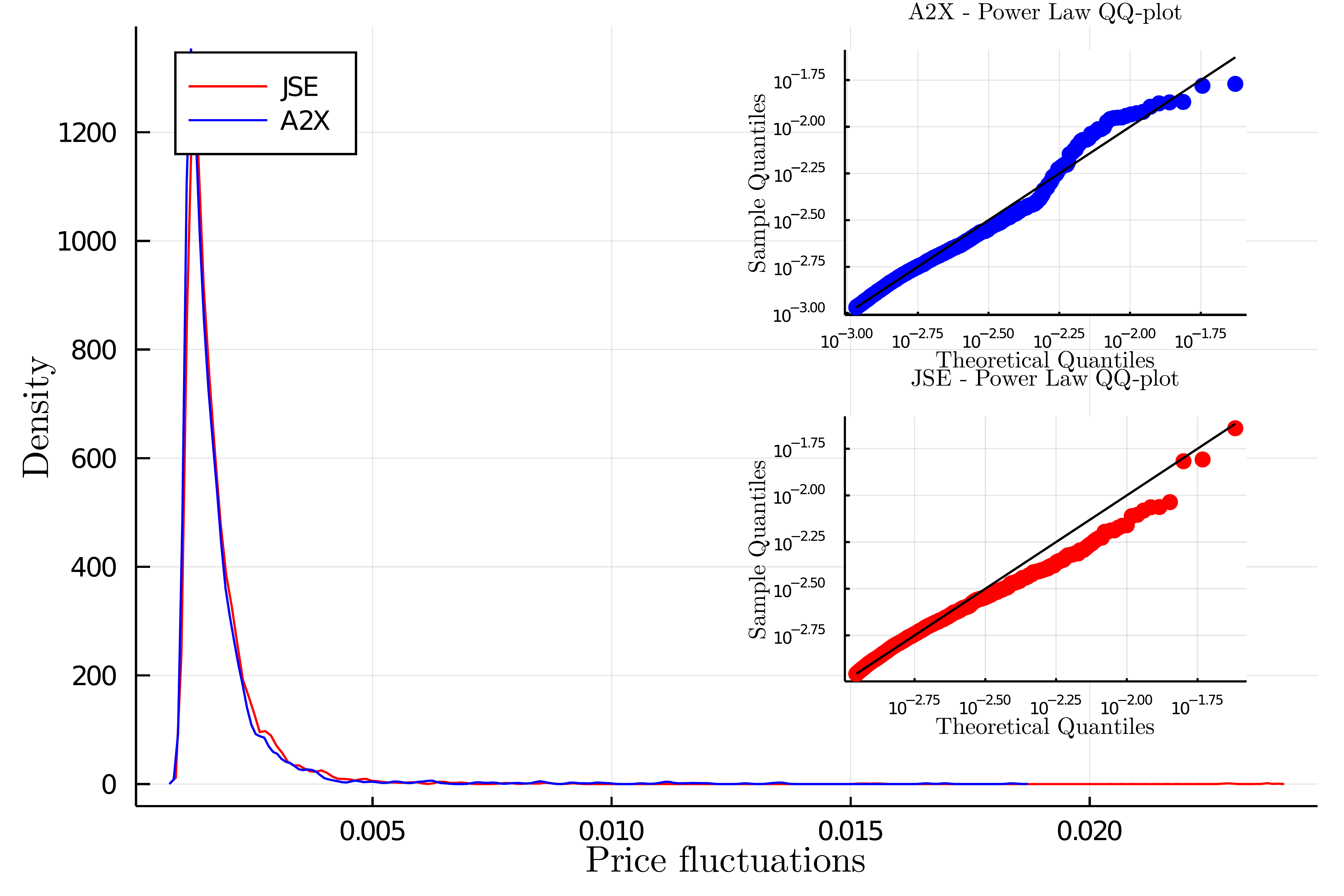}}    \\
    \subfloat[Left tail 10 min]{\label{Dist:d}\includegraphics[width=0.33\textwidth]{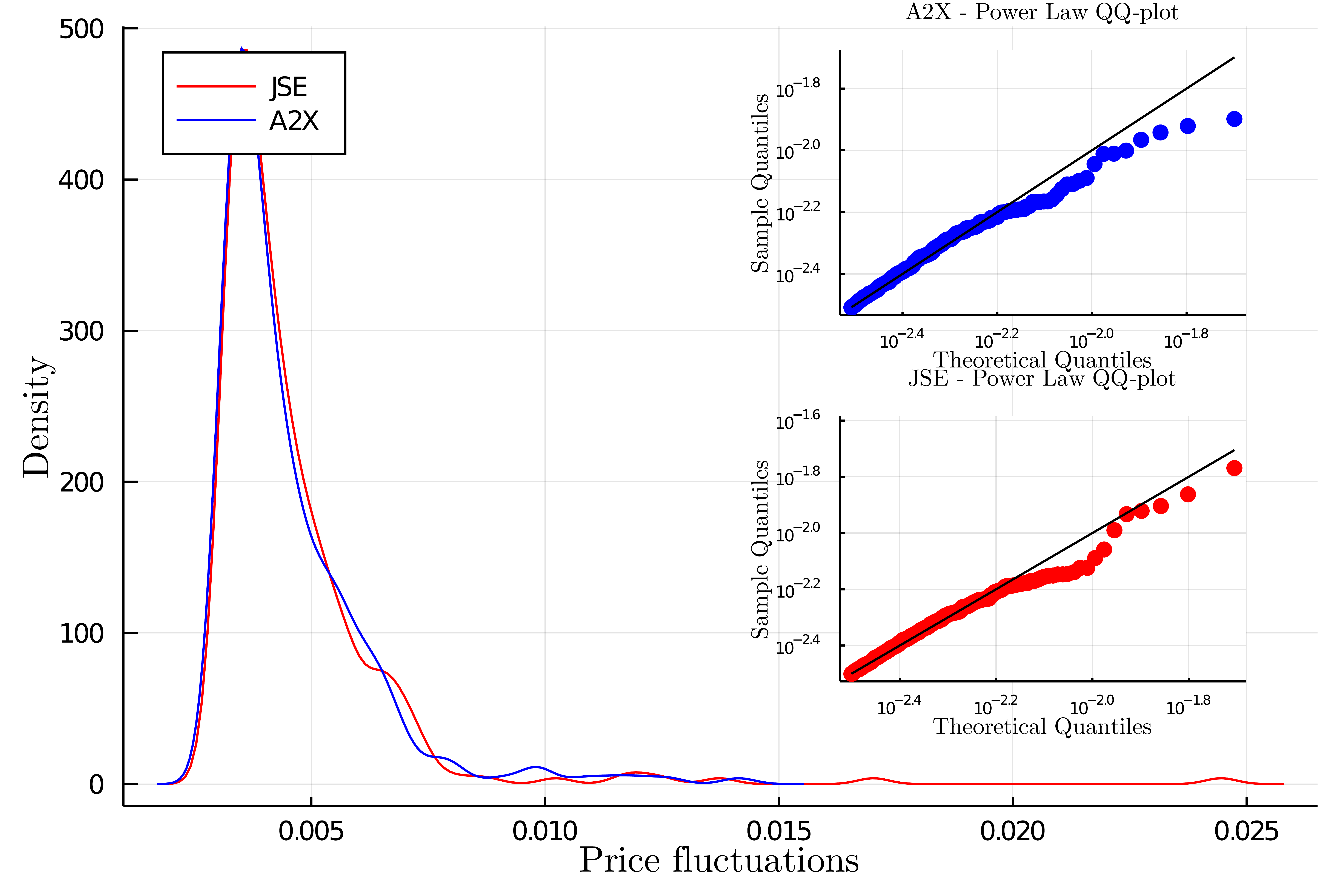}}
    \subfloat[10 min]{\label{Dist:e}\includegraphics[width=0.33\textwidth]{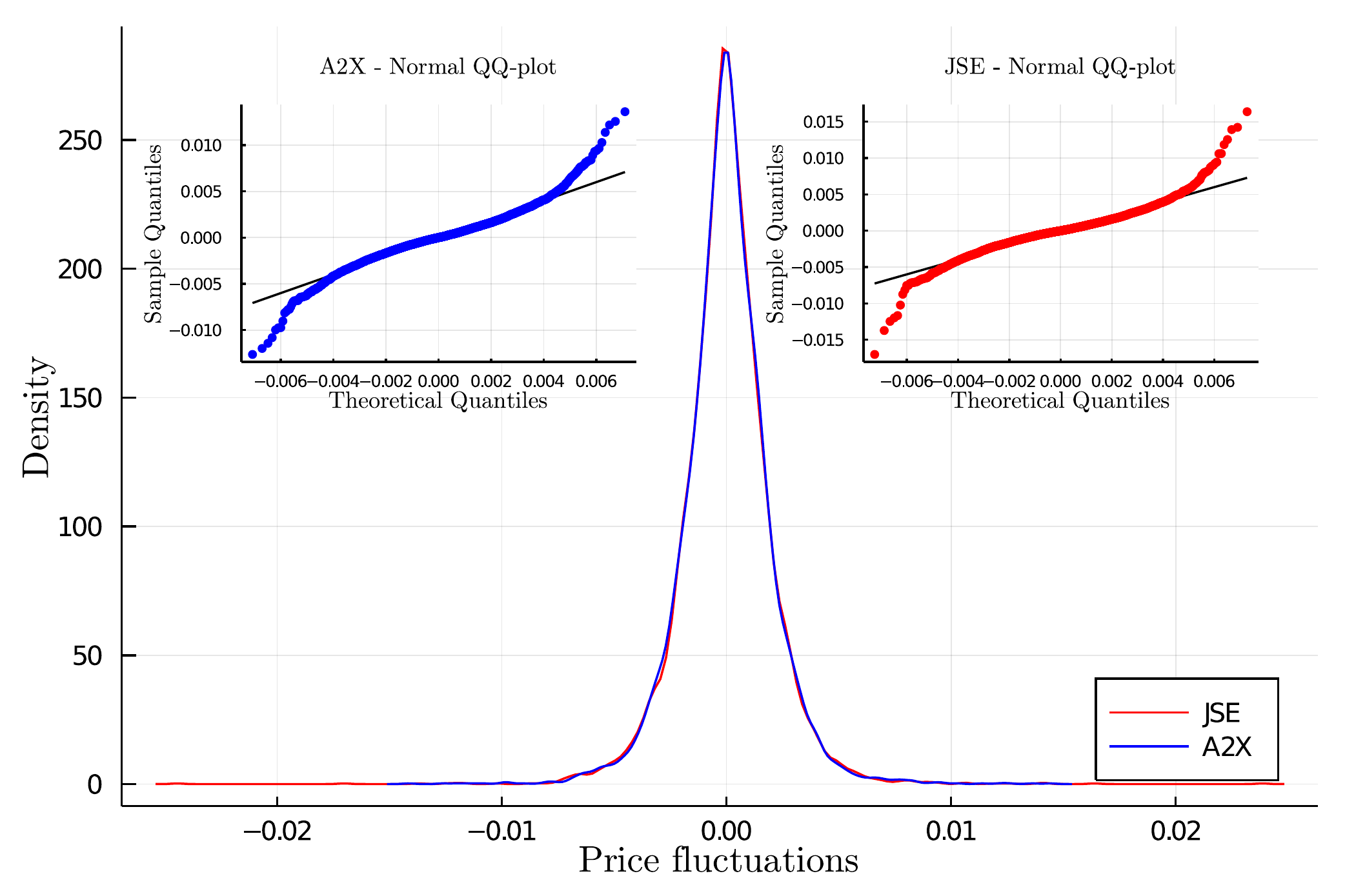}}
    \subfloat[Right tail 10 min]{\label{Dist:f}\includegraphics[width=0.33\textwidth]{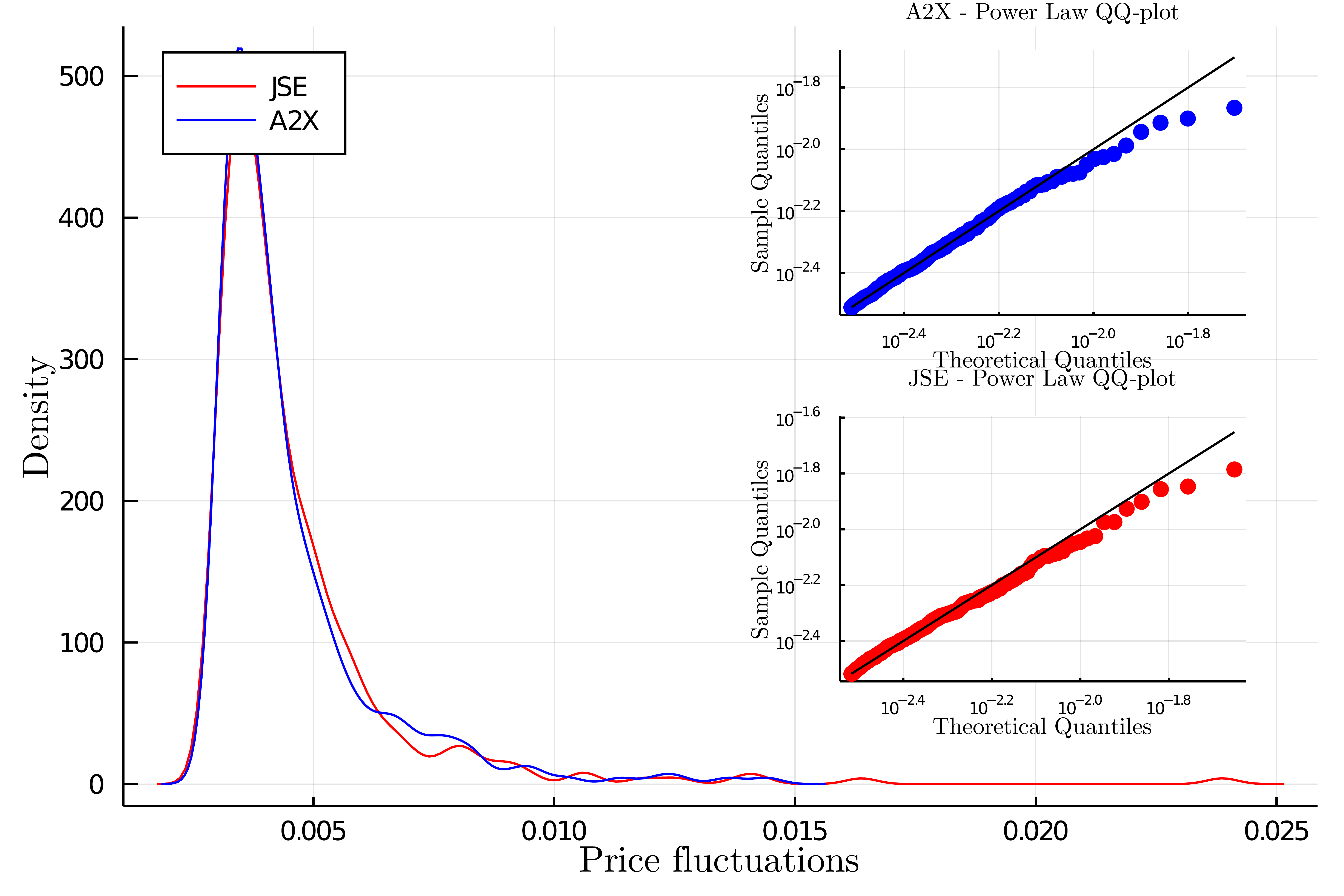}}    \\
    \subfloat[Left tail 20 min]{\label{Dist:g}\includegraphics[width=0.33\textwidth]{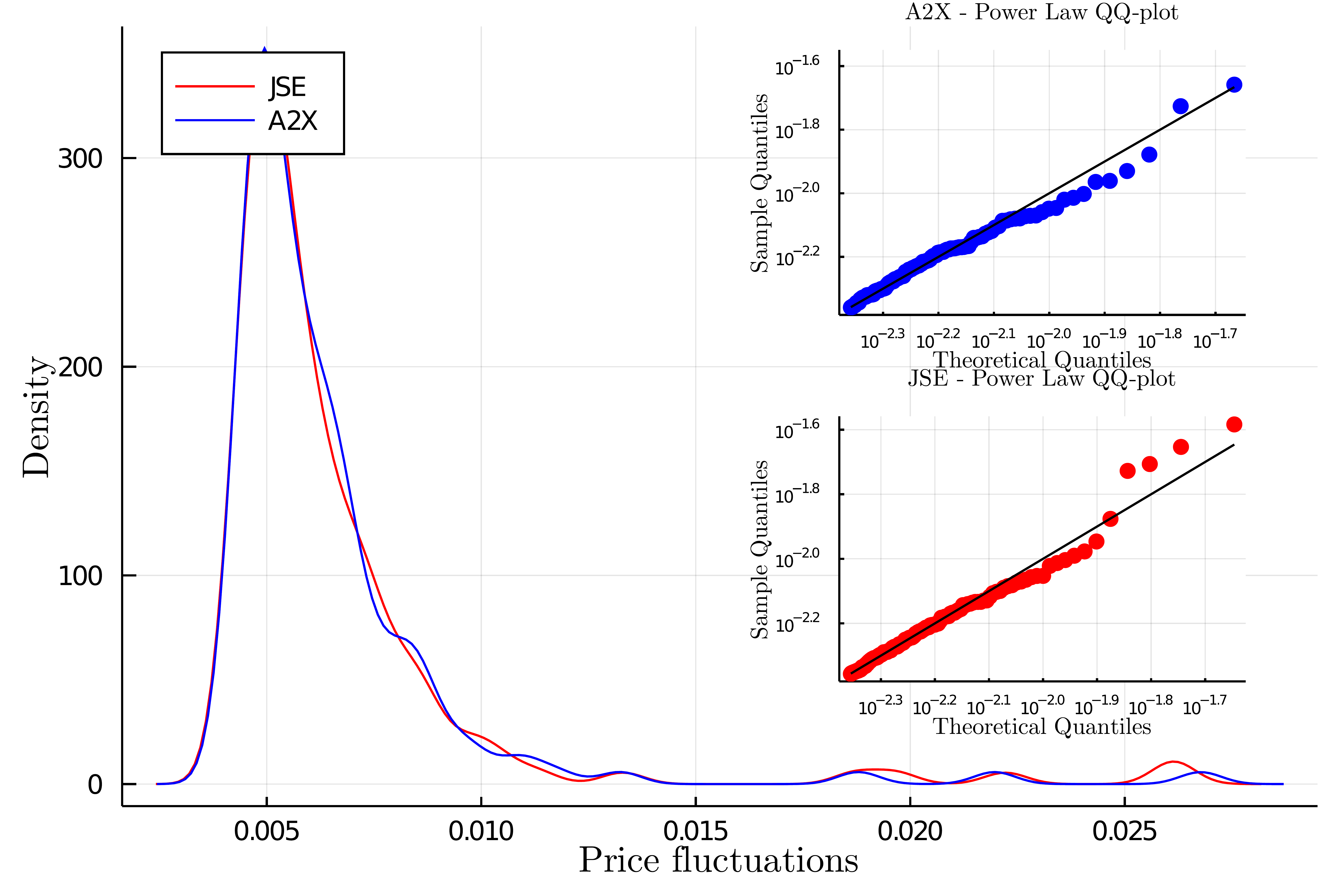}}
    \subfloat[20 min]{\label{Dist:h}\includegraphics[width=0.33\textwidth]{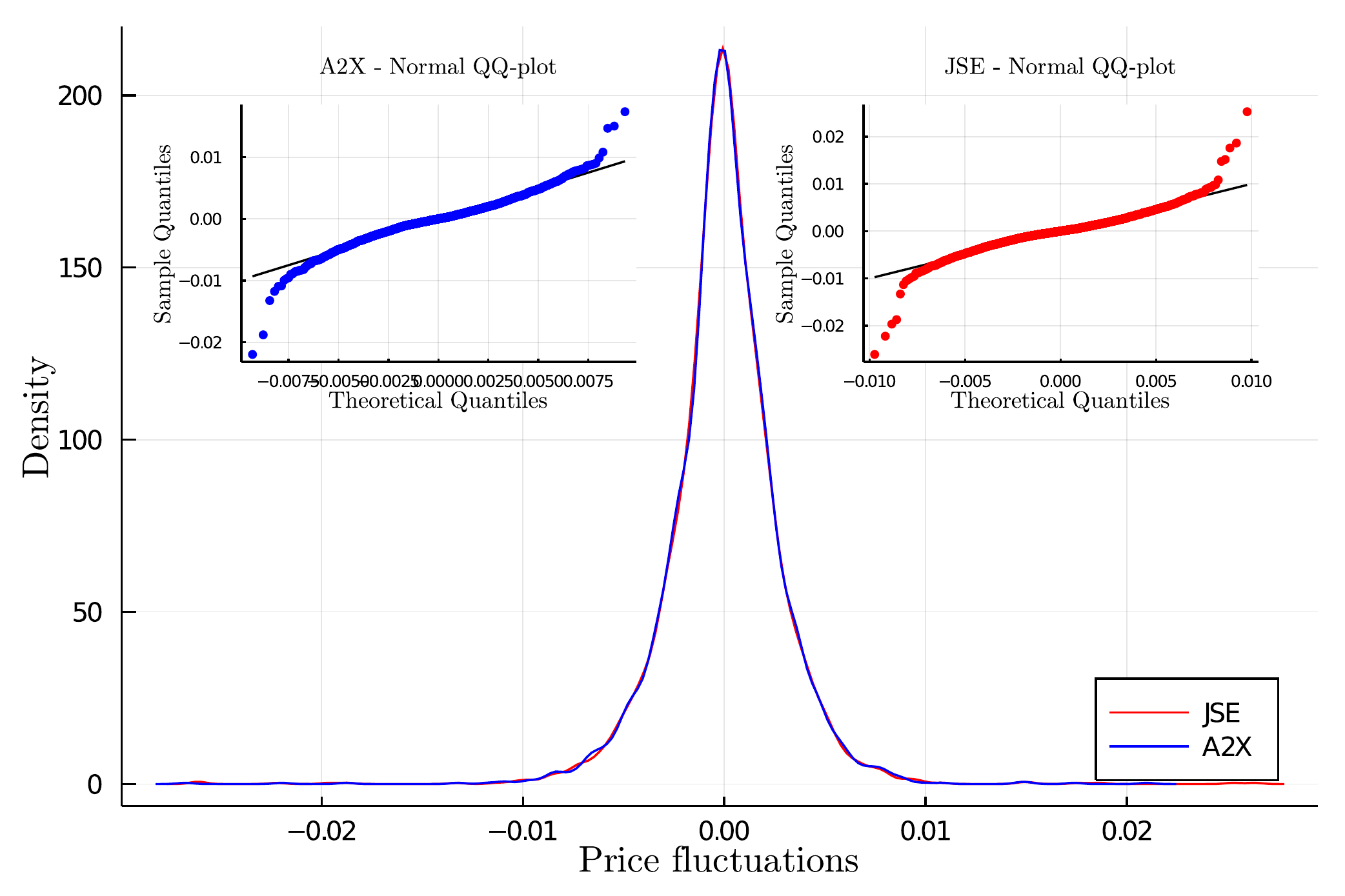}}
    \subfloat[Right tail 20 min]{\label{Dist:i}\includegraphics[width=0.33\textwidth]{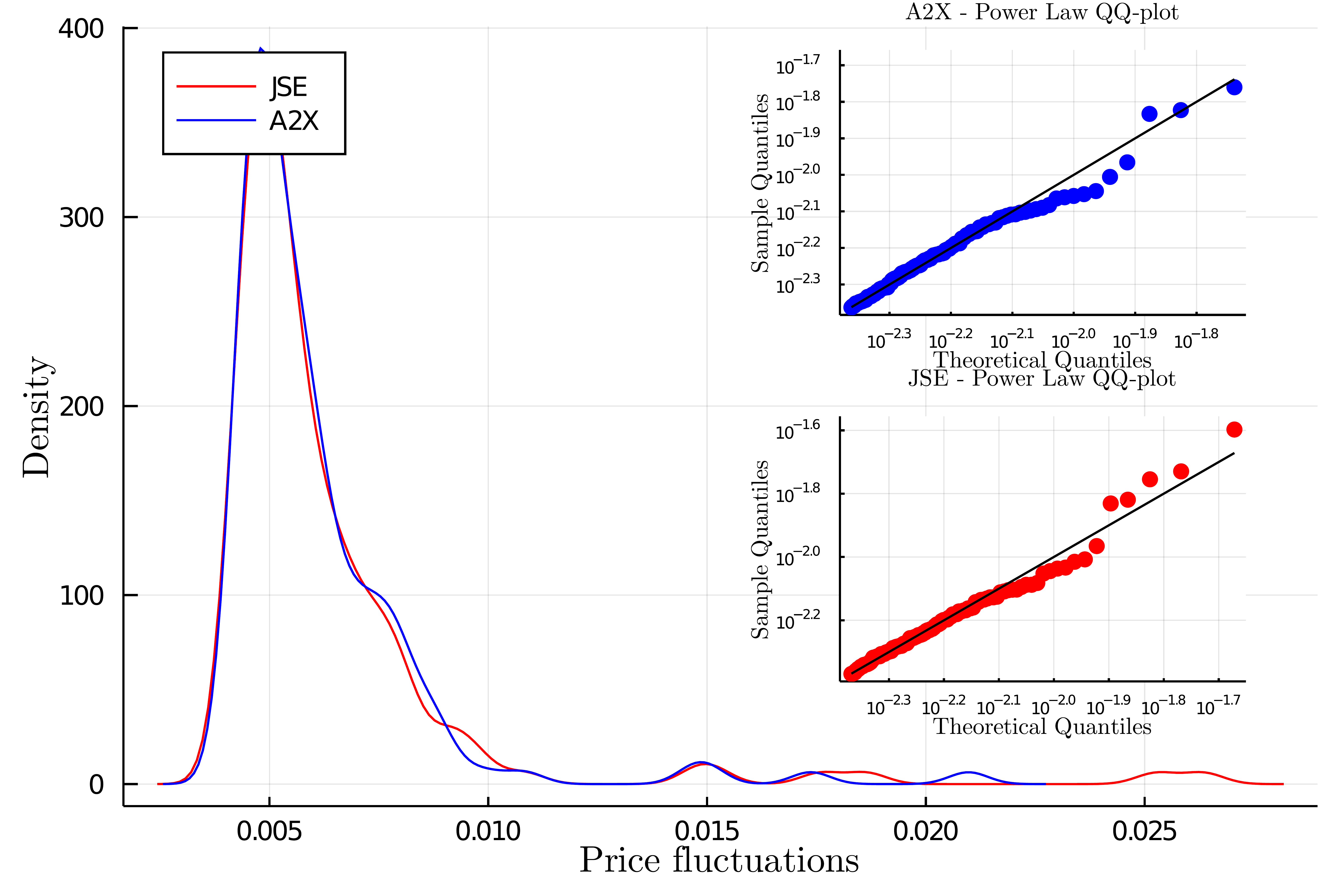}}
    \caption{Microprice bar return distributions at different bar intervals for Naspers. The left and right tails have QQ-plots fitted to a power-law distribution provided as insets whereas the full distribution has QQ-plots fitted to a Normal distribution provided as insets. From the first to third row we have 1-minute, 10-minute and 20-minute bars respectively. A2X is given in blue and JSE in red.}
\label{fig:Dist}
\end{figure*}

We then construct Open High Low Close (OHLC) bar data using the microprices. The OHLC microprice bar data is visualised using candlestick plots in \Cref{fig:bar}.\footnote{Candlestick plots can be an effective visualisation technique to summarise the information within a bar using the OHLC values because it makes it easier for humans to spot patterns from homogeneously spaced data.} Each stick is created as follows: the top (bottom) tail represents the high (low) within the bar while the edges of the box represent either the close or the open of the bar. If the stick is red then the open price is on the top edge and the close price is on the bottom edge, and if the stick is blue then the close price is on the top edge and the open price is on the bottom edge.

From the figures we see that the candlesticks are very similar between the two exchanges --- particularly as time scale increases. However, there are noticeable differences on the tails and sometimes the color of the stick, particularly when the body of the stick is extremely small, for example at 16:10 on the 10 minute bars. These differences become less noticeable over larger intervals. On longer time scales there is increased synchronisation.

The bar returns are computed using \cref{eq:MPrets} with closing to closing microprice bars. Using these returns, we investigate the auto-correlations from each exchange. \Cref{fig:baracf} plots the auto-correlations of the microprice bar returns for each exchange up to 100 lags for different bar sizes. We see that the 1-minute bar returns present a strong negative first order auto-correlation in both exchanges, suggesting a mean reverting component at this scale. The majority of the lags on A2X seem to be significant, whereas only the first few lags are significant on the JSE. It remains unclear as to why this is the case and the significance of these lags; if they are spurious relations or are significant. The auto-correlations on the 10-minute and 20-minute intervals no longer present a strong mean reverting component and there are fewer significant lags. The significance of these lags remain elusive as there are no strong theoretical reasons for observing such patterns --- again, we speculate this is due to low relative liquidity and anomalous agent behaviours.

\Cref{fig:Dist} investigates the distributions of the bar returns at different time scales. It includes the full distribution with QQ-plots fitted to a normal distribution provided as insets, and also the left and right tails with QQ-plots fitted to a power-law distribution provided as insets (presented on a log-log scale). We see that on all time scales the returns have heavy tails and are leptokurtic for both markets. We also see that both tails seem to follow a power-law distribution on all time scales. What is particularly interesting is that looking at both \Cref{fig:TickDist,fig:Dist} from smaller to larger time scales, we notice that the distribution becomes less leptokurtic and slightly more normal as the time scales increase. 

There is a meaningful difference in the distribution of the microprice returns on a tick-by-tick scale between the exchanges, particularly in the tail ends of the distributions. On the other hand, the microprice returns start to exhibit similar distributions between the exchanges when considering bar data. However, the sample quantiles for these returns remain different between the exchanges at all time scales, particularly in the tail ends of the distributions.

This contributes towards the body of empirical evidence that challenges Eugene Fama's {\it Efficient Market Hypothesis} (see \citet{JLGB2008} and the references therein). The idea of fundamental value (at these time scales) is put into question when the same asset can exhibit different behaviours depending on the exchange it is listed on. 



\section{Comparing the price impact} \label{sec:comp}

Here we compare the immediate price impact functions between the exchanges and follow \citet{LFM2003} by looking to collapse the price impact functions into master curves for each exchange, and then possible master curves for the entire SA market. We then account for each cost component contributing towards the combined cost of trading to investigate the impact of different pricing structures on the exchanges. Concretely, we will incrementally add the indirect costs to the direct costs to combined the impact of transaction costs, settlement fees, the spread and the transaction cost ceiling, but on a scale normalised to the largest trades on exchange. This will be motivated in more detail in what follows.

\begin{figure*}[htb]
    \centering
    \subfloat[JSE Buyer-initiated price impact curves]{\label{Impact:a}\includegraphics[width=0.48\textwidth]{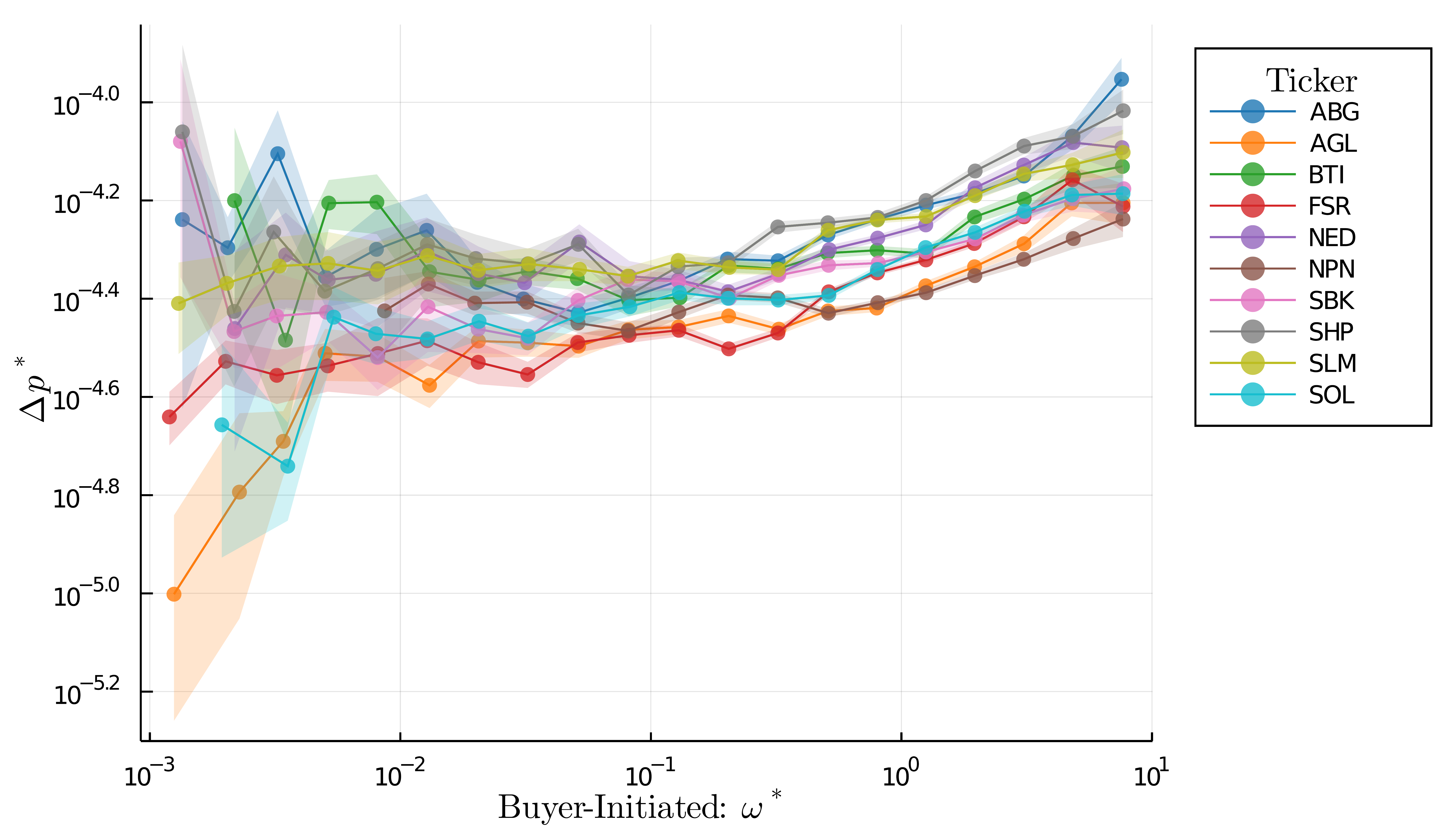}}
    \subfloat[JSE Seller-initiated price impact curves]{\label{Impact:b}\includegraphics[width=0.48\textwidth]{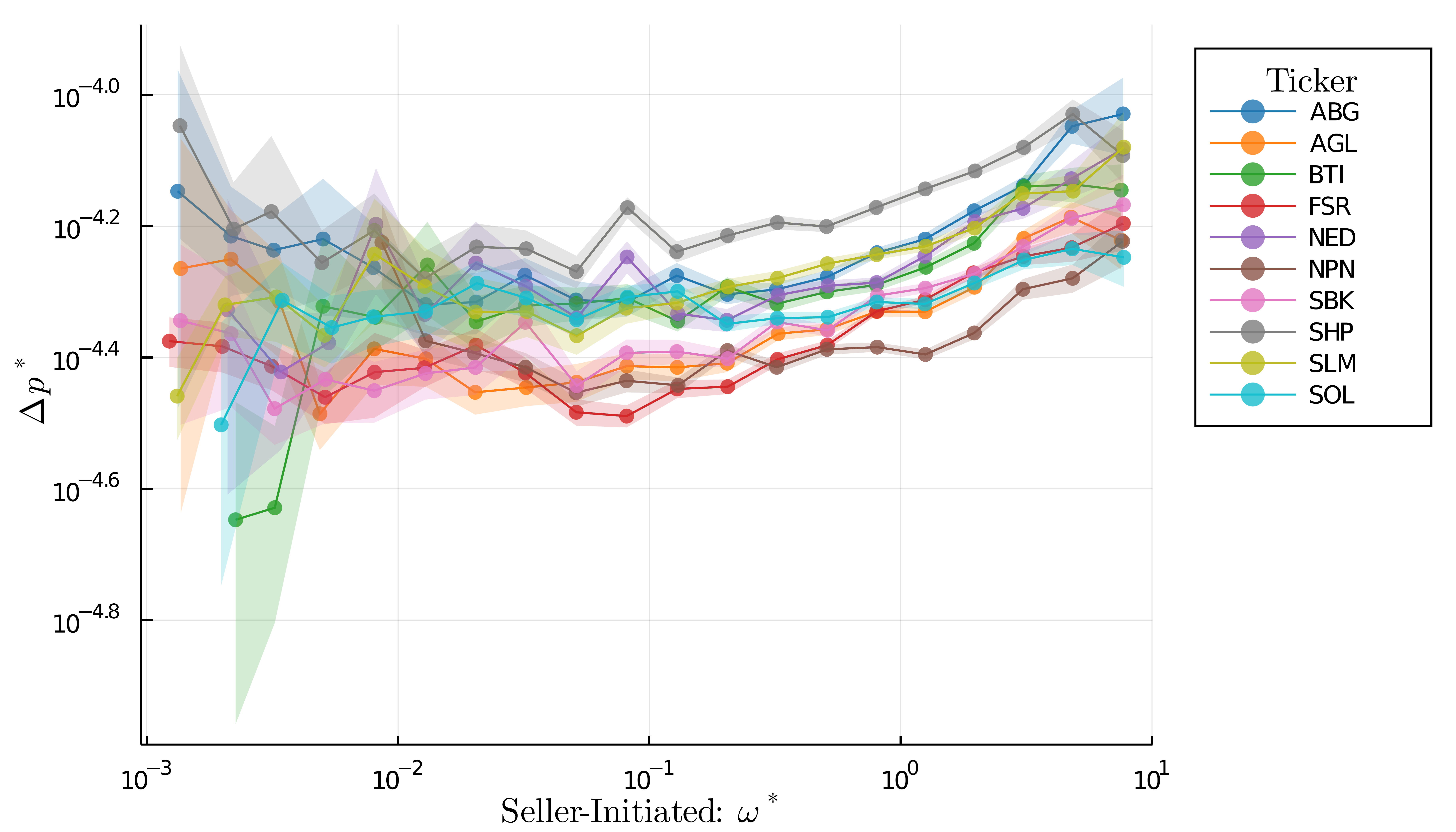}}  \\
    \subfloat[A2X Buyer-initiated price impact curves]{\label{Impact:c}\includegraphics[width=0.48\textwidth]{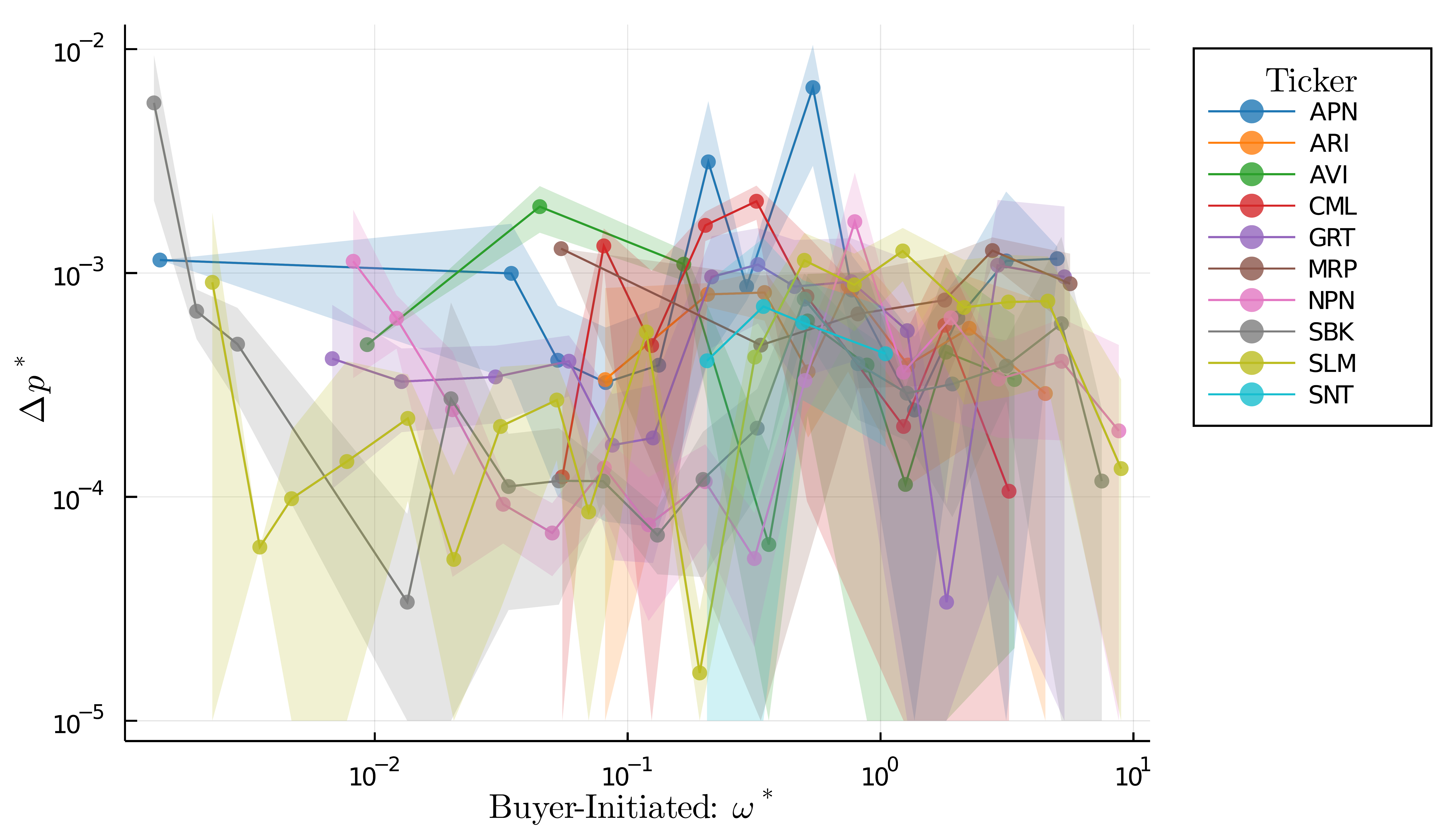}}
    \subfloat[A2X Seller-initiated price impact curves]{\label{Impact:d}\includegraphics[width=0.48\textwidth]{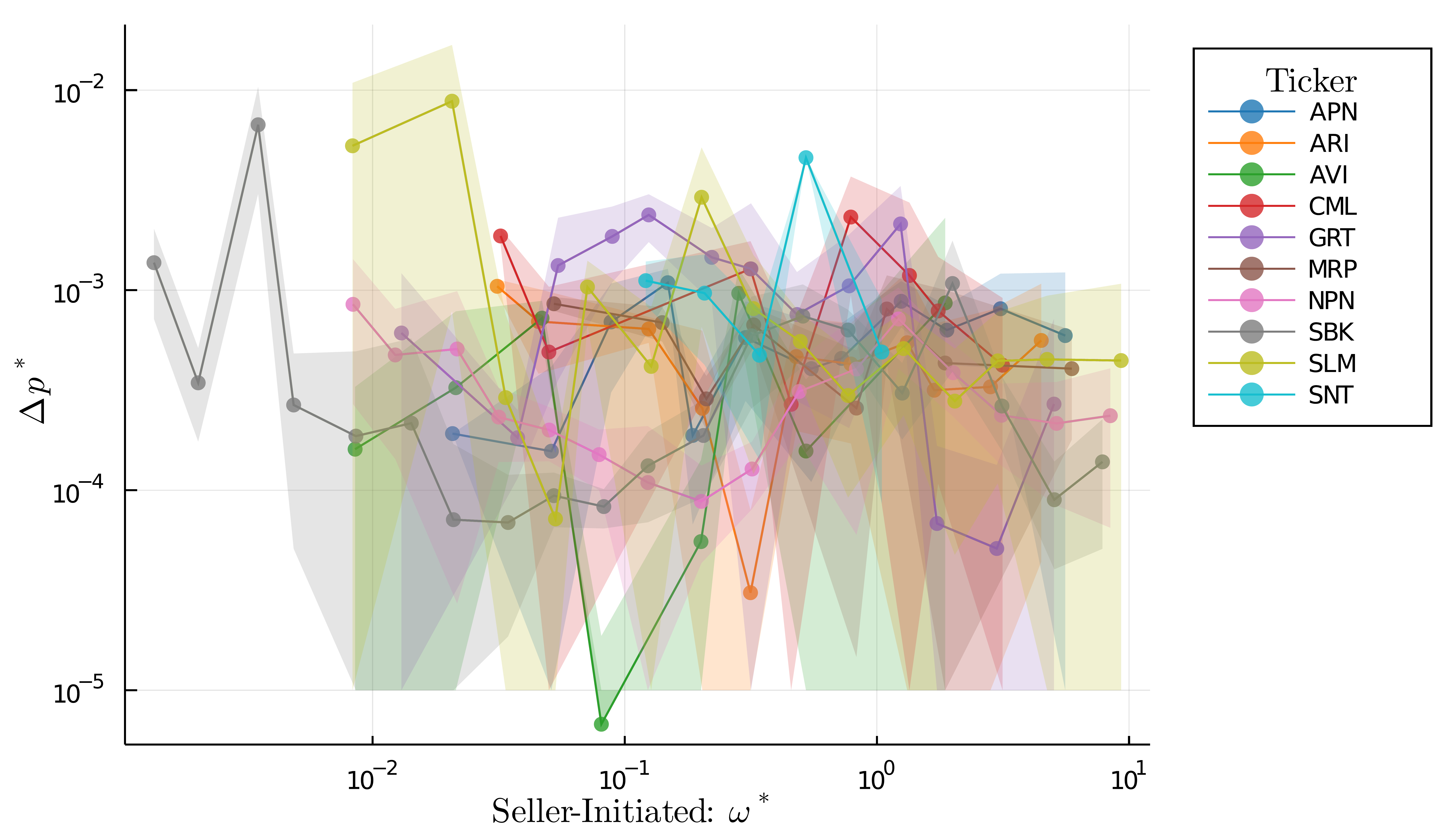}}
    \caption{Individual price impact curves for each security using normalised daily-normalised volumes with indicative measurement error bars. The error bars are constructed using 1,000 bootstrap samples for each security. The first row is the JSE and the second row is the A2X. The first column is buyer-initiated trades and the second column is seller-initiated trades.}
\label{fig:Impact}
\end{figure*}

\begin{figure*}[htb]
    \centering
    \subfloat[JSE Buyer-initiated ($\delta = 7.75 \times 10^{-5}, \gamma = 0.163163$)]{\label{Master:a}\includegraphics[width=0.48\textwidth]{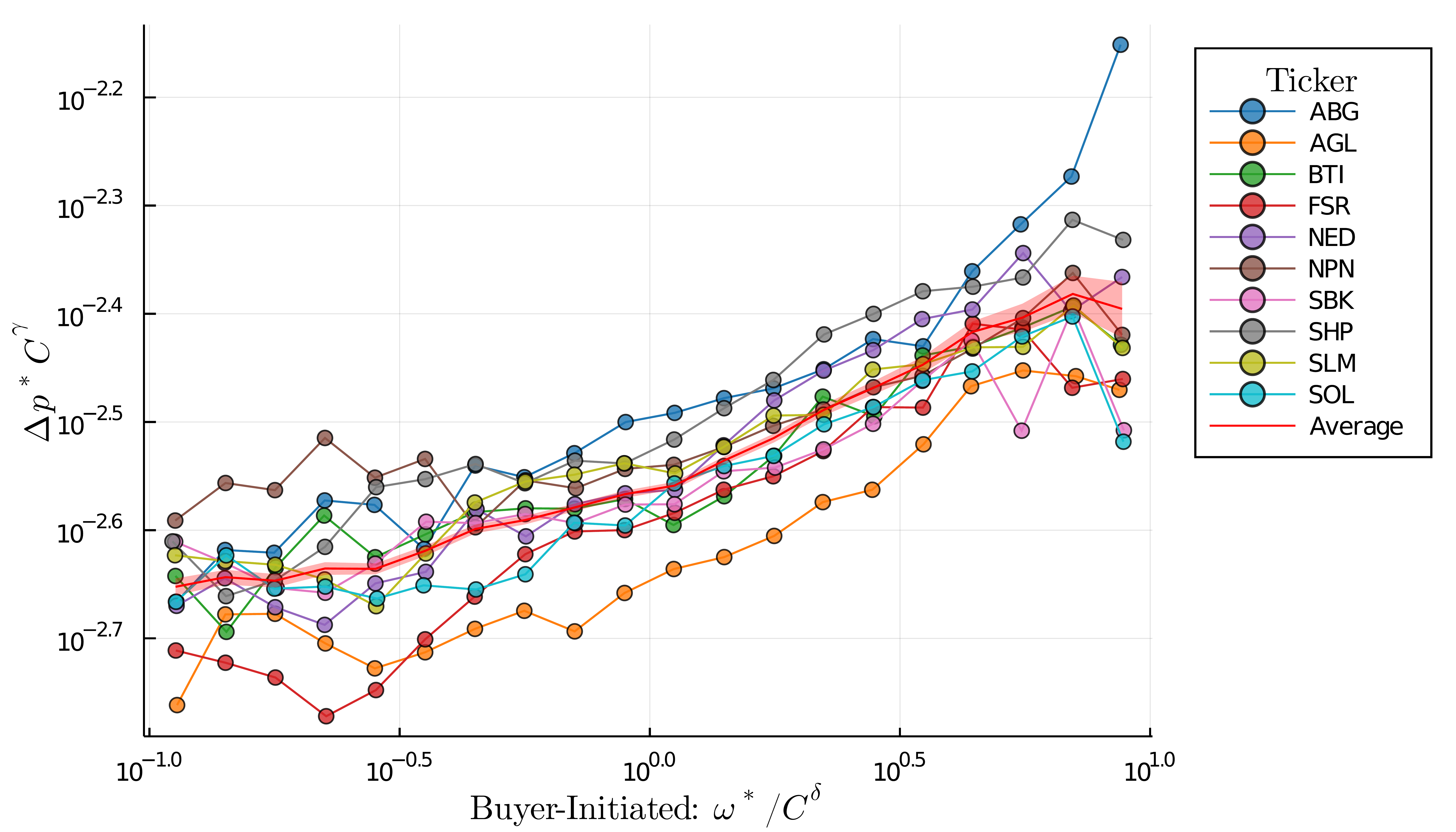}}
    \subfloat[JSE Seller-initiated ($\delta = -9.75 \times 10^{-5}, \gamma = 0.179662$)]{\label{Master:b}\includegraphics[width=0.48\textwidth]{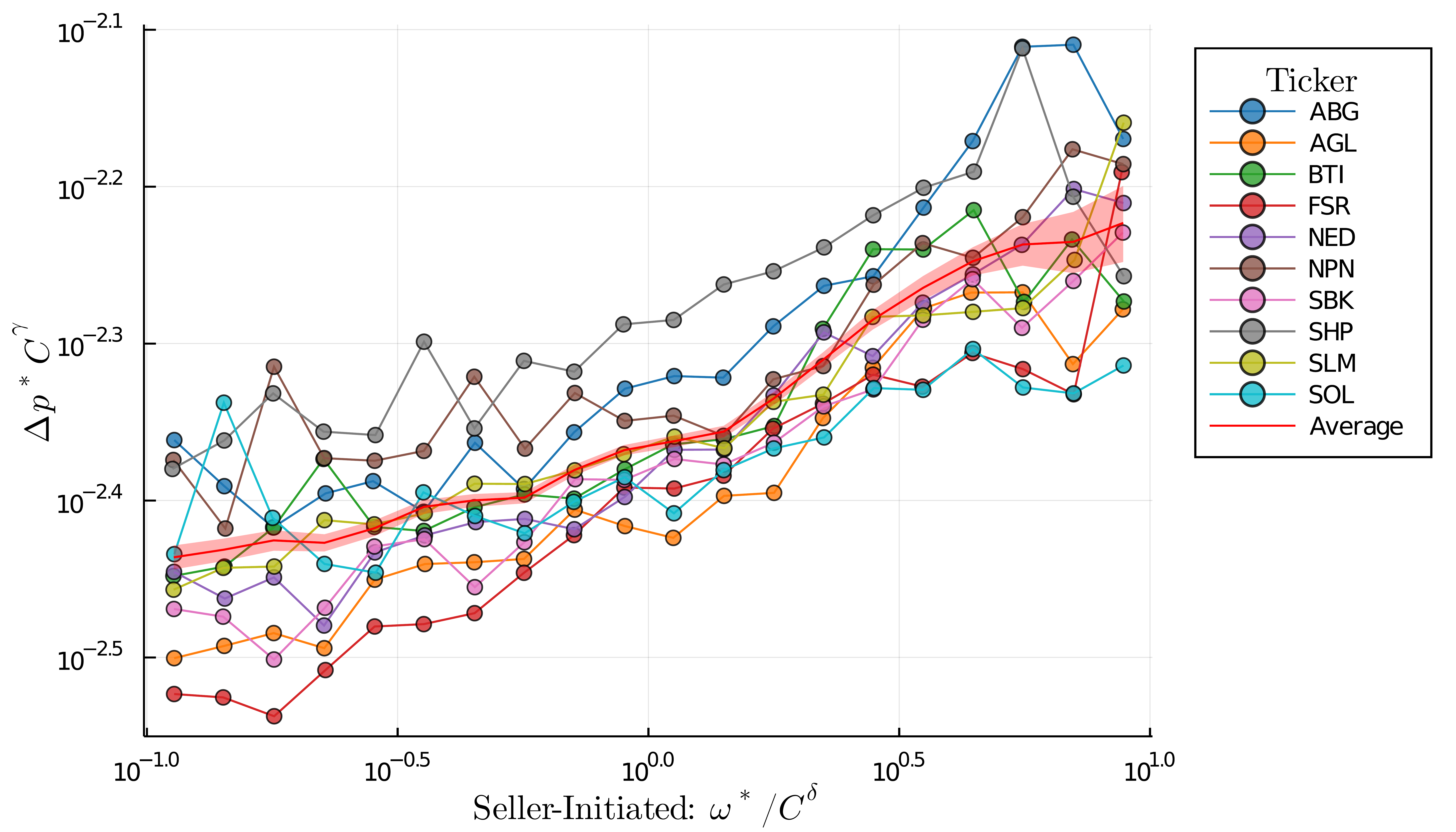}}  \\
    \subfloat[A2X Buyer-initiated ($\delta = 0.005469, \gamma = 0.057232$)]{\label{Master:c}\includegraphics[width=0.48\textwidth]{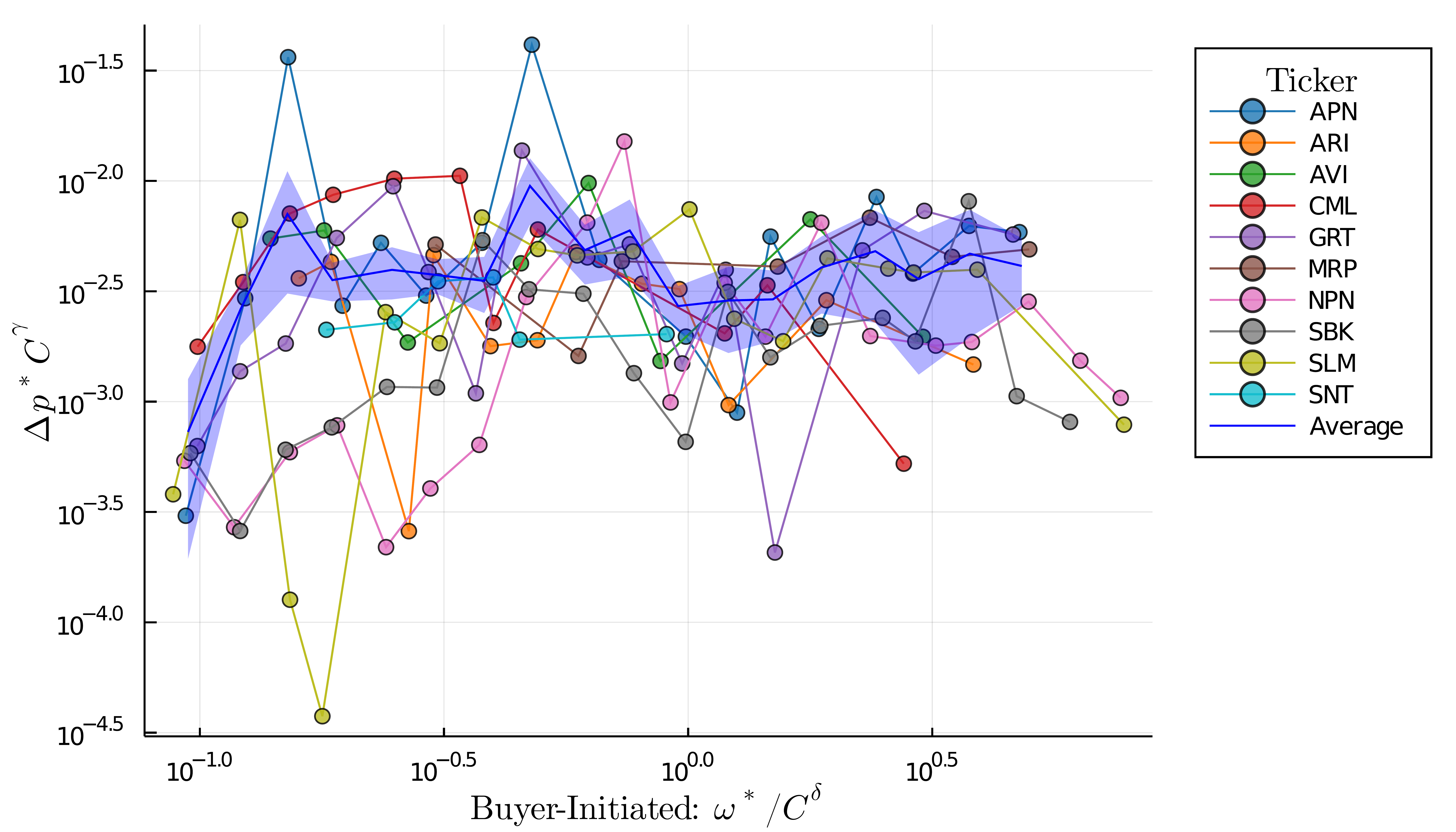}}
    \subfloat[A2X Seller-initiated ($\delta = 0.005111, \gamma = 0.396520$)]{\label{Master:d}\includegraphics[width=0.48\textwidth]{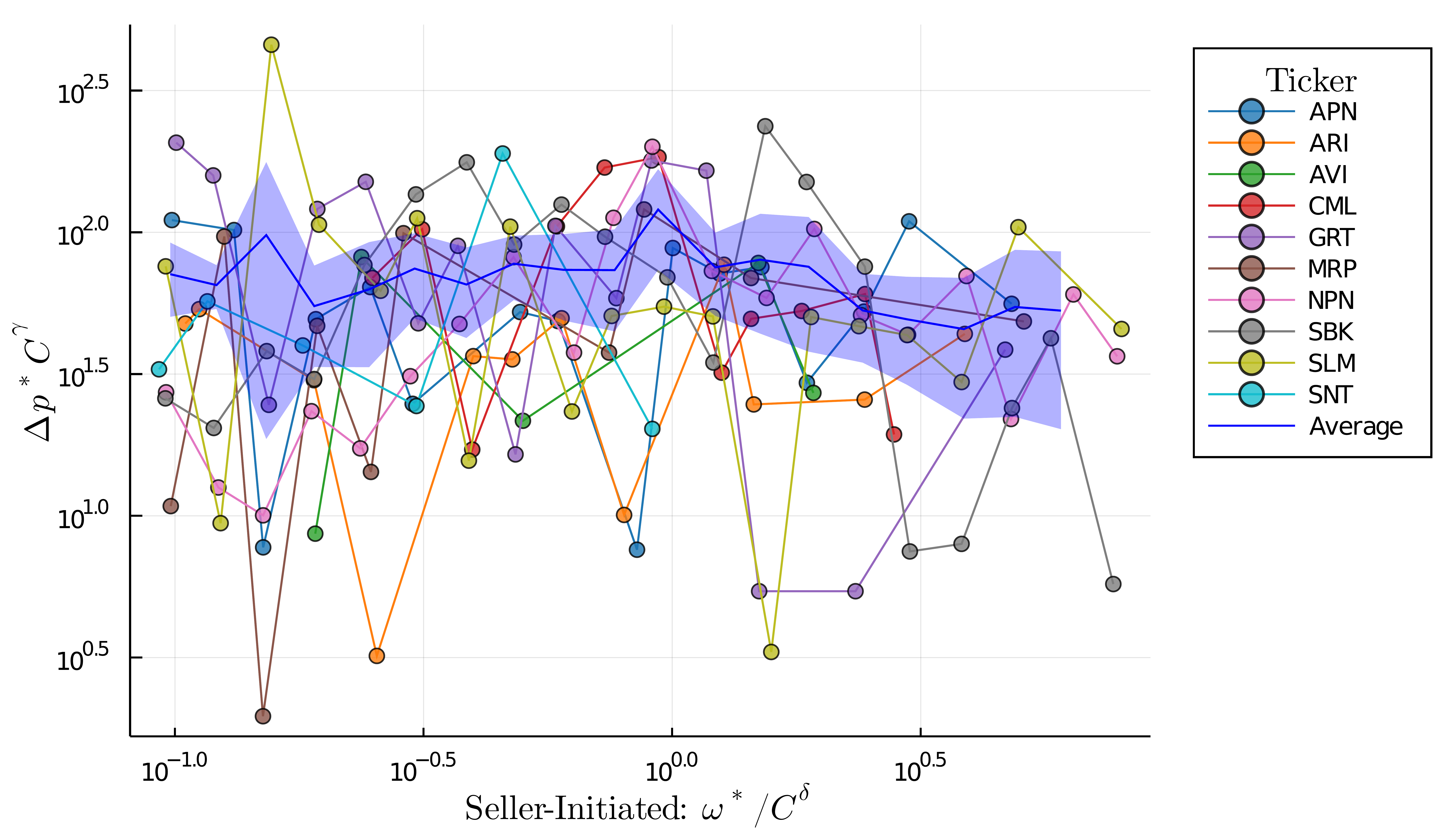}}
    \caption{Individual re-scaled price impact curves for each security collapsed following a liquidity correction using \cref{eq:master} with common $\delta$ and $\gamma$ estimated using \cref{eq:est}. Indicative master curves are estimated by averaging the re-scaled impacts across the securities on the exchange with error bars obtained from 1,000 bootstrap samples. The first row is the JSE and the second row is the A2X. The first column is buyer-initiated trades and the second column is seller-initiated trades.}
\label{fig:Master}
\end{figure*}

\begin{figure*}[htb]
    \centering
    \subfloat[Buyer-initiated, $\delta = 0.001370, \gamma = -0.726034$]{\label{SA:a}\includegraphics[width=0.48\textwidth]{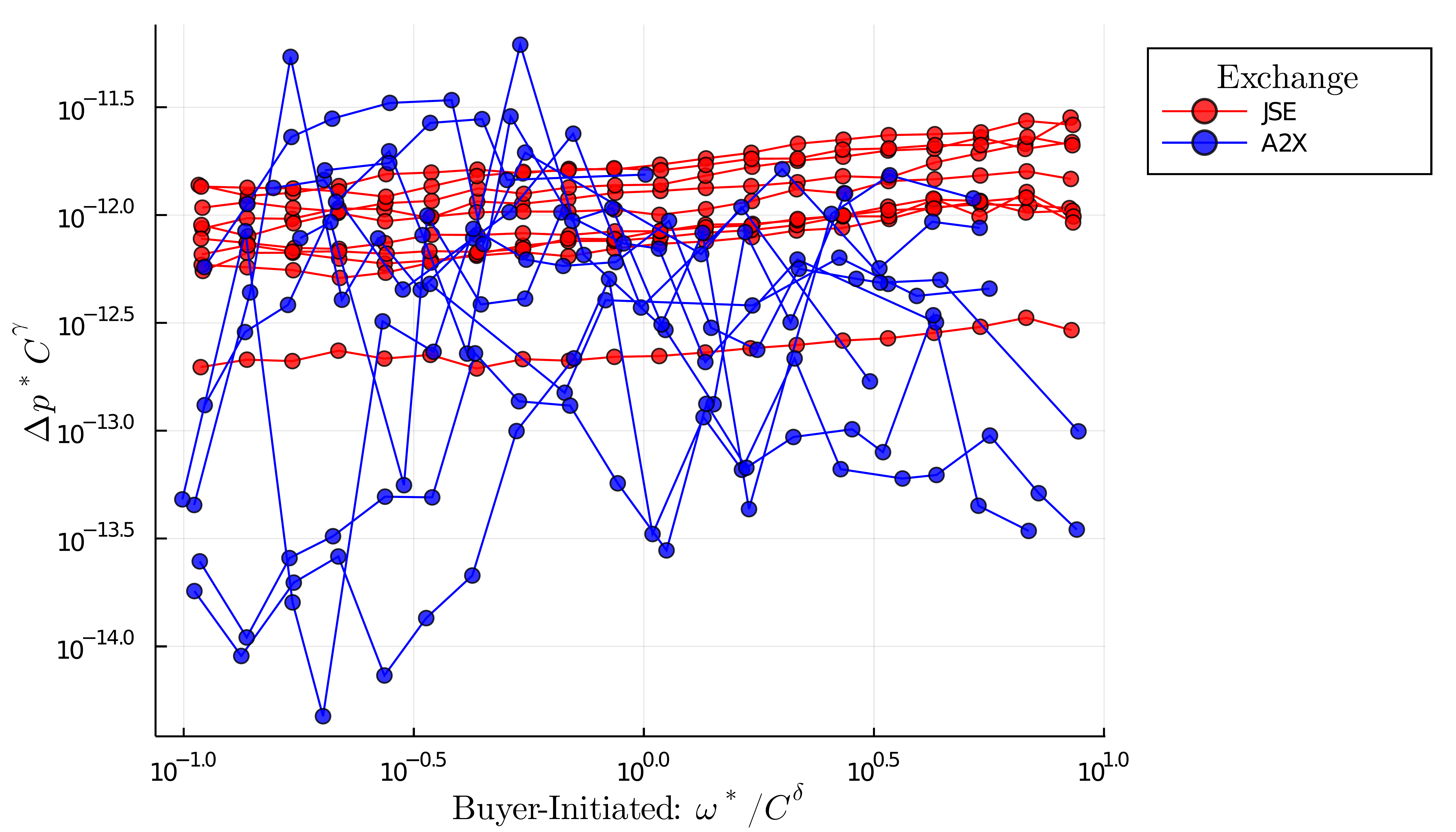}}
    \subfloat[Seller-initiated, $\delta = 0.000315, \gamma = -0.817194$]{\label{SA:b}\includegraphics[width=0.48\textwidth]{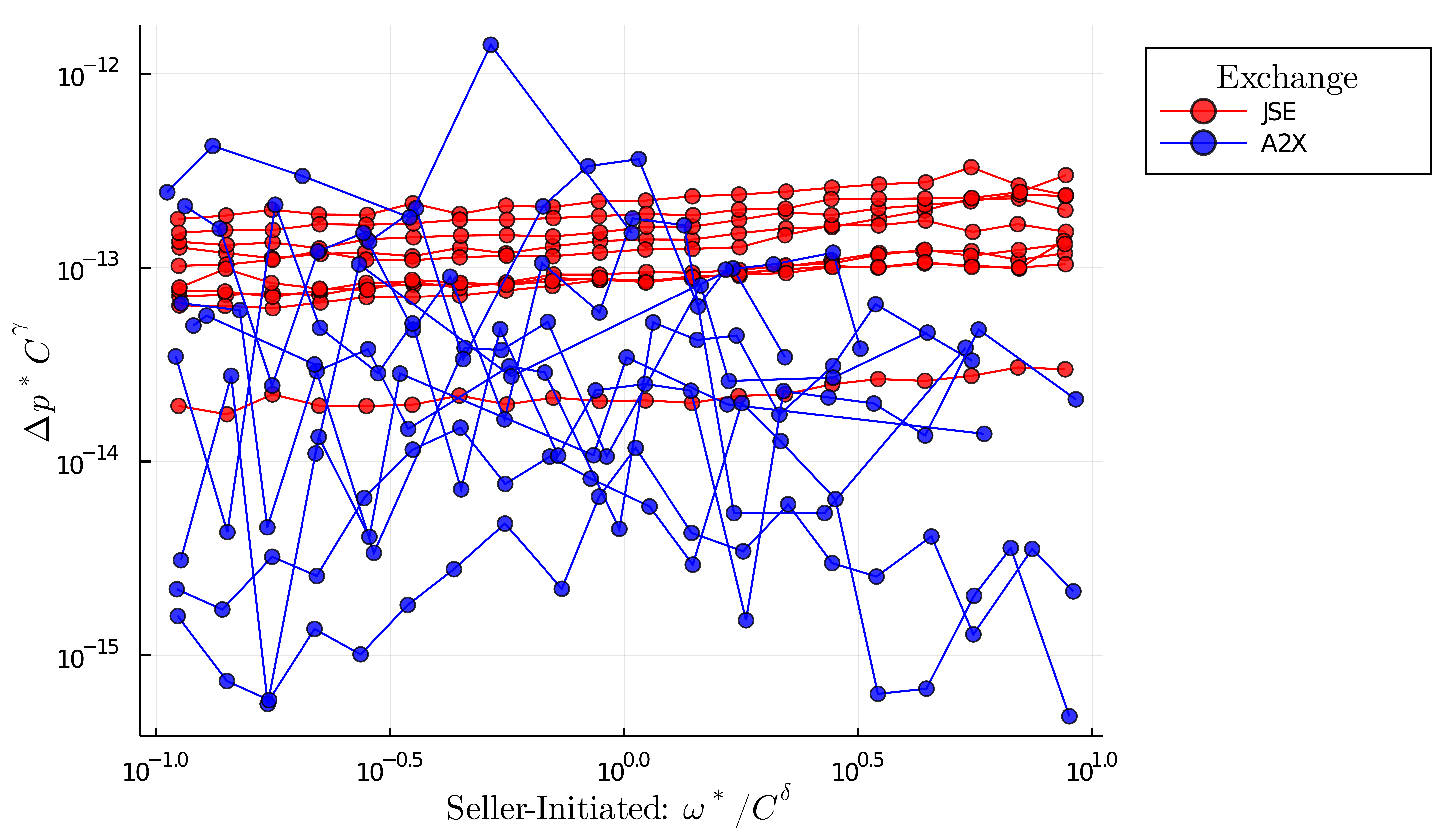}}
    \caption{Re-scaled price impact using all the equities under consideration. The collapse is achieved following a liquidity correction using \cref{eq:master} with common $\delta$ and $\gamma$ estimated using \cref{eq:est}. (a) is the buyer-initiated transactions and (b) is the seller-initiated transactions. As per the figure legends, securities from A2X are in blue and securities from the JSE are in red.}
\label{fig:SA}
\end{figure*}

\begin{figure*}[htb]
    \centering
    \subfloat[Buyer-initiated, $\delta = 0.001203, \gamma = -0.392986$]{\label{SACommon:a}\includegraphics[width=0.48\textwidth]{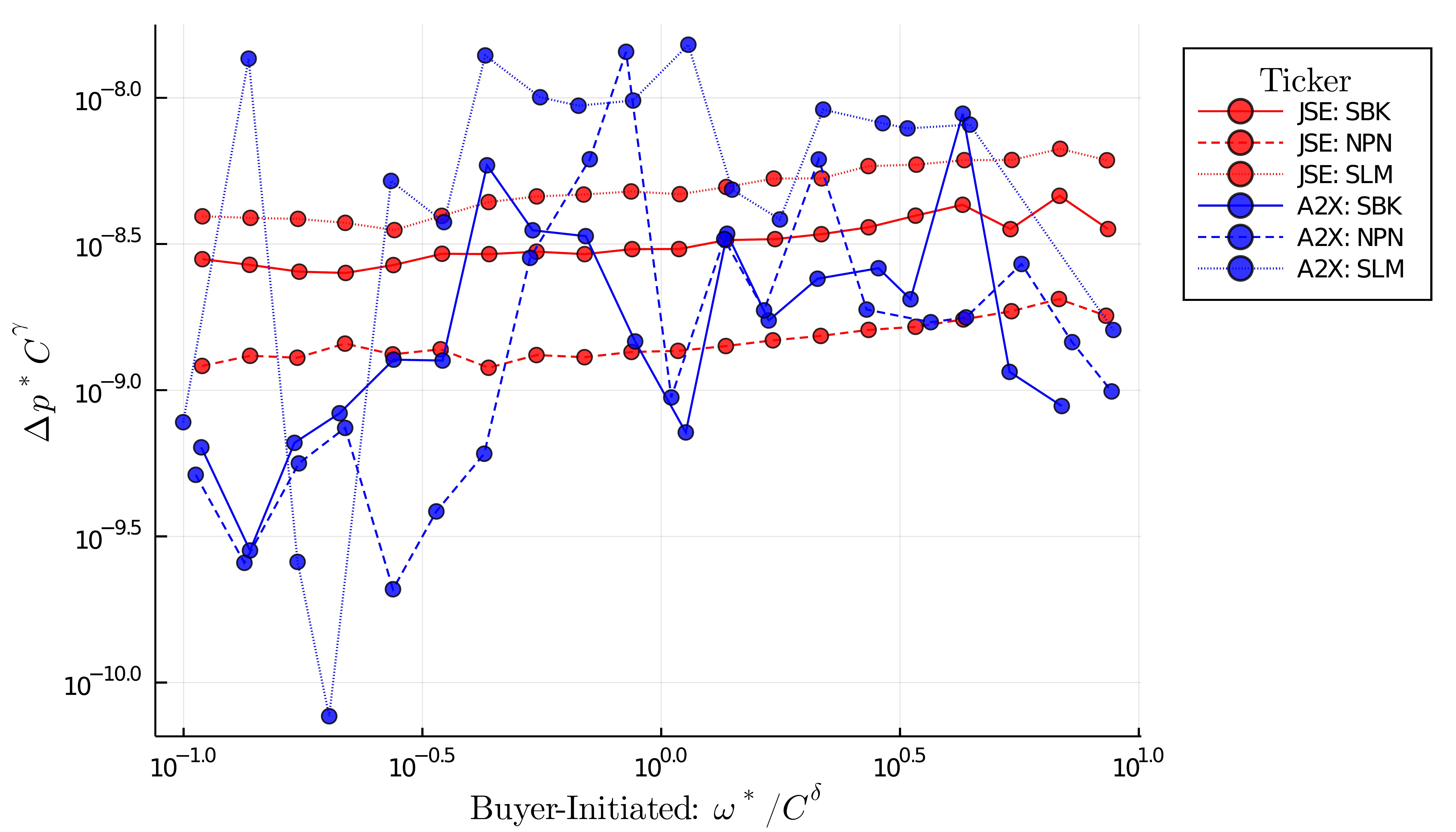}}
    \subfloat[Seller-initiated, $\delta = 0.000363, \gamma = -0.418756$]{\label{SACommon:b}\includegraphics[width=0.48\textwidth]{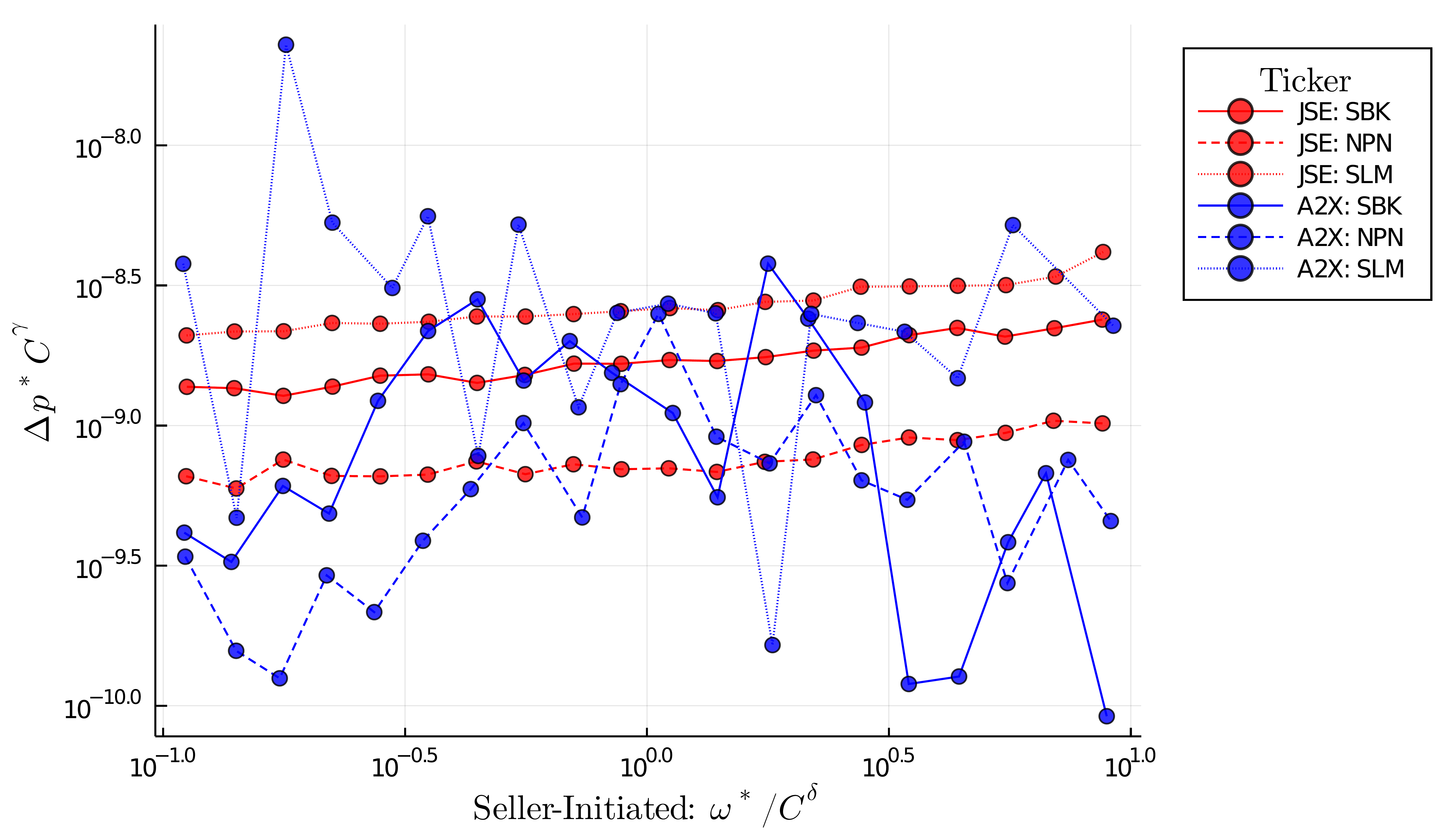}}
    \caption{Re-scaled price impact using the three common equities from each exchange. The collapse is achieved following a liquidity correction using \cref{eq:master} with common $\delta$ and $\gamma$ estimated using \cref{eq:est}. (a) is the buyer-initiated transactions and (b) is the seller-initiated transactions. As per the figure legends, securities from A2X are in blue and securities from the JSE are in red.}
\label{fig:SACommon}
\end{figure*}

\subsection{Price impact and master curves} \label{ssec:impact}

Price impact quantifies how a transaction of a given volume affects the price \cite{HHGD2016}. For each exchange, we use 10 equities to construct price impact functions. These are then used to calibrate master curves for each exchange. This then allows us to try investigate any differences in the price response from each exchange. The list of equities considered can be found in \Cref{tab:tickers} of \ref{app:tickers}. We only consider the immediate price impact.

The immediate price impact is the impact between a transaction volume and the immediate price increment which follows. Let $m_t$ be the mid-price then we define the impact of a transaction occurring at time $t$ as:
\begin{equation}
    \Delta p_{t_{k}} = \log\left( m_{t_{k+1}} \right) - \log\left( m_{t_{k}} \right),
\end{equation}
where $t_{k}$ is the time of mid-price before the transaction and $t_{k+1}$ is the time of mid-price immediately after the transaction.

\begin{remark}\label{rmk:1}
    We make the strict assumption to only use the mid-price immediately after the transaction. For the JSE data set, all transactions are immediately followed by an updated quote. For A2X, we also update the quote immediately following a transaction, however due to the lack of liquidity, the order book sometimes breaks following a transaction with no best bid or ask on offer. Therefore, we ignore the impact for these transactions.
\end{remark}

In order to compare different securities with potentially different liquidities and volumes, we follow \citet{HHGD2016} and \citet{LC2005} and normalise the trade volume as:
\begin{equation}\label{eq:PI}
    \omega_{i j}=\frac{v_{i j}}{\sum_{k=1}^{T_{j}} v_{k j}}\left[\frac{\sum_{j=1}^{N} T_{j}}{N}\right],
\end{equation}
where $\omega_{ij}$ is the normalised daily-normalised volume for trade $i$ on day $j$, $T_j$ is the number of trading events on the $j$th day and $N$ is the total number of days. The relationship between price change and transaction size is investigated separately for buyer- and seller-initiated transactions. The transactions are classified according to the Lee--Ready rule \cite{LR1991}.

\begin{remark}\label{rmk:2}
    We highlight that \cref{eq:PI} for the same equity on different exchanges does not translate to the same volume size. It is dependent on relative volume sizes with respect to the exchange it is on. However, we argue that this is a suitable approach as we will be comparing the largest (smallest) trades in A2X against the largest (smallest) trades in the JSE.
\end{remark}

In \Cref{fig:Impact} we investigate the price impact for a given volume size. We create 20 logarithmically spaced normalised volume bins between $[10^{-3}, 10^{1}]$ for each equity. For each of these volume bins we compute the average price change $\Delta p^*$ and the average normalised volume $\omega^*$ for the equity over the $N$ days. The relationship between the average price change $\Delta p^*$ and the average volume bin $\omega^*$ is then plotted on a log-log scale for buyer- and seller-initiated transactions for each equity on each exchange. 

The variability of the average impact in each security is demonstrated by applying bootstrap re-sampling to find indicative errors for the estimates. For each security, we obtain 1,000 bootstrap samples (sampled with replacement and equal in size to the observed data), impact curves are then constructed using the samples and the error bars capture 95\% of the bootstrap estimates.

Our results for the JSE are consistent with previous findings \cite{HHGD2016,LC2005,WCBH2015,Zhou2012} where there is a deviation from the expected linear relationships for smaller volumes. Moreover, for volume sizes between $[10^{-1}, 10^{1}]$, the relationship between the price change and normalised volume on the JSE seems to follow the relationship given as:
\begin{equation}\label{eq:LI}
    \Delta p^* = \frac{\text{sign} \left( \omega^* \right) |\omega^*|^{\alpha}}{\lambda},
\end{equation}
where $\lambda$ denotes a liquidity parameter. This relationship is not present for A2X, which on average maintains a relatively flat relationship between price change and normalised volume. The large measurement error intervals for A2X suggest the price impact estimates on A2X are not reliable. This is likely due to the low number of transactions on the exchange. On the other hand, the small intervals for the JSE confirms the relationship between price change and normalised volume.

Nonetheless, we look to re-scale and collapse the price impact curves into a single master curve for price impact as suggested by \citet{LFM2003}. They conjecture a power-law relation between price impact and transaction size which follows the functional form of:
\begin{equation}\label{eq:master}
    \Delta p^* \left( \omega^*, C \right) = C^{-\gamma} f \left( \omega^* C^{\delta} \right),
\end{equation}
where $C$ is used as a proxy for liquidity. We follow \citet{HHGD2016} and use the average daily value traded as the proxy rather than the average market capitalisation used by \citet{LFM2003}.\footnote{Market capitalisation as a proxy for liquidity will not work here because we have common securities across the exchanges with the same market capitalisation but with very different liquidity levels and trading volumes.} By re-scaling the price impact and volume axes, \citet{LFM2003} were able to collapse the various price impact curves into a single price impact curve known as the master curve. The parameters $\gamma$ and $\delta$ are chosen in order to best collapse all the data into the master curve.

To estimate the parameters, the normalised trade volume is first divided into $N_{\text{bins}}$. For each bin, the mean $(\mu^{(k)})$ and standard deviation $(\sigma^{(k)})$ are computed for both the re-scaled normalised volume $\left(x \rightarrow \omega^* / C^{\delta} \right)$ and the re-scaled price impact $\left(y \rightarrow \Delta p^* C^{\gamma} \right)$. The parameters are then chosen such that they minimise the average two-dimensional variance given as:
\begin{equation}\label{eq:est}
    \epsilon = \frac{1}{N_{\text{bins}}} \sum_{k=1}^{N_{\text{bins}}} \left[ \left( \frac{\sigma_x^{(k)}}{\mu_x^{(k)}} \right)^2 + \left( \frac{\sigma_y^{(k)}}{\mu_y^{(k)}} \right)^2 \right].
\end{equation}
\Cref{fig:Impact} we saw that there seemed to be an approximate linear relationship in log scale between the average price impact and average normalised trade volume for normalised trade volumes larger than $10^{-1}$ on the JSE. Therefore, the calibration of parameters is done using 20 logarithmically spaced bins between $[10^{-1}, 10^{1}]$.

\Cref{fig:Master} plots the resulting re-scaled price impact curves after calibrating the parameters separately for each exchange and for buyer- and seller-initiated transactions. The resulting parameter estimates are given in the sub-captions of the various figures. 
Additionally, the figures include estimates of the master curves for each exchange by averaging the re-scaled impacts across the securities on the exchange. To understand the variability of the master curve, we again apply bootstrap re-sampling to obtain 1,000 master curve estimates. Each bootstrap estimate is obtained by bootstrapping impact curves for each security, re-scaling the impact curves using the estimated parameters (provided in the sub-captions) and then averaging the re-scaled bootstrapped impacts across the securities. The error bars then capture 95\% of the master curve estimates.

We see the existence of a master curve for both buyer- and seller-initiated transactions on the JSE (notice the scale of the y-axis). The individual re-scaled price impact curves follow a linear relationship in the log scale. Interestingly, the re-scaled price impacts for A2X do not present a clear linear relationship in the log scale, rather the relationship looks rather flat with fluctuations. This highlights the difference in price response on the two exchanges. 

We argue that the price impacts on A2X are rather flat because the impact is larger for smaller volumes. This can be seen in \Cref{fig:Impact} where the impact on A2X for smaller volumes are an order of magnitude larger than the impact on JSE. We argue that this is due to the lack of liquidity in the order-book. This can be seen in \Cref{fig:FMD} where we visualise the full market depth of A2X. We see that the most liquid ticker on the exchange (Naspers) often only has the best bid or ask on offer and at other times has at most two to three layers on either side of the order book at much deeper levels. This means that if an incoming market order depletes the best bid or ask on offer then there will be a significant change in mid-price which results in a large impact.

We now look at the possibility of master curves for the entire South African market by calibrating the master curves using the 10 securities from each exchange. \Cref{fig:SA} plots the resulting re-scaled price impact curves after calibrating all the equities under consideration for buyer- and seller-initiated transactions. The resulting parameter estimates are given in the sub-captions of the figures.

We see that the collapse does not result in clear master curves for the South African market, rather we can very clearly see the differences in price impact between the two exchanges. This means that we cannot use the well defined master curves from the JSE to price the impact for a given transaction in the very illiquid A2X.

Since we are unable to obtain clear master curves for the entire South African market, let us look at finding master curves for the same securities between the exchanges. To this end, we consider Standard Bank (SBK), Naspers (NPN) and Sanlam (SLM) which are listed on both the JSE and A2X. \Cref{fig:SACommon} plots the resulting re-scaled price impact curves after calibrating the common equities for buyer- and seller-initiated transactions. The resulting parameter estimates are given in the sub-captions of the figures. What is interesting is that even using the same securities, master curves between the exchanges are still unattainable. There is a very clear difference in price impact between the two exchanges.

The existence of the master curves on the JSE from the single-curve collapse of the price impact functions suggests that there is indeed a common statistical rule that governs the relationship between price and volume across sectors, securities and market captialisations. However, the failure to find a common master curve for the same securities between exchanges suggests that the emergence of the common statistical rule between price and volume very much depends on the underlying market microstructure of the exchange.

Exactly what market microstructure conditions are required to lead to a common statistical rule between price and volume is not clear. However, we suspect that sufficient market participants and liquidity in the order-book is required to observe the master curves. This is because if there is little competition in the order book then there are no incentives to reduce the spread in the layers beyond the best on offer.

\begin{figure*}[htb]
    \centering
    \subfloat[Buyer-initiated]{\label{Spread:a}\includegraphics[width=0.48\textwidth]{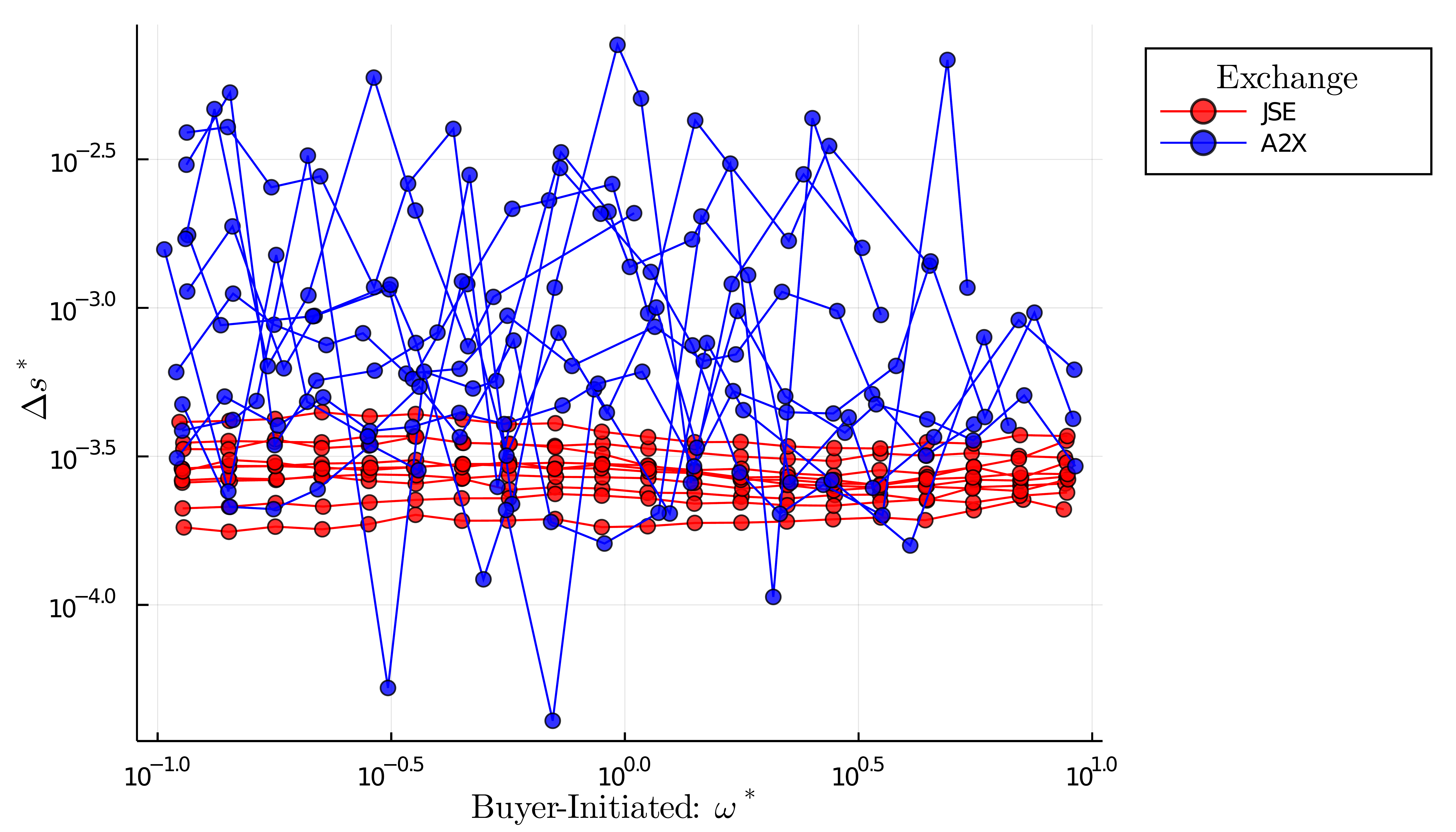}}
    \subfloat[Seller-initiated]{\label{Spread:b}\includegraphics[width=0.48\textwidth]{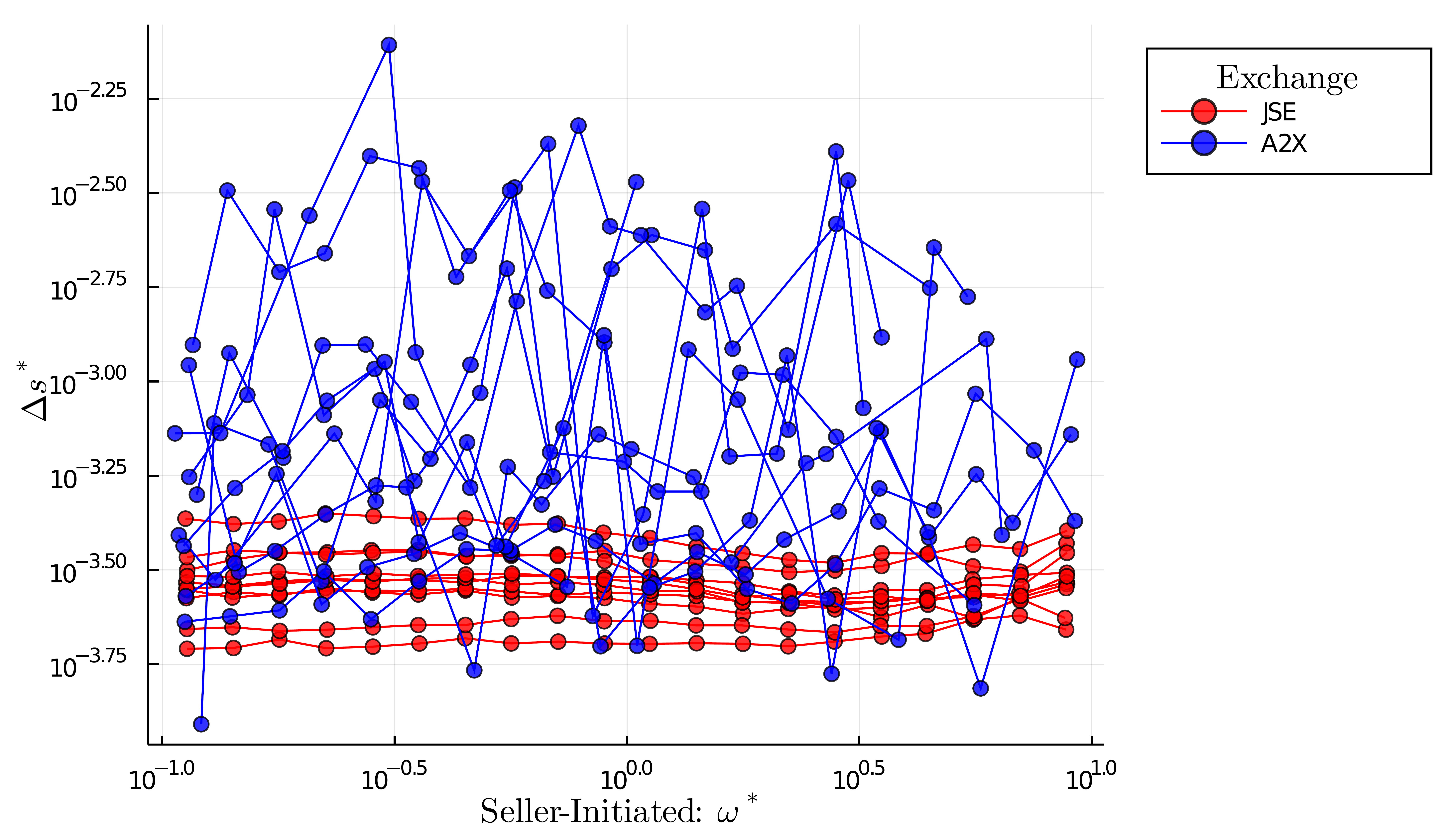}}
    \caption{Slippage cost for 10 equities from each exchange. As per the figure legends, securities from A2X are in blue and securities from the JSE are in red.}
\label{fig:Spread}
\end{figure*}

\begin{figure*}[htb]
    \centering
    \subfloat[Buyer-initiated]{\label{Direct:a}\includegraphics[width=0.48\textwidth]{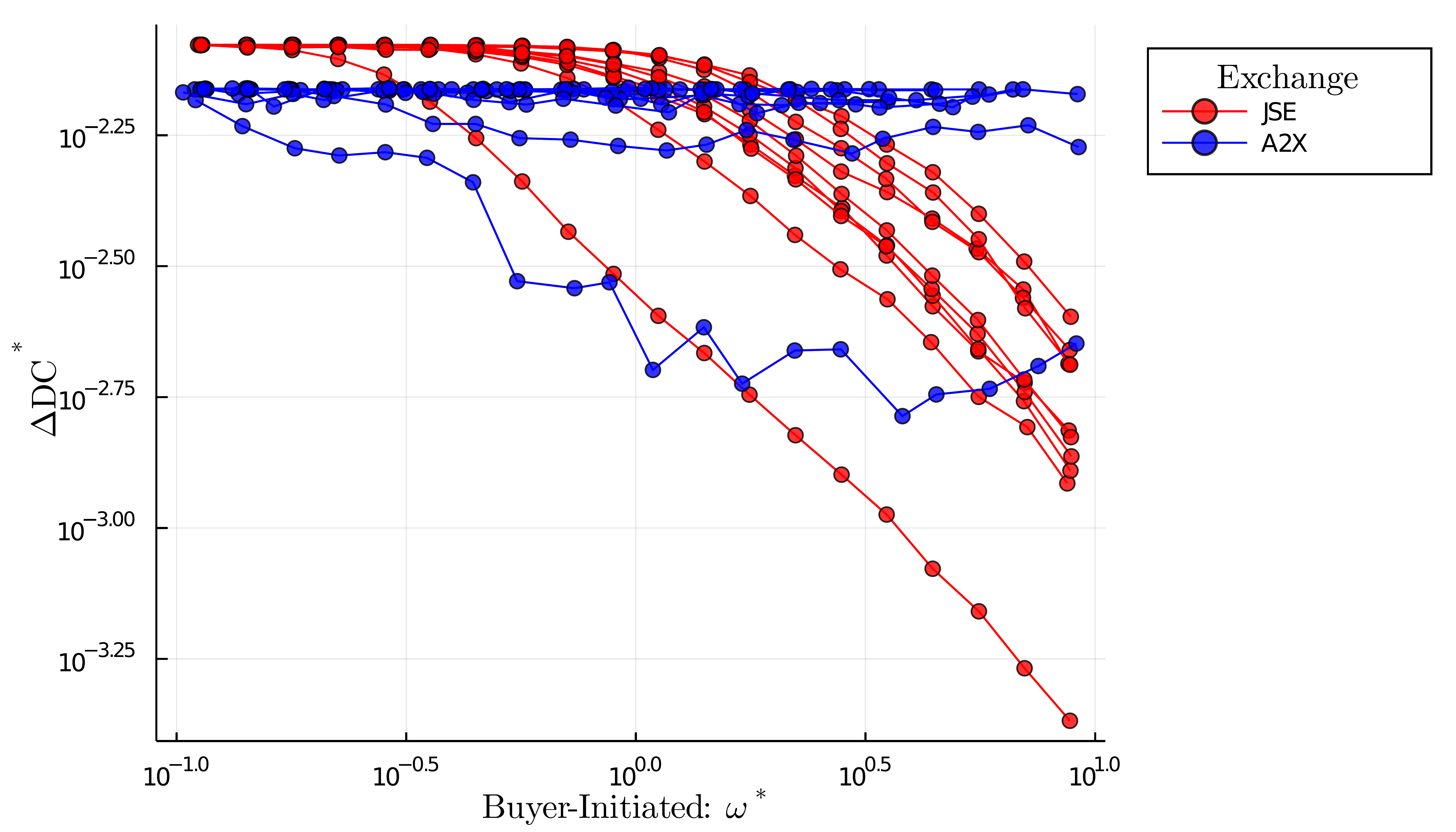}}
    \subfloat[Seller-initiated]{\label{Direct:b}\includegraphics[width=0.48\textwidth]{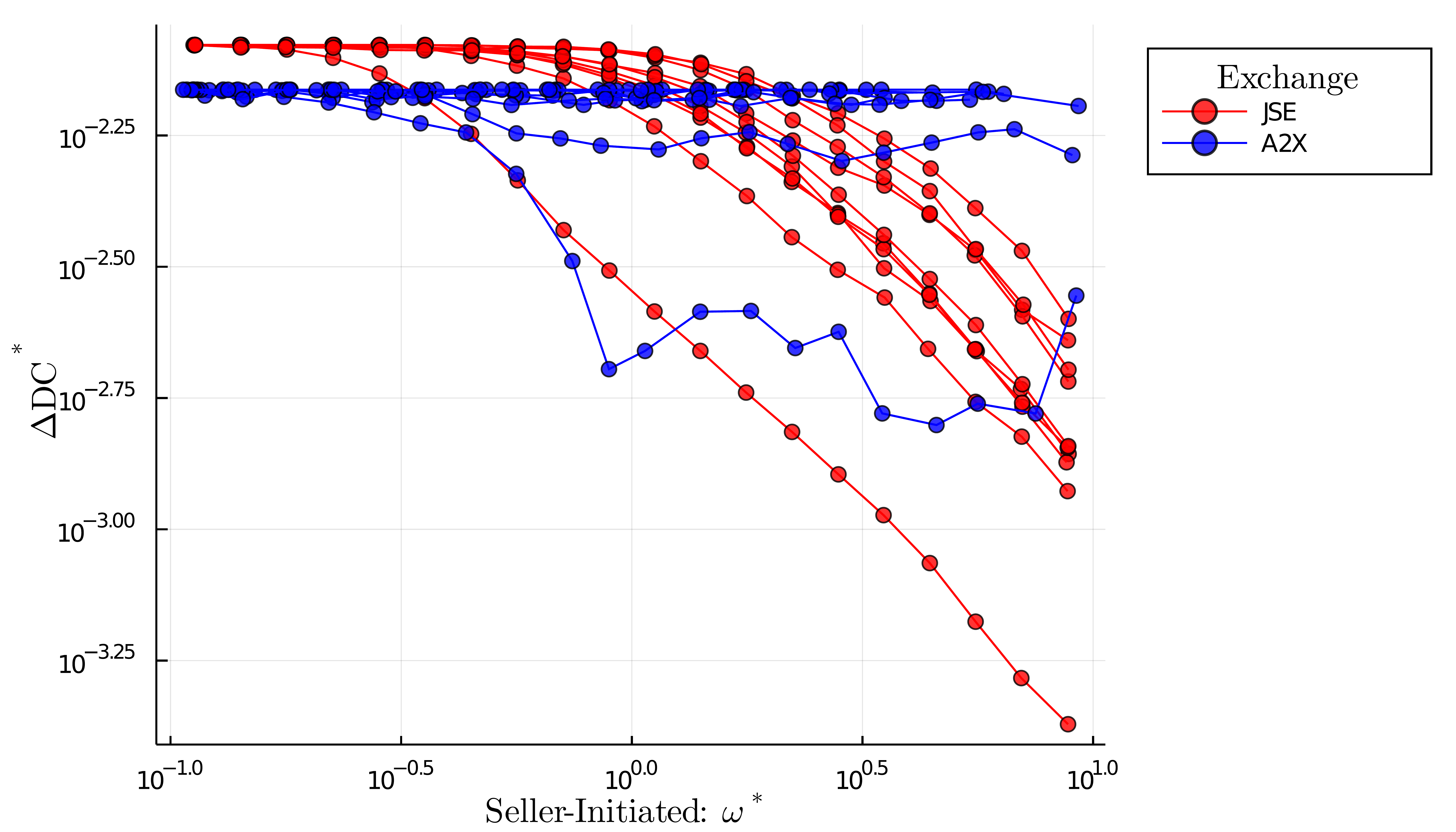}}
    \caption{Direct cost from transaction and settlement for 10 equities from each exchange. As per the figure legends, securities from A2X are in blue and securities from the JSE are in red.}
\label{fig:Direct}
\end{figure*}

\begin{figure*}[htb]
    \centering
    \subfloat[Buyer-initiated]{\label{ImpactComp:a}\includegraphics[width=0.48\textwidth]{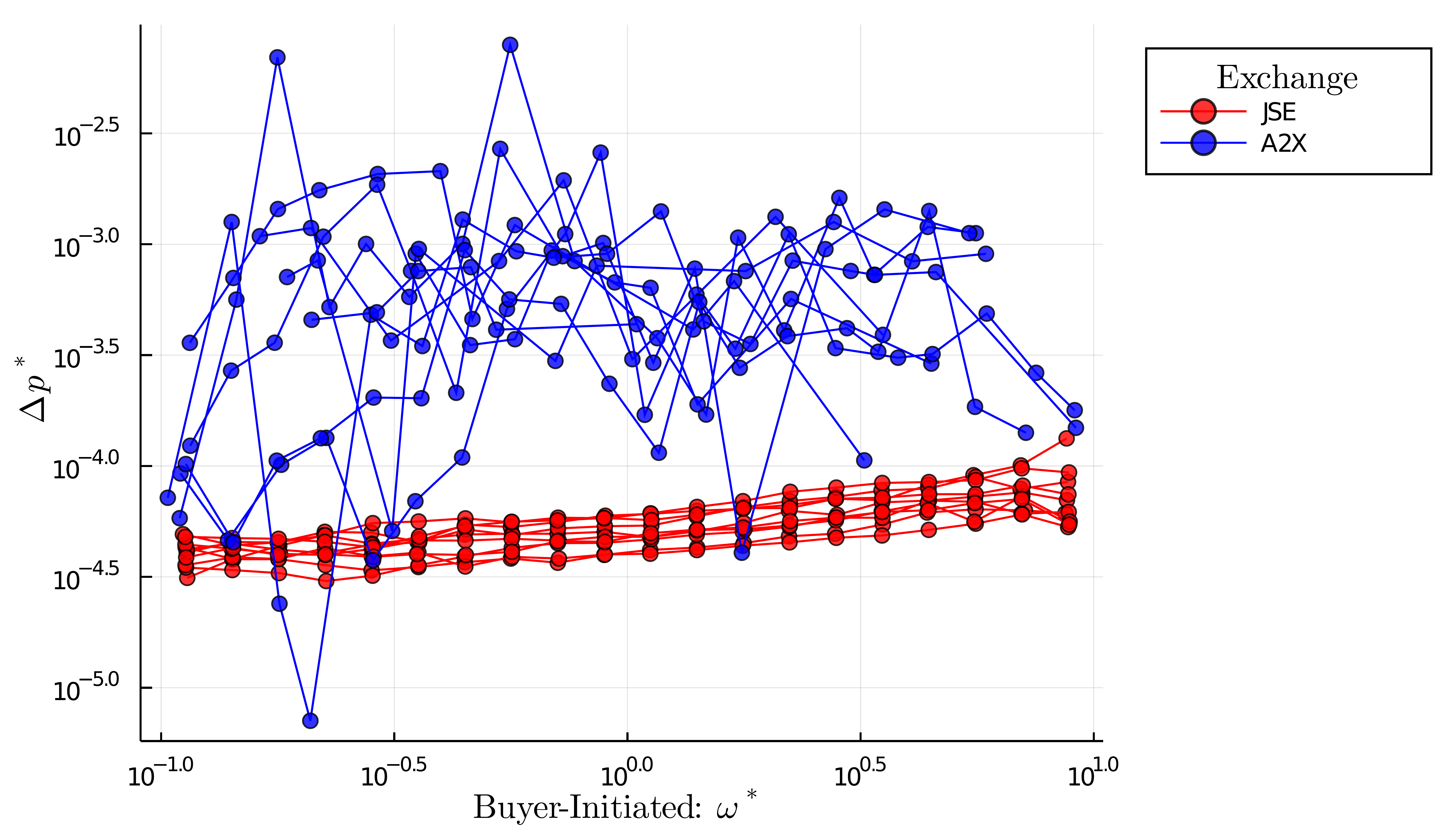}}
    \subfloat[Seller-initiated]{\label{ImpactComp:b}\includegraphics[width=0.48\textwidth]{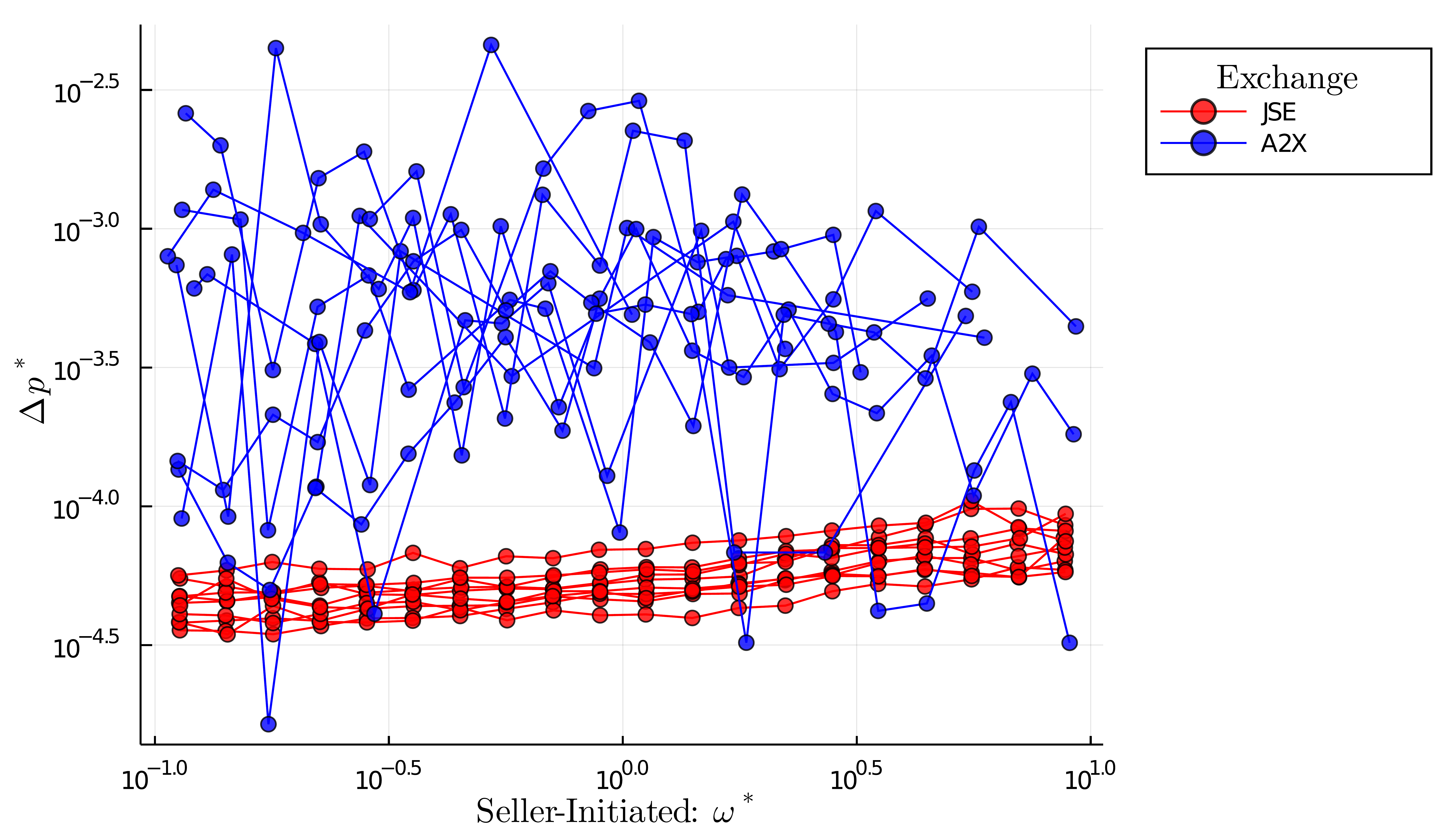}}
    \caption{Impact cost for 10 equities from each exchange. As per the figure legends, securities from A2X are in blue and securities from the JSE are in red.}
\label{fig:ImpactComp}
\end{figure*}

\begin{figure*}[htb]
    \centering
    \subfloat[Buyer-initiated]{\label{cost:a}\includegraphics[width=0.48\textwidth]{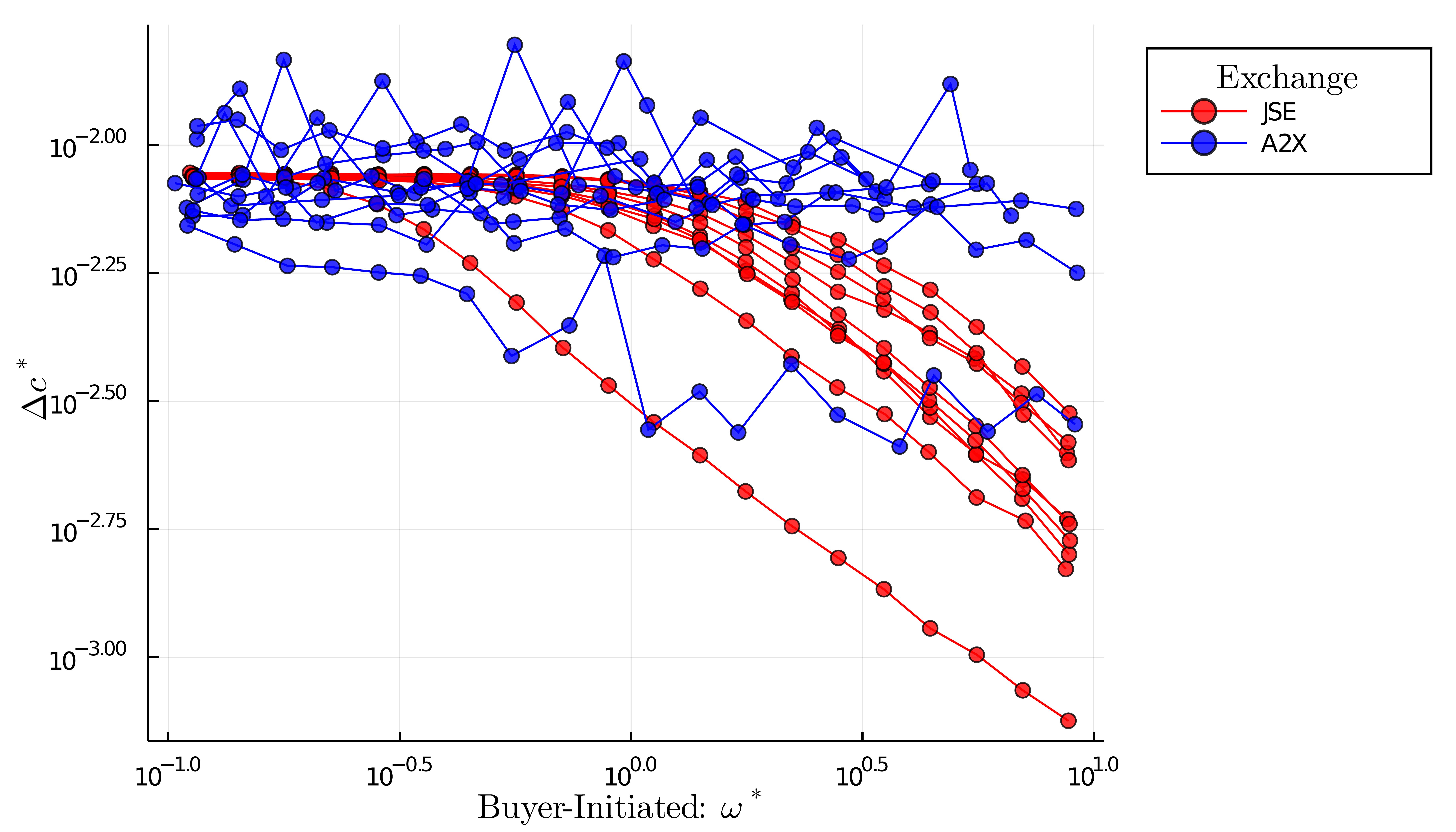}}
    \subfloat[Seller-initiated]{\label{cost:b}\includegraphics[width=0.48\textwidth]{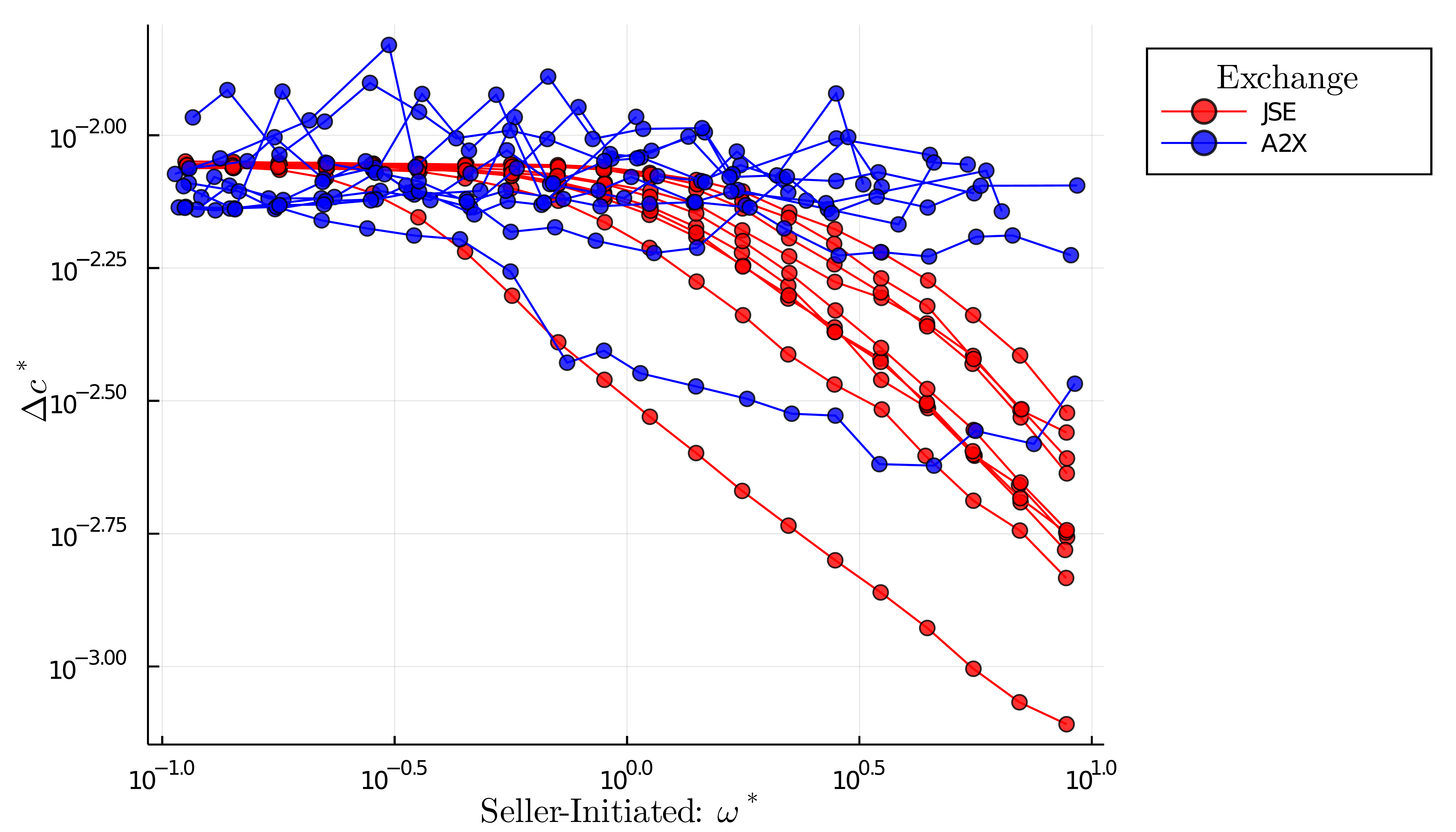}}
    \caption{Total cost for 10 equities from each exchange. As per the figure legends, securities from A2X are in blue and securities from the JSE are in red.}
\label{fig:cost}
\end{figure*}

\subsection{Cost of trading} \label{ssec:costs}

From \Cref{fig:Impact} we see that the JSE presents a cost advantage over A2X in terms of price impact. However, price impact is only one component of the total cost of trading. Other sources contributing towards the total cost come from the cost of crossing the half spread when using market orders and from direct costs such as the transaction and settlement fees charged by each exchange. Concretely, the Total Cost (TC) of a given transaction in Rands is:
\begin{equation}\label{eq:TC}
    \text{TC} = \underbrace{\text{Slippage + Impact}}_{\text{Indirect costs}} + \text{Direct Costs}.
\end{equation}
There are a few subtleties and assumptions that we make when computing \cref{eq:TC}. We are interested in computing the exact cost of trading, therefore we do not approximate the indirect costs using the square-root formula (see \citet{Gatheral2010}). Rather, we are going to try and exactly quantify the total cost per share for a transaction at a given volume. 

The first term {\it slippage} is traditionally computed using the spread. Here we compute it as the half-spread since we are not concerned with round trip trades but are rather interested in the cost relative to the mid-price prior to the transaction (as in the case with the impact). The second term {\it impact} is the absolute difference in the mid-price immediately after and prior to the transaction. The third term {\it Direct Costs} include the transaction and settlement fees charged by each exchange. Settlement fees are charged at the end of the day once the total shares bought and sold by a party is tallied up. However, we make the assumption that all trades are made by independent parties so that we can investigate the cost of any given transaction.

The Direct Costs (DC) are computed as:
\begin{equation}\label{eq:DC}
    \text{DC} = f \left( \underbrace{\text{Trade price} \times \text{volume}}_{\text{Transaction value}} \right) / \text{volume},
\end{equation}
where $f(\cdot)$ is:
\begin{equation}\label{eq:DC2}
\begin{aligned}
    f(x) 
    &= \min\{ x \cdot \text{bps}_{\text{transaction}}, \text{ceiling}_{\text{transaction}} \}  \\
    &+ \min\{ x \cdot \text{bps}_{\text{settlement}}, \text{ceiling}_{\text{settlement}} \}.
\end{aligned}
\end{equation}
There are two things to notice about \cref{eq:DC}. First, the direct cost is computed based on the value of the transaction (mid-price is not used here). Second, the direct cost for a given transaction is normalised by the number of shares in that transaction. This is because we are computing the total cost per share for a transaction.

The various cost components are given in Rand value and the magnitudes of these will vary depending on the mid-price and trade price as they move through time. Thus we require a method to normalise these components so that they are comparable. To this end, we will compute the \% change in value of these costs relative to the mid-price prior to the transaction. This gives us a slippage cost of:
\begin{equation}\label{eq:slip}
    \Delta s_{t_k} = 
    \begin{cases} 
      \log\left( m_{t_k} + \text{Slippage} \right) - \log\left( m_{t_k} \right) & \text{if BI} \\
      \log\left( m_{t_k} - \text{Slippage} \right) - \log\left( m_{t_k} \right) & \text{if SI}.
   \end{cases}
\end{equation}
Direct cost of:
\begin{equation}\label{eq:direct}
    \Delta \text{DC}_{t_k} = 
    \begin{cases} 
      \log\left( m_{t_k} + \text{DC} \right) - \log\left( m_{t_k} \right) & \text{if BI} \\
      \log\left( m_{t_k} - \text{DC} \right) - \log\left( m_{t_k} \right) & \text{if SI}.
   \end{cases}
\end{equation}
Impact cost of:
\begin{equation}\label{eq:imp}
    \Delta p_{t_k} = 
    \begin{cases} 
      \log\left( m_{t_k} + \text{Impact} \right) - \log\left( m_{t_k} \right) & \text{if BI} \\
      \log\left( m_{t_k} - \text{Impact} \right) - \log\left( m_{t_k} \right) & \text{if SI}.
   \end{cases}
\end{equation}
Finally, the total cost is:
\begin{equation}\label{eq:total}
    \Delta c_{t_k} = 
    \begin{cases} 
      \log\left( m_{t_k} + \text{TC} \right) - \log\left( m_{t_k} \right) & \text{if BI} \\
      \log\left( m_{t_k} - \text{TC} \right) - \log\left( m_{t_k} \right) & \text{if SI}.
   \end{cases}
\end{equation}
Note that \cref{eq:imp,eq:PI} are the same once \cref{eq:PI} has had the signs accounted for BI and SI where BI are Buyer-Initiated transactions and SI are Seller-Initiated transactions.\footnote{Strictly speaking, \cref{eq:slip,eq:direct,eq:imp,eq:total} are not \% change in value, but are rather approximations of it.}

To compare the various cost components of trading at a given volume size, we create 20 logarithmically spaced normalised volume bins between $[10^{-1}, 10^1]$ for each security. For each of these volume bins we compute the average change in slippage $\Delta s^*$, average change in direct costs $\Delta \text{DC}^*$, average change in impact $\Delta p^*$, average change in total costs $\Delta c^*$ and the average normalised volume $\omega^*$ for the security over the $N$ days. The relationships between the average change in each cost component ($\Delta s^*$, $\Delta \text{DC}^*$, $\Delta p^*$ and $\Delta c^*$) are plotted against the average volume bin $\omega^*$ on a log-log scale for buyer- and seller-initiated transactions for each security on each exchange. We color the securities from A2X in blue and the securities from the JSE in red for a comparison of costs between the exchanges.

\Cref{fig:Spread} compares the slippage cost between the two exchanges for buyer- and seller-initiated transactions. We see that the slippage cost on the JSE is lower than that on A2X. Additionally, we see that cost is consistent for different transaction sizes and across securities on the JSE whereas there are much larger deviations and variability for the slippage costs for different transaction sizes and across different securities on A2X. It is not unreasonable to conjecture that the larger spread on A2X is due to liquidity providers guarding against adverse selection \cite{ELO2012B} in a low liquidity environment such as A2X.

\Cref{fig:Direct} compares the direct costs between the two exchanges. We see that the lower cost for A2X prevails against the JSE for transactions with smaller volumes. However, we see that the direct costs on the JSE start to decrease as the transaction size increases. This is because there is a ceiling limit on \cref{eq:DC2} which means that \cref{eq:DC} will start to decrease as the size of the transaction continues to increase. What is interesting is that with the exception of one security on A2X, the transaction values on A2X are not large enough to reach the ceiling limit for \cref{eq:DC2}. It must be noted that this cannot be fixed even if the x-axis were in a different measurement unit, say transaction value.

\Cref{fig:ImpactComp} compares the price impact between the two exchanges. The figure is the combined version of \Cref{fig:Impact} without error bars for easier comparison. We see that the impact is significantly lower on the JSE compared to A2X. Additionally, we observe the expected linear relationship on the JSE whereas there are much larger larger deviations and variability in the price impact for different transaction sizes and across different securities on A2X.

\Cref{fig:cost} compares the total cost of trading between the two exchanges. By looking at the y-axis of \Cref{fig:Spread,fig:Direct,fig:ImpactComp} we see that direct costs have the largest contribution towards the total cost. Therefore, despite the larger cost on A2X in terms of slippage and price impact, due to the lower direct cost, we see that for $\omega^*$ less than 1 that the total cost from both exchanges are at a similar level. On the other hand, for large $\omega^*$ the ceiling limit kicks in on the JSE and thereby reducing the cost of trading. This however does not occur for A2X as the transaction values on the exchange are too small to reach the ceiling limit. 

Although the average total transaction cost in \Cref{fig:cost} seems to be similar for $\omega^*$ less than 1, we must highlight the large variability between each individual transactions on the A2X. \Cref{tab:var} reports the average standard deviation for each cost component over the securities of each exchange, where the standard deviation for each security is the variability between each transaction for buyer- and seller-initiated transactions. We see that the variability between the transactions on the JSE is very consistent, with slightly more variability for larger transactions sizes. However, we see that the variability between the transactions on the A2X is very high---sometimes being an order of magnitude larger than the variability on the JSE. The costs may be similar on average, but there is a high uncertainly on what the cost will be for a particular transaction on A2X, making it harder to predict.

Finally, it is unclear what normalisation of the x-axis can provide a better comparison between the exchanges, but following \cref{rmk:2} we argue that normalised transaction volume is still the most suitable method as we are comparing the transaction volumes between exchanges as proportional to their own volumes. It is unfortunate that the transaction values are A2X are not large enough to see the benefit of the lower ceiling limits given the dominant role direct costs plays towards the total costs of trading. Therefore, it is of interest to repeat the analysis again once A2X has matured in the future.

\setlength{\tabcolsep}{0.65em}
\renewcommand{\arraystretch}{.85}
\begin{sidewaystable}[p]
\centering
\begin{tabular}{r|rrrr|rrrr|rrrr|rrrr|rrrr}

Bin & \multicolumn{4}{c}{$\overline{\sigma(\Delta p)} [\cdot 10^3]$}                        & \multicolumn{4}{c}{$\overline{\sigma(\Delta s)} [\cdot 10^3]$}                        & \multicolumn{4}{c}{$\overline{\sigma(\Delta \text{DC})} [\cdot 10^3]$}                  & \multicolumn{4}{c}{$\overline{\sigma(\Delta c)} [\cdot 10^3]$}                         & \multicolumn{4}{c}{$\overline{\sigma(\omega)} [\cdot 10^1]$}                         \\   &
\multicolumn{2}{c}{JSE} & \multicolumn{2}{c}{A2X} & \multicolumn{2}{c}{JSE} & \multicolumn{2}{c}{A2X} & \multicolumn{2}{c}{JSE} & \multicolumn{2}{c}{A2X} & \multicolumn{2}{c}{JSE} & \multicolumn{2}{c}{A2X} & \multicolumn{2}{c}{JSE} & \multicolumn{2}{c}{A2X} \\ \hline
& BI           & SI         & BI          & SI          & BI           & SI         & BI          & SI          & BI          & SI          & BI           & SI         & BI          & SI          & BI          & SI          & BI          & SI          & BI          & SI         
 \\ 
  \hline
  1 & 0.12 & 0.13 & 0.27 & 1.02 & 0.29 & 0.29 & 1.06 & 0.65 & 0.01 & 0.01 & 0.10 & 0.02 & 0.31 & 0.31 & 1.20 & 1.04 & 0.08 & 0.07 & 0.06 & 0.06 \\ 
  2 & 0.12 & 0.13 & 0.61 & 0.54 & 0.30 & 0.30 & 1.47 & 0.83 & 0.05 & 0.05 & 0.27 & 0.07 & 0.33 & 0.33 & 1.80 & 0.71 & 0.09 & 0.09 & 0.07 & 0.06 \\ 
  3 & 0.12 & 0.13 & 1.84 & 1.41 & 0.32 & 0.30 & 0.84 & 1.54 & 0.10 & 0.09 & 0.25 & 0.12 & 0.37 & 0.35 & 2.63 & 2.53 & 0.12 & 0.12 & 0.09 & 0.10 \\ 
  4 & 0.13 & 0.12 & 0.64 & 0.87 & 0.30 & 0.30 & 0.70 & 1.07 & 0.17 & 0.17 & 0.27 & 0.13 & 0.39 & 0.40 & 1.03 & 1.47 & 0.15 & 0.15 & 0.13 & 0.12 \\ 
  5 & 0.12 & 0.13 & 0.85 & 1.54 & 0.30 & 0.30 & 1.80 & 0.84 & 0.28 & 0.28 & 0.32 & 0.22 & 0.47 & 0.47 & 2.31 & 1.80 & 0.19 & 0.19 & 0.14 & 0.17 \\ 
  6 & 0.12 & 0.13 & 0.70 & 1.29 & 0.31 & 0.29 & 0.99 & 0.81 & 0.41 & 0.41 & 0.38 & 0.22 & 0.56 & 0.56 & 1.33 & 1.51 & 0.24 & 0.24 & 0.19 & 0.18 \\ 
  7 & 0.13 & 0.12 & 0.94 & 1.13 & 0.30 & 0.29 & 1.60 & 0.98 & 0.52 & 0.55 & 0.41 & 0.35 & 0.64 & 0.67 & 1.93 & 1.45 & 0.30 & 0.30 & 0.21 & 0.24 \\ 
  8 & 0.12 & 0.13 & 2.06 & 0.81 & 0.29 & 0.29 & 0.68 & 1.58 & 0.64 & 0.67 & 0.36 & 0.42 & 0.74 & 0.76 & 2.18 & 2.09 & 0.37 & 0.37 & 0.33 & 0.36 \\ 
  9 & 0.12 & 0.13 & 1.59 & 1.12 & 0.28 & 0.28 & 1.01 & 1.26 & 0.76 & 0.77 & 0.49 & 0.44 & 0.83 & 0.85 & 2.01 & 1.93 & 0.47 & 0.47 & 0.42 & 0.38 \\ 
  10 & 0.12 & 0.13 & 1.63 & 0.94 & 0.28 & 0.28 & 1.20 & 1.51 & 0.88 & 0.89 & 0.42 & 0.29 & 0.94 & 0.97 & 1.81 & 1.86 & 0.59 & 0.59 & 0.45 & 0.43 \\ 
  11 & 0.12 & 0.13 & 0.49 & 1.58 & 0.26 & 0.27 & 1.54 & 1.05 & 1.03 & 1.04 & 0.26 & 0.32 & 1.08 & 1.11 & 1.55 & 2.05 & 0.75 & 0.75 & 0.52 & 0.53 \\ 
  12 & 0.13 & 0.13 & 0.73 & 0.94 & 0.26 & 0.26 & 1.12 & 0.99 & 1.20 & 1.20 & 0.37 & 0.46 & 1.25 & 1.26 & 1.56 & 1.26 & 0.94 & 0.94 & 0.89 & 0.74 \\ 
  13 & 0.13 & 0.13 & 0.67 & 1.01 & 0.25 & 0.25 & 1.17 & 0.73 & 1.33 & 1.32 & 0.28 & 0.32 & 1.37 & 1.39 & 1.31 & 1.26 & 1.17 & 1.18 & 1.06 & 0.96 \\ 
  14 & 0.15 & 0.13 & 1.24 & 1.07 & 0.25 & 0.27 & 0.58 & 0.90 & 1.38 & 1.37 & 0.48 & 0.54 & 1.42 & 1.43 & 1.31 & 1.55 & 1.47 & 1.48 & 1.28 & 1.31 \\ 
  15 & 0.14 & 0.15 & 1.10 & 0.61 & 0.26 & 0.26 & 0.39 & 1.37 & 1.34 & 1.33 & 0.33 & 0.26 & 1.38 & 1.39 & 1.25 & 1.44 & 1.87 & 1.86 & 1.00 & 1.51 \\ 
  16 & 0.15 & 0.15 & 0.87 & 0.54 & 0.26 & 0.28 & 0.54 & 0.76 & 1.31 & 1.28 & 0.25 & 0.28 & 1.35 & 1.34 & 0.95 & 0.93 & 2.34 & 2.33 & 1.37 & 1.54 \\ 
  17 & 0.16 & 0.15 & 1.24 & 0.47 & 0.26 & 0.29 & 0.95 & 1.09 & 1.18 & 1.21 & 0.40 & 0.32 & 1.22 & 1.27 & 1.60 & 1.19 & 2.93 & 2.92 & 1.85 & 1.88 \\ 
  18 & 0.16 & 0.16 & 0.66 & 0.66 & 0.29 & 0.34 & 0.91 & 1.02 & 1.09 & 1.07 & 0.33 & 0.20 & 1.14 & 1.15 & 1.14 & 1.41 & 3.71 & 3.72 & 2.10 & 2.16 \\ 
  19 & 0.17 & 0.16 & 0.34 & 0.34 & 0.33 & 0.34 & 0.44 & 0.48 & 0.91 & 0.91 & 0.66 & 0.36 & 0.98 & 1.00 & 0.55 & 0.40 & 4.73 & 4.67 & 2.68 & 2.47 \\ 
  20 & 0.16 & 0.16 & 0.46 & 0.44 & 0.32 & 0.39 & 0.73 & 1.02 & 0.74 & 0.74 & 0.73 & 0.84 & 0.82 & 0.87 & 0.90 & 1.37 & 5.81 & 5.81 & 5.28 & 2.98 \\ 
   \hline
\end{tabular}
\caption{The table reports the average standard deviation for each cost component over the securities of each exchange. The standard deviation for each security is the variability of \cref{eq:slip,eq:direct,eq:imp,eq:total,eq:PI} for buyer-(BI) and seller-(SI) initiated transactions.}
\label{tab:var}
\end{sidewaystable}

\section{Conclusions} \label{sec:conc}

We compared stylised facts for the same security listed on different exchanges and find that securities have different distributional properties in their microprice returns at the tick-by-tick scale. However, these differences vanish when considering bar data when moving to larger measurement time scales.

We compared price response dynamics between the two exchanges by considering 10 equities from each exchange. We found the existence of the master curve on the Johannesburg Stock Exchange from the single-curve collapse of the price impact functions. This suggests there is indeed a common statistical rule that can govern the relationship between price and volume across sectors, securities and market liquidity. However, the failure to find a master curve mapping to aggregate the exchanges using the same securities is not surprising and suggests that we cannot use well defined master curves from the JSE to price the impact for a given transaction on the illiquid A2X; a simple liquidity proxy scaling is insufficient. This may be because the emergence of universal statistical relationships between price changes and volume depends on broader set of market agent activity, market constraints and liquidity features --- however, it is clear that the liquidity differences are not well proxied by either market capitalisation, nor value traded.  

We systematically accounted for the total cost of trading by directly computing the slippage cost, price impact and direct costs for each transaction. We compared the various cost components between the exchanges and found that A2X has larger indirect costs associated with slippage and price impact, but because of its lower direct costs we found that both exchanges have similar trading costs for small to medium sized transactions (relative to their own volume). However, due to the additional liquidity and larger transaction values on the Johannesburg Stock Exchange, the cost of trading for very large transactions is lower than A2X because the ceiling limit comes into effect. Due to the limited liquidity and smaller transaction values on A2X, the exchange has yet to benefit from its lower ceiling limit in the direct costs. However, it will be interesting to perform the analysis again once A2X has matured for two reasons in particular. First, to calibrate master curves for the entire South African market and second, to illustrate the benefit of lower direct costs of A2X.

The question of the value of hidden orders remains a concern. The market structure and linkages between the hidden order mechanics and the Lit market order mechanics remains unclear --- in particular the value of using Hidden orders on a combined cost basis. Hidden orders executing against the Lit order-book would be problematic from the perspective of allowing for predictable price impact; while hidden orders executing against a hidden double can imply the emergence of two interacting price impact curves that may then be linked through induced order-flow mechanics rather than price impact itself. 

At this stage, because of the low liquidity and intermittent price discovery, A2X may well be better suited for: 1.) facilitation trading where large pension funds are trying to reduce the burden of cost on fund holders when an institutional buyer and seller are known to each other and are merely crossing inventory, and 2.) opportunistic day-traders playing with very small volumes. The cost advantage could be meaningful for both agents. However, for typical Lit Market trading, the JSE has substantially better and more predictable liquidity. Hence for day-to-day pension fund re-balancing, given the prevailing cost structures and the liquidity, fund re-balancing is probably best left to execution during the daily end of day closing auctions on the JSE. 

\section*{Reproducibility of the Research}
Test data compromising of market data messages from A2X and vendor datasets from Bloomberg Pro are provided for replication and algorithm verification \cite{data}. The Julia code and instructions for verification can be found in our GitHub site \cite{code}.

\section*{Acknowledgements}
The authors would like to thank Aveshen Pillay, Dieter Hendricks, Alexa Orton and Etienne Pienaar for helpful comments and critique. The historic snap-shot of part of the A2X transaction message data was kindly provided to us by A2X. Patrick Chang would like to acknowledge the support of the Manuel \& Luby Washkansky Scholarship and the South African Statistical Association [grant number 127931]. 

\balance


\onecolumn
\appendix
\section{A2X data cleaning}\label{app:A2X}

We discuss the workflow to process raw file message data into Trade and Quote (TAQ) data files for analysis using historical market data from the A2X exchange.
The A2X data set comprises trade history data where the A2X market data servers monitor trading activity on the system and convert these events into market data messages.

We are given raw A2X market data from the continuous trading sessions for the period 2017-09-20 - 2020-03-18. The data we have is the real time market data feed. We do not have messages from the snapshot feed or from the replay service (\emph{msgType}: 10 -- 15). The replay service has no bearing on the work we want to do, but the lack of snapshot feed means that we do not have information regarding the opening and closing auctions. This data is anonymised, so that the messages do not include any information identifying the trading members involved.

The raw data comes in the form of zipped files separated by date which are comprised of lines of string messages containing all the information about all securities that were active during that day ordered by time. Messages fall under defined categories, each with different fields, based on the specific event that occurred---not all of which are important for our purposes. An example of the format of these files is shown in the snippet below.

\begin{Verbatim}[fontsize=\tiny]
    1583827086138193000 (2020-03-10T07:58:06.138193) 1
    32000 MdOrderCancel:{MdHeader:{msgType:3,length:20,seqNo:204774},OrderCancel:{securityId:24,orderId:92564,timestamp:1583827086138161000}}
    1583827086138422000 (2020-03-10T07:58:06.138422) 1
    35000 MdOrderModify:{MdHeader:{msgType:4,length:32,seqNo:204775},OrderModify:{securityId:24,quantity:240,price:2497000000,orderId:28123,timestamp:1583827086138387000}}
    1583827086155788000 (2020-03-10T07:58:06.155788) 1
    34000 MdOrderModify:{MdHeader:{msgType:4,length:32,seqNo:204776},OrderModify:{securityId:24,quantity:240,price:2512500000,orderId:28110,timestamp:1583827086155754000}}
    1583827086179468000 (2020-03-10T07:58:06.179468) 1
    33000 MdOrderAdd:{MdHeader:{msgType:2,length:33,seqNo:204777},OrderAdd:{securityId:24,side:SELL,quantity:191,limitPrice:2507400000,orderId:92575,timestamp:1583827086179435000}}
    1583827086181382000 (2020-03-10T07:58:06.181382) 1
    39000 MdOrderCancel:{MdHeader:{msgType:3,length:20,seqNo:204778},OrderCancel:{securityId:24,orderId:92567,timestamp:1583827086181343000}}
    1583827086290870000 (2020-03-10T07:58:06.290870) 1
    46000 MdOrderAdd:{MdHeader:{msgType:2,length:33,seqNo:204779},OrderAdd:{securityId:21,side:SELL,quantity:203,limitPrice:1481700000,orderId:92576,timestamp:1583827086290824000}}
\end{Verbatim}

In exploring this dataset we are first required to clean and compact the raw data files into trade and quote data. The data cleaning process is divided into three phases and is demonstrated in what follows.

\subsection*{Phase 1: Converting data into usable messages} \label{subsec:phase1}
In terms of the structure and format of individual messages, timestamps are given in Unix time, that is, the nanoseconds since the Unix epoch 00:00:00 UTC 1970-01-01. To obtain South African time we use UTC + 2 hours. This time then corresponds to A2X's continuous trading hours between 09:00 and 16:50. The categories under which each message falls are defined by the field \emph{msgType} taking on integer values from 1 to 15 corresponding to the different message types. The count of these message packets for each day for the period between 2019-01-01 and 2019-07-15 are shown in \Cref{table:frequencies}. Considering the information pertaining the securities and the market itself, at the start of each day, before market open, tick-table information (\emph{msgType}: 7) is published on the feed to identify the securities that are traded on A2X and the tick size for which orders can be placed. Similarly, a security definition message (\emph{msgType}: 8) is published before market open, providing relevant information regarding each security's Id and currency. Additionally, a security status message (\emph{msgType}: 9) may be published when the trading status (for example, ``active'', ``suspended'' or ``halted''), or the trading session (for example, ``Closed'', ``Continuous trading open'' or ``Continuous trading closed'') of a particular security has changed. The above messages do not play a particular important role when extracting the useful information but give context to how the data feed is structured.

\begin{table}[!h]
    \setlength{\tabcolsep}{3pt}
    \centering
    \small
    \begin{tabular}{lccccccc} \toprule
         & Order & Order & Order & Trade & Tick & Security & Security \\
         & add & cancel & modify & & table data & definition & status \\ \midrule
         Count & 4,435,738 (20.65\%) & 4,428,798 (20.62\%) & 12,593,071 (58.62\%) & 12,438 (0.06\%) & 193 (0\%) & 3,653 (0.02\%) & 7,642 (0.04\%) \\ \bottomrule
    \end{tabular}
    \caption{Total event frequencies from the raw market data feed for the period between 2019-01-01 and 2019-07-15.}
    \label{table:frequencies}
\end{table}

The important messages which are relevant to us are \emph{msgType}: 2 - 6 and their fields are summarised in \Cref{table:messages} below. Blank/heartbeat messages contain no information and are published in one second intervals when no event occurs. Note that there is an additional field for added market orders, namely ``tradeType'', which indicates if a trade was executed against a visible order quantity or hidden order quantity. We ignore the hidden order book (dark pool). Order and Trade reference numbers are assigned by A2X and are unique for the day. Note that a particular order reference can appear multiple times on the market data stream and always represents the same order within the trading system.

\setlength{\tabcolsep}{0.5em}
\begin{table}[!h]
    \centering
    \small
    \begin{tabular}{lcccccccc} \toprule
         Event & msgType & timestamp & securityId & orderRef & price & quantity & side & tradeRef  \\
         & $\langle$ \textit{Int} $\rangle$ & $\langle$ \textit{Int}, \textit{Unix} $\rangle$ & $\langle$ \textit{Int} $\rangle$ & $\langle$ \textit{Int} $\rangle$ & $\langle$ \textit{Int}, \textit{ZAC}$\cdot 10^5$$\rangle$ & $\langle$ \textit{Int} $\rangle$ & $\langle$ \textit{String} $\rangle$ & $\langle$ \textit{Int} $\rangle$ \\ \midrule
         Add LO & 2 & \checkmark & \checkmark & \checkmark & \checkmark & \checkmark & \checkmark & - \\
         Cancel LO & 3 & \checkmark & \checkmark & \checkmark & - & - & - & - \\
         Modify LO & 4 & \checkmark & \checkmark & \checkmark & \checkmark & \checkmark & - & - \\
         Add MO (trade) & 5 & \checkmark & \checkmark & \checkmark & \checkmark & \checkmark & & \checkmark \\
         Cancel MO (trade bust) & 6 & \checkmark & \checkmark & - & \checkmark & \checkmark & - & \checkmark \\ \bottomrule
    \end{tabular}
    \caption{Relevant fields of market data messages.}
    \label{table:messages}
\end{table}

The procedure to convert the raw messages from each day into usable messages is described in Algorithm~\ref{algo:phase1}. This procedure creates a dataframe of cleaned and usable messages for each security over the entire period under consideration.

One needs to be careful of the fact that A2X is a new exchange with new securities being listed or de-listed. Therefore, creating a dataframe for the entire history of each security must be done with care so as to ensure that the exact date at which securities become active is taken into account.

\begin{algorithm}[!h]
    \small
    \SetAlgoLined
    \DontPrintSemicolon
    \KwIn{Raw data feed}
    \KwResult{Usable messages for the entire history}
    Initialise a master dictionary to separate and hold the order book data of each security for the period 2017-09-20 to 2020-03-18\;
    \For{each day in 2017-09-20 to 2020-03-18}{
        Remove message type 1 and the redundant information regarding the time of the message (reducing the data set 1/200 th of its original size).\;
        Find the active securities in this day\;
        Initialise a dataframe format for all relevant events and all securities\;
        \For{each line in data}{
            Extract the fields in each line and reformat\;
            Push to dataframe\;
        }
        Initialise a dictionary of dataframes with keys corresponding to all active securities\;
        \For{each security in all active securities}{
            Separate all securities and place them in the dictionary positions which corresponds to each security's key\;
        }
        Concatenate each security's cleaned messages data for this day to the master dictionary\;
    }
    
    \caption{Cleaning the raw market data messages.}
    \label{algo:phase1}
\end{algorithm}

This process has converted the raw message feed into a dictionary of dataframes containing easy-to-use message data. The fact that we have the entire continuous trading feed means that we could potentially construct the entire order-book. For the sake of comparison against JSE L1LOB data, we only focus on building the top-of-book.

\subsection*{Phase 2: Building the top-of-book}\label{subsec:phase2}

\begin{algorithm}[!h]
    \small
    \SetAlgoLined
    \DontPrintSemicolon
    \KwIn{Cleaned message data}
    \KwResult{L1LOB file for the entire history}
    Initialise  a master dictionary to seperate and hold the level 1 order book data of each security\;
    \For{all securities}{
        \For{each day in the entire history}{
            Extract the day's cleaned message data\;
            Initialise a bid and ask dictionary keeping track of all active bid and ask orders with keys corresponding to the order reference numbers\;
            \For{each line in cleaned message data of the day}{
                \If{Order add}{
                    \eIf{SELL}{
                        Obtain the current best ask\;
                        Add the order to the ask side dictionary\;
                        Compare the new order against the best ask. If the new order is better than the best ask then update the L1 order book. Otherwise the L1 order book does not update\;
                    }{
                        Obtain the current best bid\;
                        Add the order to the bid side dictionary\;
                        Compare the new order against the best bid. If the new order is better than the best bid then update the L1 order book. Otherwise the L1 order book does not update\;
                    }
                }
                \If{Order cancel}{
                    Check if the order reference to be cancelled is from the bid or ask dictionary\;
                    Check if the order to cancel is the best bid or ask\;
                    If the order to cancel is not the best, remove the order from the bid dictionary and keep the best as is\;
                    If the order to cancel is the best, check if there are more orders in the book. If there are more orders in the book then find the new best and update the L1 order book. Otherwise, if there are no more orders in the book then update the new best as NaN\;
                    
                }
                \If{Order modify}{
                    Obtain the order reference that needs to be modified and the appropriate price level and volume for modification\;
                    Check if the order reference to be modified is from the bid or ask dictionary\;
                    Check if the order to be modified is the best bid or ask\;
                    If the order to be modified is not the best, modify the order and compare it against the best. If the order to be modified is better than the best then update the L1 order book. Otherwise the L1 order book does not update\;
                    If the order to be modified is the best, modify the order, get the new best and update the L1 order book\;
                }
                \If{Market order}{
                    Check if the order reference is from the bid or ask dictionary. Then classify the trade appropriately and update the trade in the L1 order book\;
                    Compute the volume left in either the best bid or ask. Specifically, the quantity in the order reference minus the trade quantity\;
                    If the volume is non-zero then modify the quantity in the best bid or ask and update the best bid or ask appropriately\;
                    If the volume is zero then remove the best bid or ask from the dictionary and update the best bid or ask appropriately with either a new best bid or ask (post trade), or NaN if there are no more bids or asks left in the dictionary (post trade).\footnote{The trades larger than the quantity of the best on offer gets split into two trades, thus we need not worry about adjusting the value of deeper orders from a single trade within a single logic check.}\;
                }
            }
        }
    }
    \caption{Building the top-of-book.}
    \label{algo:phase2}
\end{algorithm}

Once we have converted the raw message feed into usable messages in a data-frame format, the next step is to use these messages to create the order book. Note that because we have every message of the order book, this means that we can actually create the full market depth which is the entire order book with every price level and quantity for a bid and ask that has been submitted. Creating this full market depth is not difficult, the difficult aspect of it is to find an appropriate format to store the information. Here our focus on building the top-of-book. The top-of-book is also known as the level 1 (L1) order book. It is a real-time stream of the best bid, best ask, trades and each of their respective volume. 

There are some key differences when building the top-of-book versus the full market depth. The full market depth gets updated with every message submitted. This is because each of the messages modifies the order book in some shape or form, thus warranting the order book to be updated. In contrast, the top-of-book will only get updated whenever an event affects the current best bid or ask in any shape of form. Therefore, as an example, if a new bid order arrives below the best bid, the full market depth will be updated while the top-of-book will have no update because the new bid does not affect the best bid.

After some preliminary checking, sometimes after an order gets cancelled, one side of the order book can be left completely empty with no orders. When this is the case, the resulting best bid or ask (depending on the side of the order book) will have a price level and volume of NaN. It is important to print these events as well. This is because the top-of-book is a real-time feed of the best bid or ask currently on offer. Thus if one side of the book is empty, then there cannot be a best bid or ask on offer. 

The trick to build the top-of-book is to use dictionaries. We create a dictionary for the bids and a separate dictionary for the asks. Since each order has an associated order reference, this can be used as the key for the dictionary. This allows us to keep track of which order to modify or cancel. Moreover, each trade has an associated order reference as well. This means we know the true trade sign of the trade. If the trade matches the order reference in the ask dictionary then the trade was a buyer initiated trade; likewise if the trade matches the order reference in the bid dictionary then the trade was a seller initiated trade. With this in place, we just need to loop through each message and modify either the bid or ask dictionary, perform some logic checks and update the order book if needed. The various logic checks are presented in Algorithm~\ref{algo:phase2}.

An aspect that makes dealing with trade messages slightly more challenging than the rest is that a trade requires two updates to the top-of-book. First is the trade itself, but second is the impact a trade has on the best bid or ask on offer. Since a trade will alter either the best bid or ask on offer (depending on the trade sign), the appropriate best bid or ask needs to be updated as well. Since A2X's order-book is very shallow (see \ref{app:vis}), sometimes a trade can deplete all the orders in one side of the order book. When this is the case, the appropriate best bid or ask must be set to NaN to indicate that there are no best bid or ask on offer.

\subsection*{Phase 3: Useful data from the top-of-book}\label{subsec:phase3}
The final step of the data cleaning is to use the cleaned ordered sequence of Trade and Quote data of the top-of-book to further extract useful information. This includes computing the mid-price, mid-price change, microprice, and trade inter-arrival times. 

Let us first define some useful notations. Let $b_t$ be the best bid and $a_t$ be the best ask at time $t$. Let their associated volume be $v_t^b$ and $v_t^a$ respectively. Moreover, let the transaction price be $p_{ij}$ and the associated volume be $v_{ij}$ where the indices $i$ and $j$ is the $i$th transaction on the $j$th day. Note that each of these $ij$-th event have a one-to-one mapping to calendar time $t_{i,j}$.

We compute the trade inter-arrival be $\tau_{kj} = t_{k+1,j} - t_{k,j}$. Note that the inter-arrivals are not computed across the days, {\it viz.} we do not consider the inter-arrival between first trade of a new day and the last trade of the previous day.

To build the mid-price $m_t$ and microprice $S_t$, we can exploit the previous tick interpolation. This is because the best bid or ask is currently on offer at any given time (even though the its associated event occurred some time ago), this is why we emphasised that NaNs should be reported if one side of the order book is empty. This informs us if there are breaks in the mid-price or microprice due to the order book being empty. Therefore, we compute the mid-price and microprice by:
\begin{enumerate}
    \item Looping through the top-of-book TAQ data, and identify the event type.
    \item If the current event is a bid (ask) then go back to the previous events and identify the latest best ask (bid) and extract their current price and volume on offer, even if it is NaN.
    \item Compute:
    \begin{align*}
        m_{t} = \frac{1}{2} \left( b_{t} + a_{t} \right) \ \ \text{and}   \ \
        S_{t} = \frac{v_{t}^a}{v_{t}^a + v_{t}^b} a_{t} + \frac{v_{t}^b}{v_{t}^a + v_{t}^b} b_{t},
    \end{align*}
 \end{enumerate}
where the mid-price and microprice are piece-wise constant with jumps whenever there is an update to the best bid or ask. Moreover, the mid-price and microprice can have periods with no value if at least one side of the book is empty.

Since A2X has very low liquidity in the order book, we have to make some choices when computing the mid-price change. We compute the mid-price change as:
$$
\Delta p_{t_{k}} = \log\left( m_{t_{k+1}} \right) - \log\left( m_{t_{k}} \right),
$$
where $t_{k}$ is the time of mid-price before the transaction and $t_{k+1}$ is the time of mid-price immediately after the transaction. The problem is that sometimes when a trade occurs it finishes an entire side of the order book which means there are no more bids or asks left in one side of the order book. This means that the mid-price immediately after the transaction is non-existent. To deal with this, we ignore the mid-price changes for these transactions.

Using the methods described above, we can extend the top-of-book to contain additional information. The resulting dataframe to be used hereafter can be seen in \Cref{table:l1lobjse}. Note that for now we have yet to compute the mid-price change. This can be computed at a later stage using the information in \Cref{table:l1lobjse}.

\begin{table}[!h]
    \setlength{\tabcolsep}{2pt}
    \centering
    \scriptsize
    \begin{tabular}{lccccccccccc} \toprule
         TimeStamp & EventType & Bid & BidVol & Ask & AskVol & Trade & TradeVol & TradeSign & MicroPrice & MidPrice & InterArrivals \\ \midrule
         2018-12-27T09:00:01.026 & ASK & NaN & NaN & 2.99499e10 & 85.0 & NaN & NaN & - & NaN & NaN & NaN \\
         2018-12-27T09:00:01.026 & BID & 2.96759e10 & 85.0 & NaN & NaN & NaN & NaN & - & 2.98129e10 & 2.98129e10 & NaN \\
         2018-12-27T09:00:11.024 & BID & 2.75102e10 & 92.0 & NaN & NaN & NaN & NaN & - & 2.86818e10 & 2.873005e10 & NaN \\
         2018-12-27T09:00:11.024 & ASK & NaN & NaN & 2.77892e10 & 92.0 & NaN & NaN & - & 2.76497e10 & 2.76497e10 & NaN \\
         2018-12-27T09:00:29.569 & ASK & NaN & NaN & NaN & NaN & NaN & NaN & - & NaN & NaN & NaN \\
         2018-12-27T09:00:29.569 & BID & 2.96001e10 & 86.0 & NaN & NaN & NaN & NaN & - & NaN & NaN & NaN \\ \bottomrule
    \end{tabular}
    \caption{Snippet of the format of cleaned Naspers L1LOB data.}
    \label{table:l1lobjse}
\end{table}

\section{JSE data cleaning}\label{app:JSE}

We obtained commercial data vendor data-sets from Bloomberg Pro using the University of Cape Town library services. The data is the top-of-book information from the Johannesburg Stock Exchange. The data contains auction information and various trade types that are irrelevant to the analysis. Therefore, majority of the processing in this case is extracting the relevant information for the analysis. \Cref{table:jseraw} provides a snippet of the Naspers TAQ data from Bloomberg Pro.

The data set is from 2018-12-31 to 2019-07-15. We remove 2018-12-31 from our analysis as the trading day only lasts from 09:00 to 12:00 whereas a normal trading day of the JSE is from 09:00 to 16:50 with the closing auction happening between 16:50 and 17:00. Therefore, we only retain the continuous trading information of each day between 09:00 to 16:50. This removes majority of the unwanted trade types such as after-hour trades (LT), correction of previous days published off book trade (LC) and an indicative auction price based on the volume maximising auction algorithm used to determine the auction uncrossing price (IP) \cite{jserules}. The only trade types we want to retain is automated trades (AT).

\begin{table}[!h]
    \setlength{\tabcolsep}{2pt}
    \centering
    \scriptsize
    \begin{tabular}{lcccc} \toprule
         times & type & value & size & condcode \\ \midrule
         2018-12-31T07:00:00.0 & BID & 279278.25 & 400.0 & - \\
         2018-12-31T07:00:00.0 & ASK & 286046.38 & 10.0 & - \\
         2018-12-31T07:00:00.0 & ASK & 286046.38 & 30.0 & - \\
         2018-12-31T08:30:00.0 & TRADE & 0.0,0.0 & IP \\
         2018-12-31T08:32:30.0 & BID & 279278.25 & 413.0 & - \\
         2018-12-31T08:32:30.0 & ASK & 281198.16 & 3.0 & - \\ \bottomrule
    \end{tabular}
    \caption{Snippet of the format of raw Naspers L1LOB data.}
    \label{table:jseraw}
\end{table}

Although the data only contains time stamps up to the second unlike A2X where we had time stamps to the nanosecond, we found that the ordering of events seems to be correct. After careful inspection, we found that every single automated trade is immediately followed by an update in the order book with the appropriate change in volume if the trade did not deplete the current best bid or ask. Therefore, there is no need to worry about same time quote and trades.

Using this filtered top-of-book TAQ data, we further extract useful information such as mid-price, microprice, trade inter-arrivals and trade signs. The features are computed as before with the exception that the trade sign must be inferred through the Lee--Ready classification rule \cite{LR1991}. \Cref{table:JSECleaned} provides a snippet of the processed Naspers TAQ data to be used hereafter with the various additional features computed.

\begin{table}[!h]
    \setlength{\tabcolsep}{2pt}
    \centering
    \scriptsize
    \begin{tabular}{lccccccccccc} \toprule
         TimeStamp & EventType & Bid & BidVol & Ask & AskVol & Trade & TradeVol & MicroPrice & MidPrice & InterArrivals & TradeSign \\ \midrule
         2019-01-02T09:00:00.0 & BID & 279258.84 & 1195.0 & NaN & NaN & NaN & NaN & NaN & NaN & NaN & - \\
         2019-01-02T09:00:04.0 & BID & 279258.84 & 1196.0 & NaN & NaN & NaN & NaN & NaN & NaN & NaN & - \\
         2019-01-02T09:00:35.0 & ASK & NaN & NaN & 261475.48 & 2869.0 & NaN & NaN & 266707.68 & 270367.16 & NaN & - \\
         2019-01-02T09:00:35.0 & BID & 279258.84 & 1193.0 & NaN & NaN & NaN & NaN & 266698.41 & 270367.16 & NaN & - \\
         2019-01-02T09:00:35.0 & ASK & NaN & NaN & 261475.48 & 2868.0 & NaN & NaN & 266699.70 & 270367.16 & NaN & - \\
         2019-01-02T09:00:35.0 & BID & 279258.84 & 1192.0 & NaN & NaN & NaN & NaN & 266696.60 & 270367.16 & NaN & - \\ \bottomrule
    \end{tabular}
    \caption{Snippet of the format of cleaned Naspers L1LOB data. \label{table:JSECleaned}}
\end{table}

\section{Equities considered}\label{app:tickers}

\setlength{\tabcolsep}{1.6em}
\begin{table*}[!h]
\centering
\begin{tabular}{lc|lc} \toprule
\multicolumn{2}{c}{A2X}                           & \multicolumn{2}{c}{JSE}                           \\ \cmidrule{1-2} \cmidrule{3-4}
Security name                     & Security code & Security name                     & Security code \\ \midrule
Aspen Pharmacare Holdings Ltd     & APN           & Absa Group Ltd                    & ABG           \\
African Rainbow Min Ltd           & ARI           & Anglo American Plc                & AGL           \\
AVI Ltd                           & AVI           & British American Tobacco Plc      & BTI           \\
Coronation Fund Managers Ltd      & CML           & FirstRand Ltd                     & FSR           \\
Growthpoint Prop Ltd              & GRT           & Nedbank Group Ltd                 & NED           \\
Mr Price Group Ltd                & MRP           & Naspers Ltd                       & NPN           \\
Naspers Ltd                       & NPN           & Standard Bank Group Ltd           & SBK           \\
Standard Bank Group Ltd           & SBK           & Shoprite Holdings Ltd             & SHP           \\
Sanlam Ltd                        & SLM           & Sanlam Ltd                        & SLM           \\
Santam Ltd                        & SNT           & Sasol Ltd                         & SOL           \\ \bottomrule
\end{tabular}
\caption{Equities considered from A2X and the JSE.}
\label{tab:tickers}
\end{table*}

\Cref{tab:tickers} lists the various equities from each exchange used in the investigation of \Cref{sec:comp}. The data set for all the JSE equities are for the period from 2019-01-02 to 2019-07-15. Therefore, we only look for equities on the A2X with data within the same period. However, quite a few of the equities from A2X which were listed before 2019-01-02 have very little activity, we could not find 10 equities with ``sufficient''\footnote{Quite a few equities which were listed before 2019-01-02 only had kilobytes of data for over a year of trading.} data that all started before 2019-01-02. Therefore, three of the equities we used were only listed between 2019-01-02 and 2019-07-15. These were: Santam listed on 2019-02-01, Aspen Pharmacare listed on 2019-04-01 and Mr Price Group listed on 2019-05-02. For each of these equities, we only use data from the start of their listing until 2019-07-15 when investigating the price impact.

\section{Trade classification} \label{sec:class}

Since we have the true classification of the various transactions on A2X. This gives us a very unique opportunity to test out various trade classification rules and their ability to infer the correct trade sign. To this end, let us consider three types of classification rules: the quote rule \cite{Hasbrouck1988}, the tick rule \cite{MMB1989}, and the Lee--Ready rule \cite{LR1991}.

\citet{EMO2000} provide a nice summary of the various rules. The quote rule classifies the trade in relation to the mid-price. A trade is classified as buyer-initiated (seller-initiated) if the trade price is above (below) the mid-price prior to the trade. Trades which are executed at the mid-price are not classified. The tick rule classifies the trade in relation to previous trades. A trade is classified as buyer-initiated (seller-initiated) if the trade is above (below) the previous trade. If there is no price change from the previous trade then the trade is classified against the last trade that was different using the same logic.\footnote{In the implementation, the first trade of each day is not classified and if we cannot find a last trade that was different then the trade is not classified.} Finally, the \citet{LR1991} rule is a combination of the quote rule and tick rule. \citet{Theissen2001} provides a neat description of the Lee--Ready rule broken into three steps:
\begin{itemize}
    \item[1.] Trades that are above (below) the mid-price are classified as buyer-initiated (seller-initiated).
    \item[2.] Trades that occur at the mid-price but is higher (lower) than the previous trade are classified as buyer-initiated (seller-initiated).
    \item[3.] Trades that occur at a price that equals both the mid-price and previous transaction but is higher (lower) than the last different transaction are classified as buyer-initiated (seller-initiated).
\end{itemize}
Here the first step is just the quote rule while the second and third step is the tick rule.

We apply the various rules to the full history of three securities: Naspers (NPN), Standard Bank (SBK), and Sanlam (SLM). We obtain the number of transactions that occurred, the number of correct classifications and the number of transactions that could not be classified. \Cref{tab:class} reports the percentage of correct classification over the number of classifiable transactions, {\it i.e.} number of transactions that occurred minus the number of transactions that could not be classified. We see that the Lee--Ready rule performs the best followed by the quote rule and then the tick rule. The Lee--Ready rule only marginally outperforms the quote rule because using the quote rule there was only one trade from SBK and SLM that was unable to be classified. This was resolved using steps 2 and 3 from the Lee--Ready rule. 

\begin{table}[!h]
    \centering
    \begin{tabular}{l|ccc} \toprule
             & Quote Rule & Tick Rule & Lee--Ready Rule \\ \midrule
         NPN & 99.41\%    & 51.58\%   & 99.41\% \\
         SBK & 97.10\%    & 45.37\%   & 97.12\% \\
         SLM & 99.14\%    & 48.25\%   & 99.16\% \\ \bottomrule
    \end{tabular}
    \caption{Comparison of classification rules. The table reports the percentage of correct classifications.}
    \label{tab:class}
\end{table}

The quote rule here performs better than what was previously reported in the literature (see \citet{EMO2000} and \citet{Theissen2001} for a discussion on the literature). We conjecture the reason the quote rule performs well here may be attributed towards an accurate recording of the sequence of events in the data set, our data set records events up to the nanosecond which means we have an accurate account of the mid-price prior to the trade. Moreover, when creating the top-of-book TAQ data we update the impact a trade has on the best bid or ask immediately. This means trades in close succession have the accurate mid-price before each trade. Contrast this to the suggestion by \citet{LR1991} to match transactions with mid-prices that were in effect five seconds before the reported transaction time. This was suggested because transaction prices on the New York Stock Exchange were likely to be reported with a short delay.

What is surprising is how poorly the tick rule performs. We conjecture that this is due to the lack of transactions on the A2X market. The tick rule performs well when trades occur together in close succession. However, trades on the A2X market usually occur in small bursts but with long inter-arrivals between these bursts. This allows the mid-price to significantly shift during the long inter-arrivals which can affect the ability of the tick rule to classify trades. As an example, suppose the current trade price is at R110.00, if the gap between the next trade is long then the mid-price could move to say R100.00. Now if a buyer-initiated trade occurs and results in a transaction price of R101.00, the tick rule would classify this as a seller-initiated trade when in fact it is a buyer-initiated trade. These occurrences could be a possible explanation as to why the tick rule performs so poorly.

\section{Shallowness of A2X order book} \label{app:vis}

We look at the low levels of liquidity in the order books of A2X by visualising the full order book and investigating what percentage of transactions break the order book {\it viz.} the transactions that deplete all the orders in one side of the order book causing there to be no bids or asks on offer.

\begin{figure*}[htb]
    \centering
    \subfloat[Naspers from 13:50 to 14:00]{\label{FMD:a}\includegraphics[width=0.48\textwidth]{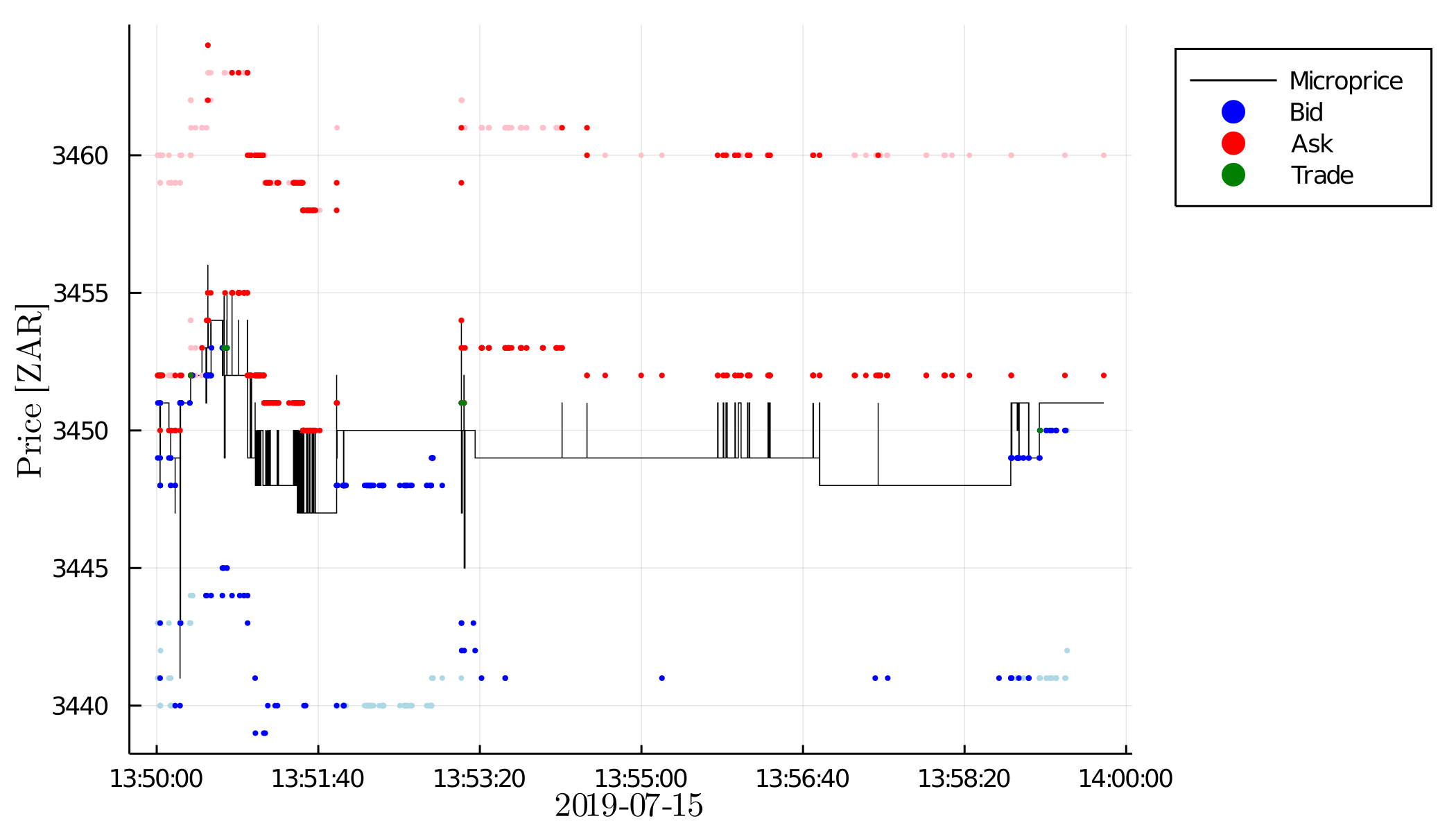}}
    \subfloat[Naspers from 14:30 to 14:40]{\label{FMD:b}\includegraphics[width=0.48\textwidth]{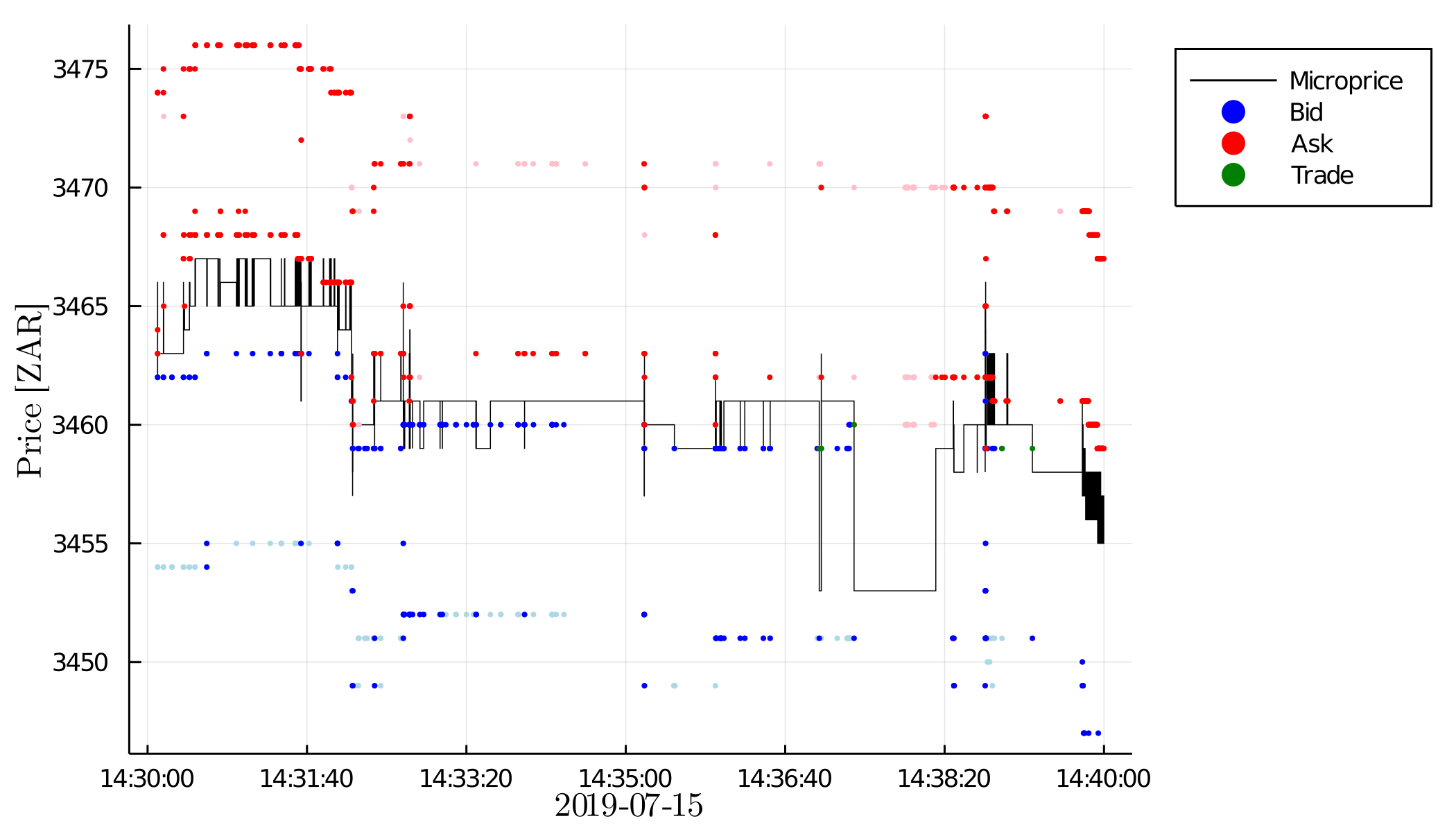}}
    \caption{Visualising the full market depth for Naspers on 2019-07-15 between (a) 13:50 to 14:00 and (b) 14:30 to 14:40. As per the figure legend, the black line is the microprice, the blue bubbles are the bids, the red bubbles are the asks, and the green bubbles are the trades. The best bids and asks are the deeper blue and red bubbles respectively, and the remaining bids and asks in the order book are in lighter blue and red bubbles respectively.}
\label{fig:FMD}
\end{figure*}

Since we have all the messages regarding the order book, this gives us a unique opportunity to visualise the full market depth and not just the top-of-book. However, finding a method to store the full order book is difficult, therefore the solution is to plot the content of the bid and ask dictionary as we are looping through the messages. To this end, the appropriate dictionary is enumerated into the plot whenever any event updates the dictionary. This includes order add, order modify, order cancel and trade. Contrast this to the top-of-book where the best bid and ask are only updated when an event affects and changes the best on offer. 

\Cref{fig:FMD} visualises the full market depth for Naspers on 2019-07-15 between (a) 13:50 to 14:00 and (b) 14:30 to 14:40. We visualise the best bid and ask using the deeper blue and red bubbles respectively. The remaining bids and asks in the order book are in lighter blue and red bubbles respectively. We see that the most liquid ticker on the exchange (Naspers) often only has the best bid or ask on offer and at other times has at most two to three layers on either side of the order book at much deeper levels. This means that if an incoming market order depletes the best bid or ask on offer then there will be a significant change in mid-price which results in a large impact.

\begin{table}[!h]
    \setlength{\tabcolsep}{4pt}
    \small
    \centering
    \begin{tabular}{l|cccccccccc} \toprule
           Security & NPN & SLM & SNT & GRT & CML & APN & AVI & MRP & ARI & SBK\\ \midrule
         Percentage & 13.81\% & 11.16\% & 32.74\% & 13.22\% & 16.7\% & 28.66\% & 25.92\% & 19.24\% & 13.72\% & 15.86\% \\ \bottomrule
    \end{tabular}
    \caption{Percentage of trades that consume the entirety of either the bid or ask side of the order book.}
    \label{tab:aggressive}
\end{table}

As measure of the low levels of liquidity in the order books of A2X, we look at the percentage of transactions that break the order book. These are transactions that consume the entirety of either the bid or ask side of the order book. Transactions can only break an order book if there is only the best bid or ask on offer and the transaction consumes all of it, thus leaving one side of the order book empty.

\Cref{tab:aggressive} reports the percentage of transactions from each security on A2X that break the order book between the period of 2019-01-02 to 2019-07-15. We see that all securities have at least 10\% of their transactions breaking the order book, with some securities even having 32.74\% of transactions breaking the order book. This highlights the low liquidity levels on the order books of A2X. Moreover, the volume of the transactions are not particularly large. We saw in \Cref{fig:Direct} that the largest transactions on A2X in terms of volume were not large enough to reach the ceiling limit in the direct costs.

\section{Miscellaneous comparisons} \label{app:comp}

\subsection*{Order-flow} \label{sssec:taq}

\begin{figure*}[!h]
    \centering
    \subfloat[A2X]{\label{fig:OrderFlow:a}\includegraphics[width=0.48\textwidth]{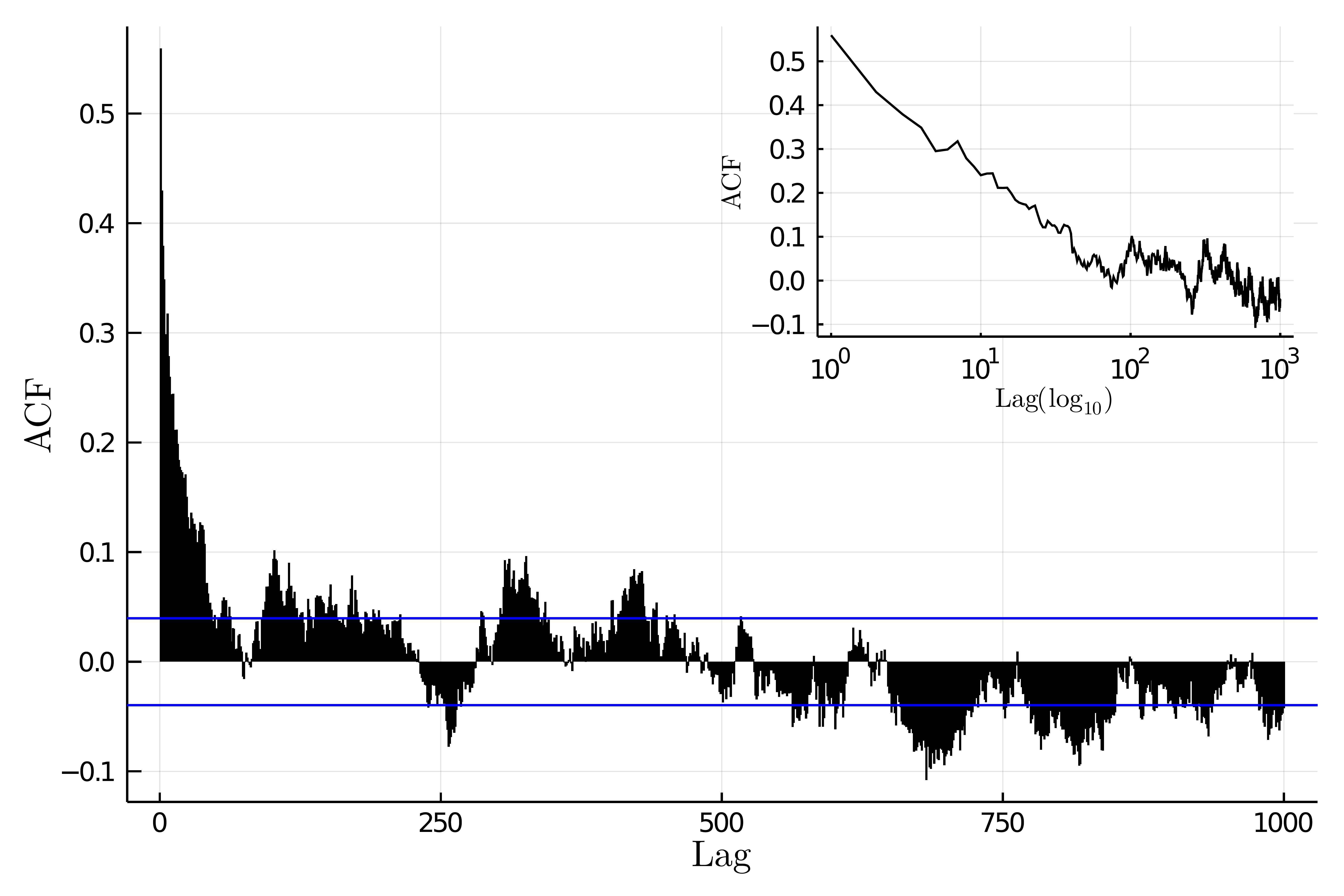}}
    \subfloat[JSE]{\label{fig:OrderFlow:b}\includegraphics[width=0.48\textwidth]{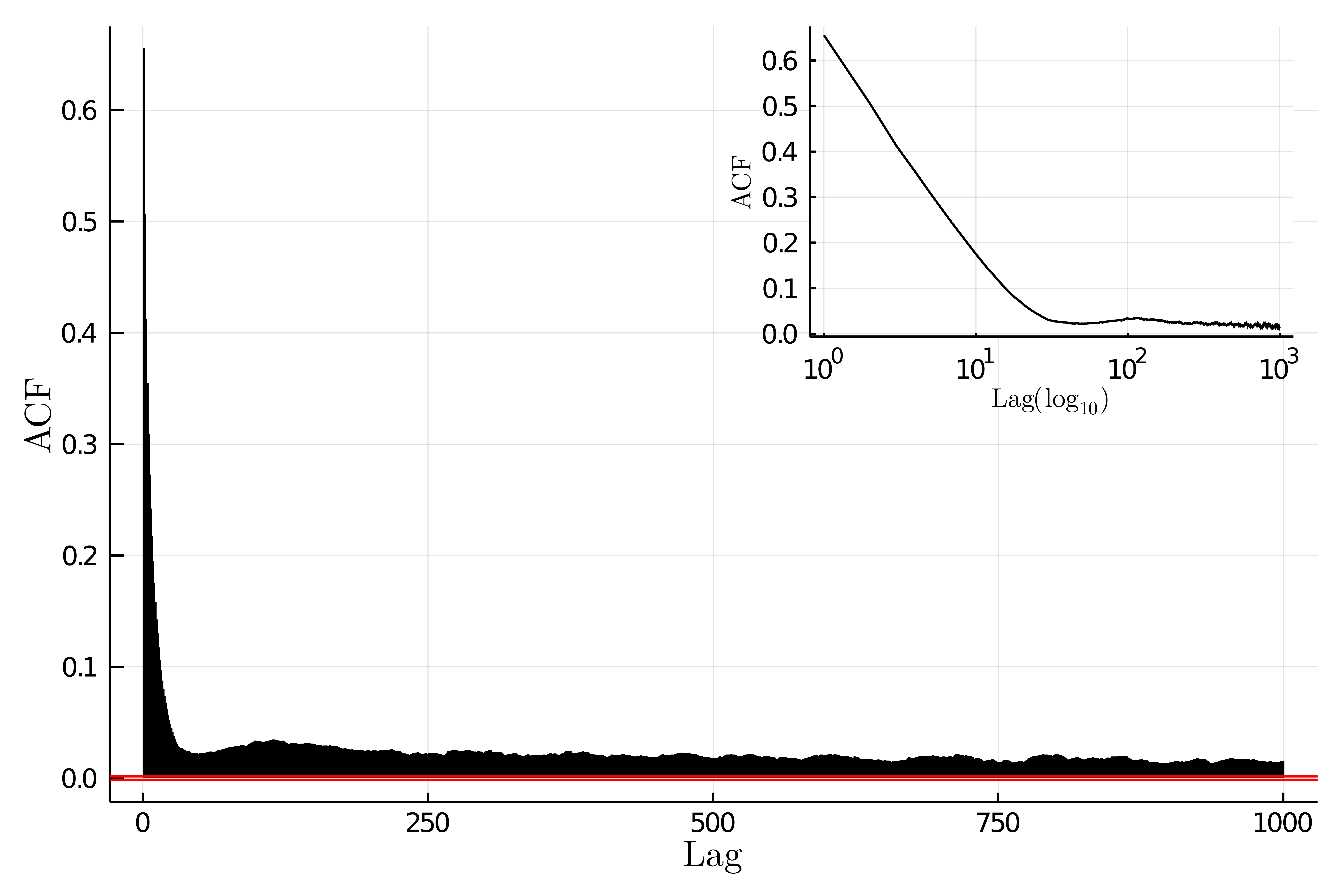}} 
    \caption{Order-flow auto-correlation for Naspers on the (a) A2X exchange and the (b) Johannesburg Stock Exchange. The trade classification is performed using the Lee--Ready classification and we compute up to 1,000 lags between the trades. Provided as insets are the same auto-correlation plots with the lags presented on a $\log_{10}$ scale.}
\label{fig:OrderFlow}
\end{figure*}

One of the high frequency stylised facts is that the order-flow in equity markets are persistent. This has been observed in many different markets (see \citet{TPLF2015} and the references therein). The order-flow is the process defined assuming a value of $+1$ for buyer initiated trades and $-1$ for seller initiated trades \cite{TPLF2015}. Persistence in the order-flow is in the sense that buy order tend to be followed by more buy orders and sell orders tend to be followed by more sell orders \cite{TPLF2015}. Possible causes for this are herding behaviours (positive correlation between the behaviour of different investors) and order splitting (positive auto-correlation in the behaviour of single investors) as pointed out in \citet{TPLF2015}. They were able to further demonstrate on time scales less than a few hours that the main cause is due to order splitting rather than herding. We however cannot investigate this further because our data sets do not contain membership identifiers, thus we cannot determine which member initiated which trade. What we can do is recover the stylised facts.

\Cref{fig:OrderFlow} plots the order-flow auto-correlations for Naspers on the (a) A2X exchange and the (b) Johannesburg Stock Exchange. The trade classification is performed using the Lee--Ready classification \cite{LR1991} and we compute up to 1,000 lags between the trades. Furthermore, the same auto-correlation plots with lags presented on a $\log_{10}$ scale are provided as insets.

We must highlight the difference in liquidity between the exchanges. Over the same 133 tradings days for which the auto-correlations are computed, JSE had a total of 1,456,553 trades whereas A2X had a total of 2,447 trades for the same equity. Assuming that the number of trades are uniform across the days then that means JSE has an average of 10,951 trades per day whereas A2X has an average of 18 trades per day. This is particularly important when interpreting the auto-correlation as 1,000 lags is around 54 days of trading on the A2X and around 43 minutes of trading on the JSE. Nonetheless, we see that both exchanges exhibit a persistent order-flow.

\subsection*{Intraday seasonality} \label{ssec:calendar}

\begin{figure*}[!h]
    \centering
    \subfloat[JSE normalised transaction volume]{\label{Seasonality:a}\includegraphics[width=0.32\textwidth]{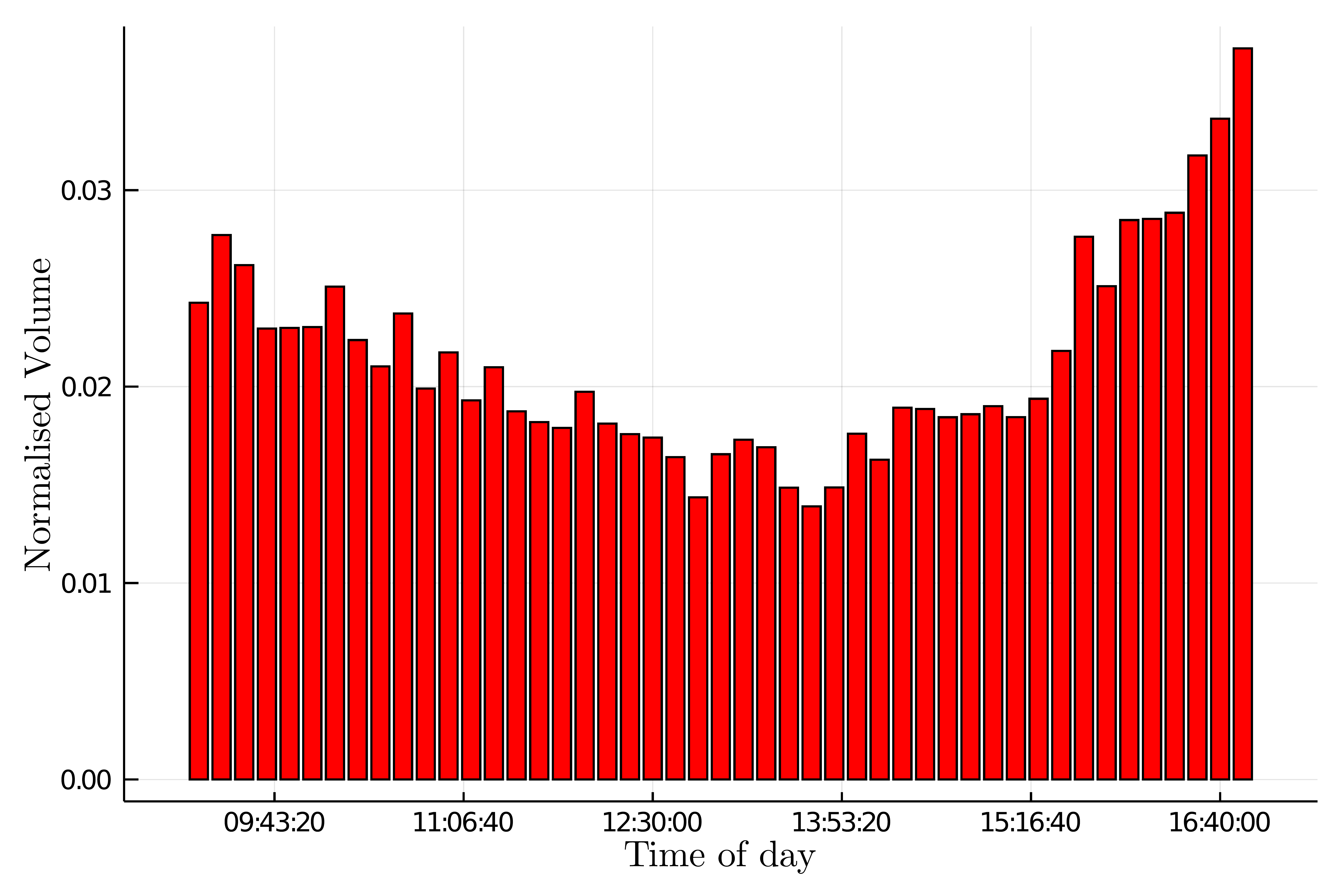}}
    \subfloat[JSE normalised absolute returns]{\label{Seasonality:b}\includegraphics[width=0.32\textwidth]{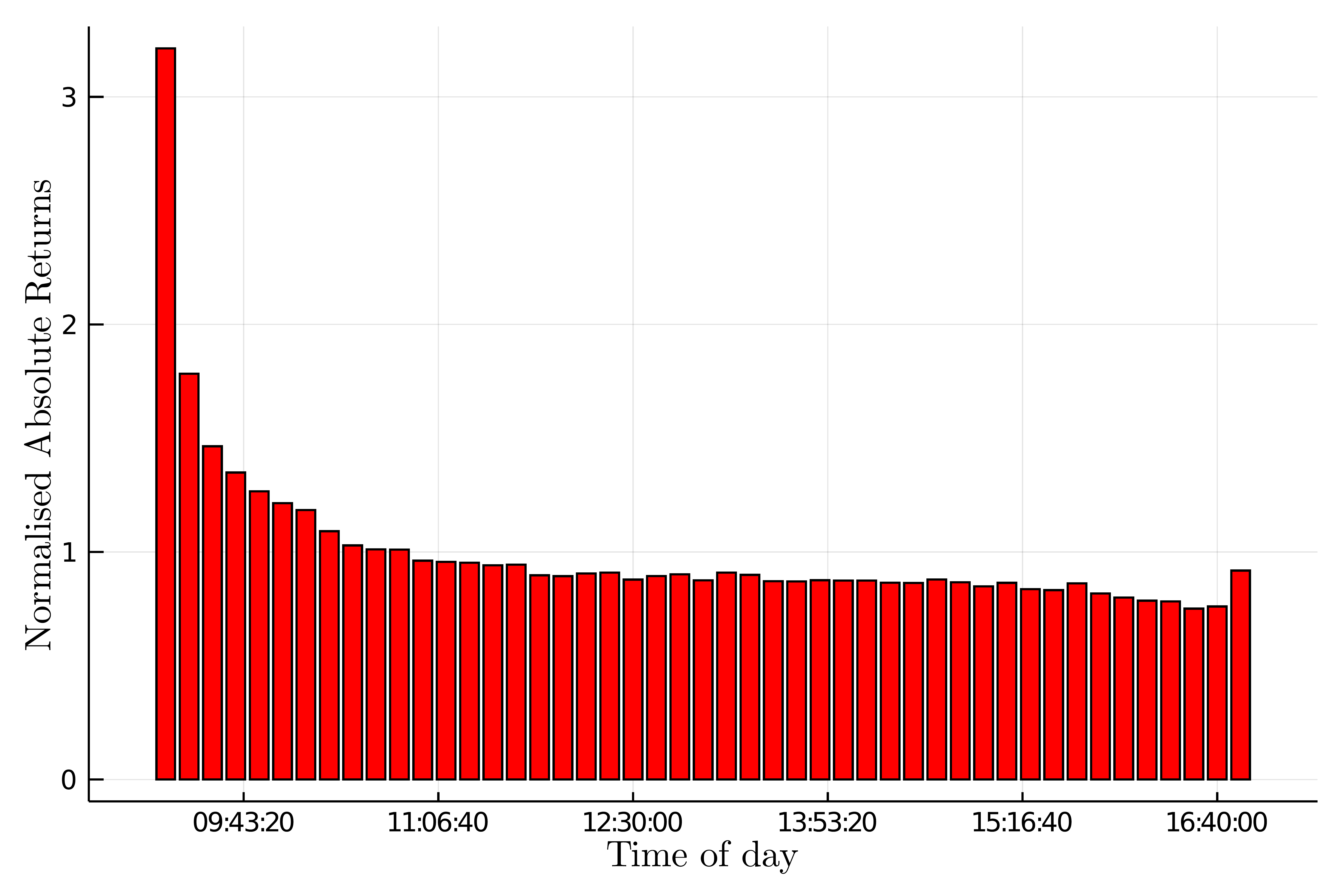}}
    \subfloat[JSE normalised spread]{\label{Seasonality:c}\includegraphics[width=0.32\textwidth]{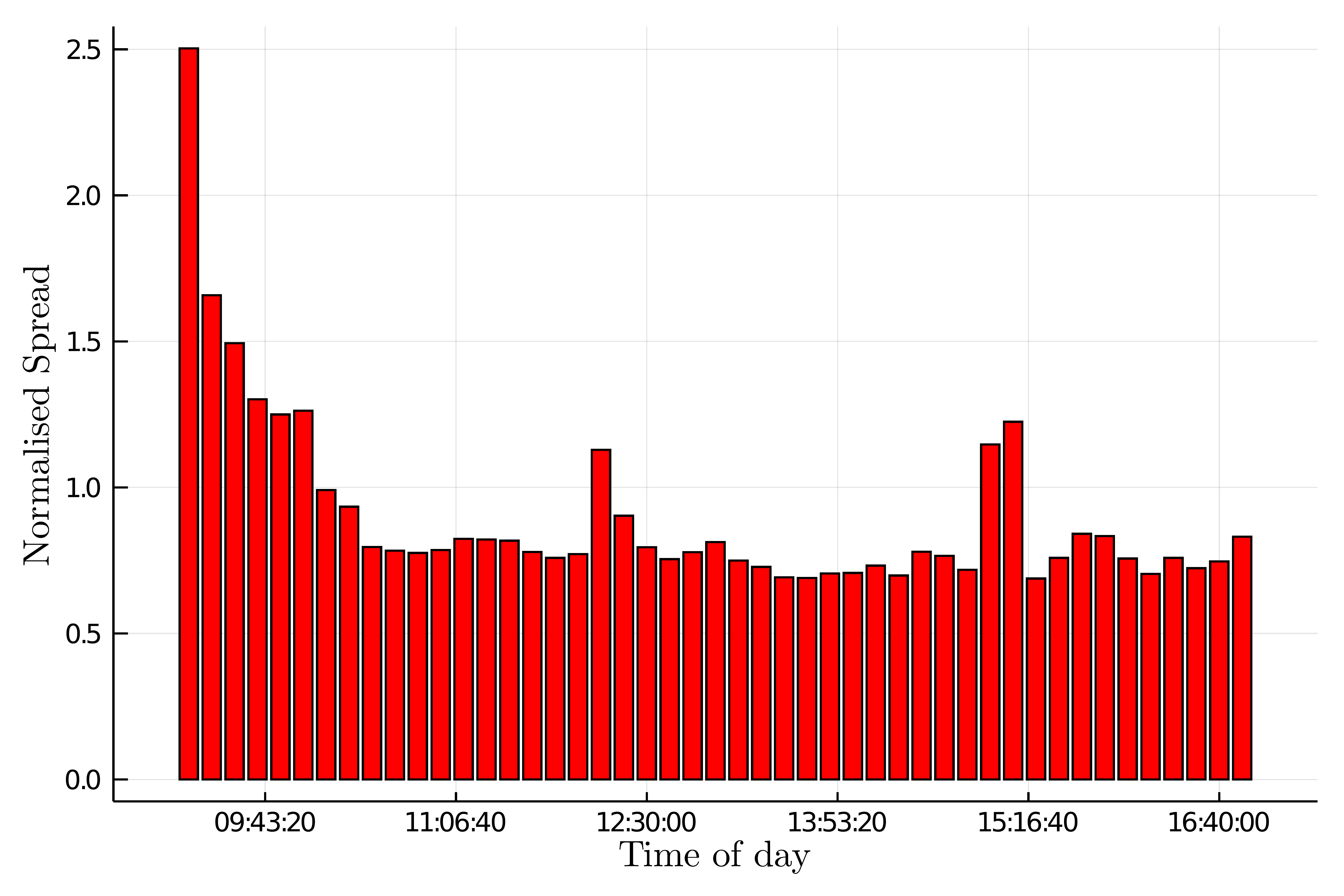}}  \\
    \subfloat[A2X normalised transaction volume]{\label{Seasonality:d}\includegraphics[width=0.32\textwidth]{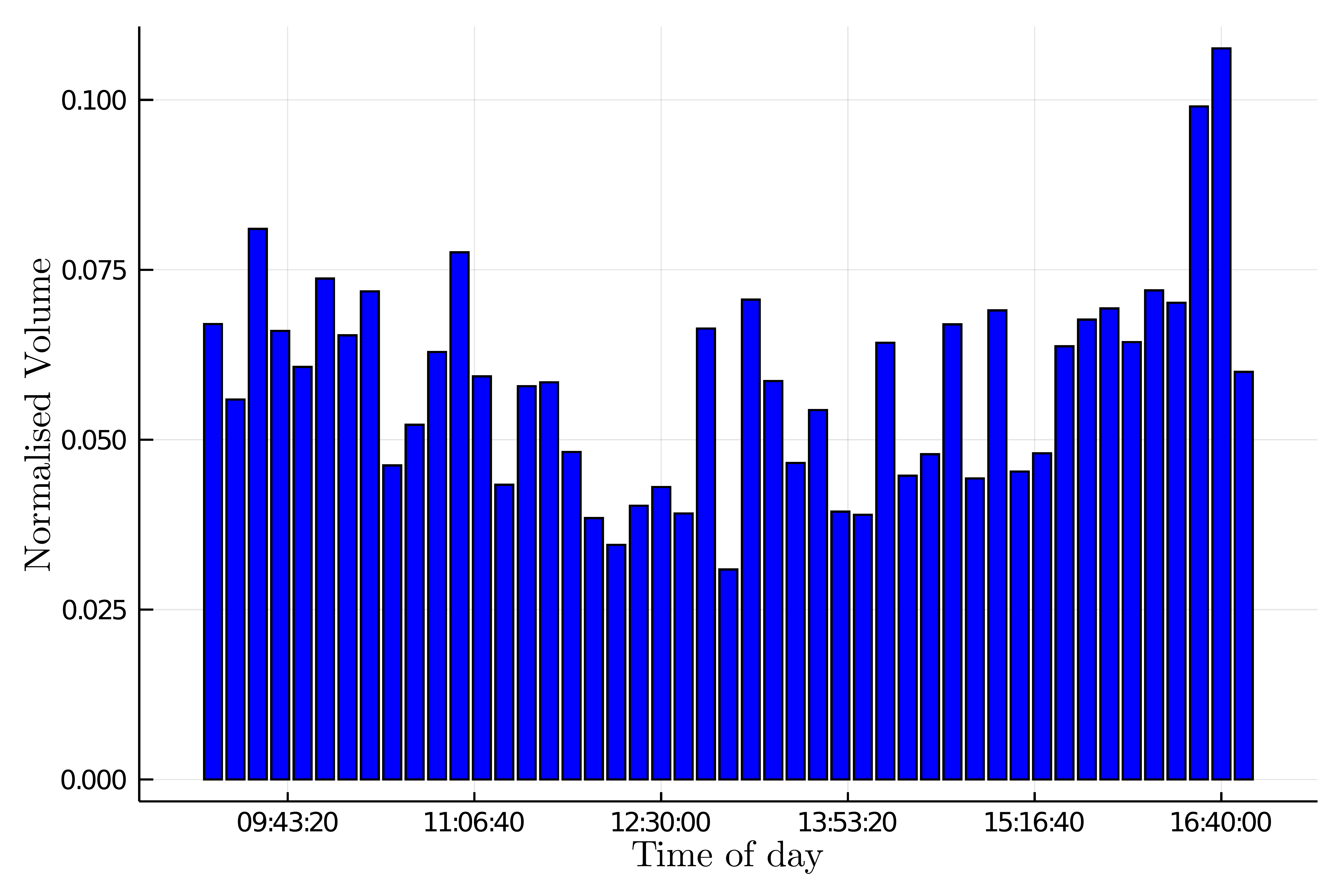}}
    \subfloat[A2X normalised absolute returns]{\label{Seasonality:e}\includegraphics[width=0.32\textwidth]{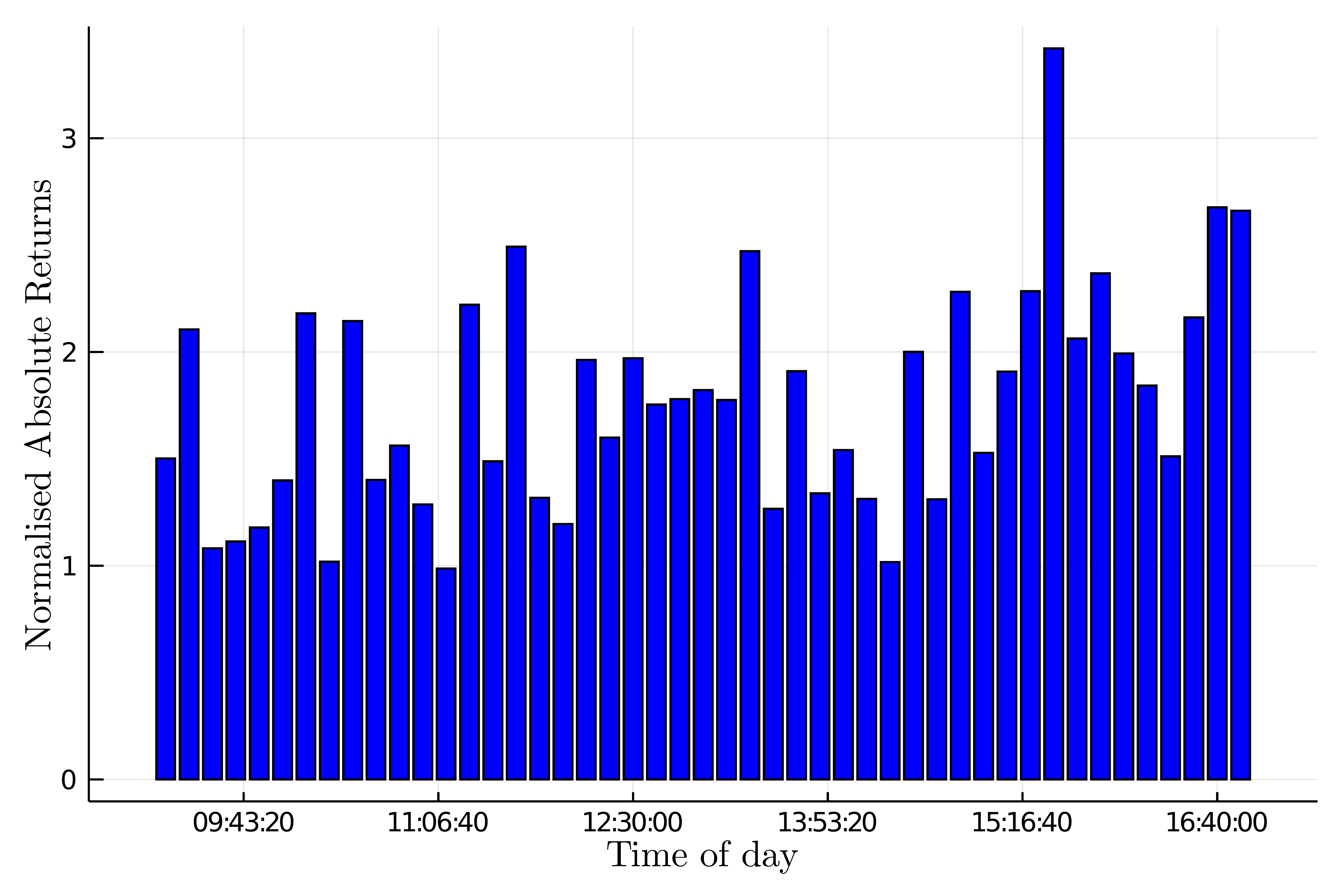}}
    \subfloat[A2X normalised spread]{\label{Seasonality:f}\includegraphics[width=0.32\textwidth]{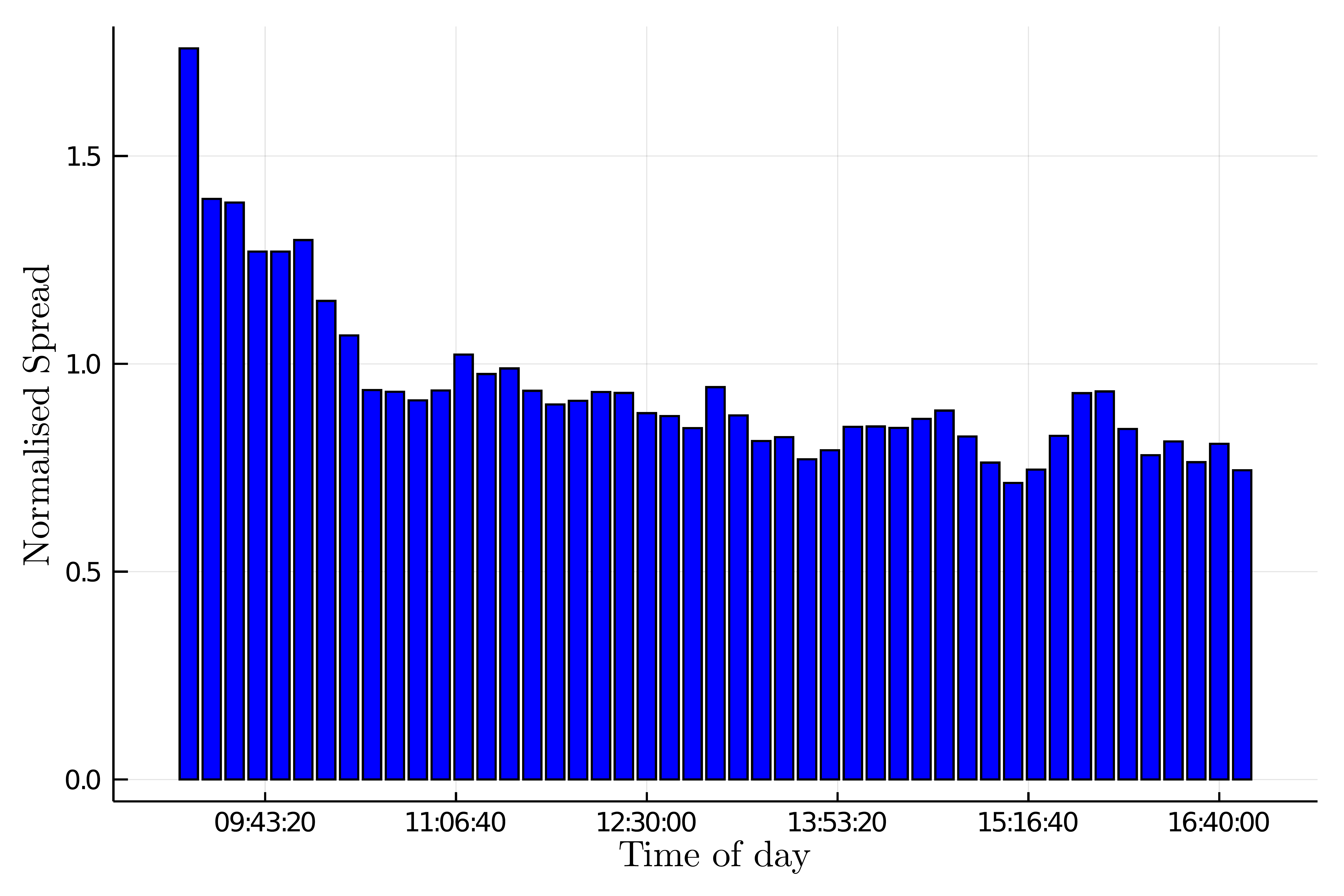}}
    \caption{Intraday seasonality as aggregated across 10 securities from each exchange.}
    \label{fig:Seasonality}
\end{figure*}

We compare the intraday seasonality of volume, returns (as a proxy for volatility) and spreads for all securities in each exchange. The process of creating the seasonality curves in \Cref{fig:Seasonality} is as follows: we break each day into 10-minute buckets. For each bucket and each day, normalise the volume/return/spread by the total daily volume/return/spread and then get the average. Thereafter, average the volume/return/spread within each bin across all days (so averaging is done two-fold) and plot the result as a function of time of day.

More important, however, is how individual seasonality curves are constructed for each market---that is, how volume/returns/spreads are aggregated across securities in each exchange. Returns are handled by computing the lag-1 difference of the log consecutive transaction prices of all securities and grouping them all together (this is done due to the low number of transactions on A2X, to utilise more transactions). Thereafter, the seasonality curves are constructed according to the process above on the combined returns as normalised by the total average daily volume traded.

Spread curves are handled in a similar way in that tick-by-tick spreads are first computed and combined across securities and then normalised and averaged within each bin. When it comes to the volume curves, we deviate from these two processes: 10-minute bars are first computed for each security in each exchange and are then combined. To account for the significant difference in volume from each security, we take a weighted average of transaction volumes in each bin as weighted by the number of trades in each bin.\footnote{The approach taken is slightly unconventional as we are trying to investigate the seasonality of the entire exchange. This is usually only done for one asset.}

In past empirical studies, it has been found that intraday volumes generally have a U-shape or volume smile \cite{PCJ2015}. This pattern is obvious for the JSE but less so for A2X. We see that the average normalised volume for all stocks is large at the beginning of the day, and it gradually slows down until around 13:00, at which time there is a small surge in activity. The 13:00 surge slowly builds up and accelerates during the last half hour of the trading day, peaking at the close. On the other hand, the A2X exchange exhibits random peaks throughout the day with the characteristic surge at the close.

In terms of return seasonality, we use it as a proxy for volatility. In past empirical studies, a left-slanted intraday volatility smile is found \cite{PCJ2015}. The pattern is seen for the JSE but not for A2X. Rather, volatility is found to peak randomly with the largest peaks occurring at the close.

Spreads measure the execution costs of small transactions by measuring how close the price of a trade is to the market price. Here the market price is the mid-price and the quote spread is calculated as the difference between the ask and bid ($s_t = a_t - b_t$) \cite{PCJ2015}. In some cases we found negative spreads, thus we take the absolute value of the spreads. We see that for both markets, quoted spreads are initially high and decline rapidly during the first half-hour of trading. They become mostly constant throughout the remainder of the day (with the exception of two spikes for all securities on the JSE). The intraday spread for both exchanges are consistent with what is found in \citet{PCJ2015}.

In general, we see that the JSE recovers the stylised facts of intraday seasonality whereas A2X only recovers the stylised facts around volume and spread.

\end{document}